\documentclass[a4paper,fleqn,10pt]{article}
\usepackage{amsmath}
\usepackage{amsfonts}
\usepackage{graphicx}
\graphicspath{{./plots/}}
\usepackage[merge,numbers,compress]{natbib}
\usepackage{booktabs}
\usepackage{array}
\usepackage{xcolor}
\usepackage{todonotes}
\usepackage{xspace}
\usepackage{dcolumn}
\usepackage{placeins}
\usepackage{tikz}
\usepackage[compat=1.1.0]{tikz-feynman}
\usepackage{makecell}
\usepackage[colorlinks=true,citecolor=blue!50!black,linkcolor=black]{hyperref}
\usepackage{caption}
\usepackage{soul}
\usepackage
[subrefformat=parens,position=top,skip=-15pt,margin=15pt,justification=justified,singlelinecheck=false]
{subcaption}
\usepackage{authblk}

\usepackage{mciteplus}
\mciteErrorOnUnknownfalse
\usepackage{geometry}
\usepackage{sectsty}
\usepackage{siunitx}
\let\spreprint\empty
\newcommand{\preprint}[1]{\def\spreprint{\protect#1}}
\let\sinstitute\empty

\makeatletter
\renewcommand{\maketitle}{\begingroup
  \null\thispagestyle{empty}%
    \ifx\spreprint\empty
      \vskip 5ex
    \else
      \flushright\large\spreprint\vskip 2ex
    \fi
    \vskip 5ex
    \flushleft
      {\sffamily\bfseries\huge\@title}\vskip 2ex
      \@author\vskip 2ex
      \ifx\sinstitute\empty
      \else
        {\small\sinstitute}
      \fi
    \vskip 5ex
  \endgroup
}
\makeatother

\renewenvironment{abstract}{\begin{center}
  {\large\sffamily\bfseries Abstract: }
  \begin{minipage}[t]{0.75\textwidth}
}{\end{minipage}\end{center}\vskip 10ex}

\allsectionsfont{\normalfont\sffamily\bfseries}
\subsectionfont{\normalfont\sffamily\bfseries}
\subsubsectionfont{\normalfont\sffamily\bfseries}

\makeatletter
\DeclareRobustCommand*{\bfseries}{%
  \not@math@alphabet\bfseries\mathbf
  \fontseries\bfdefault\selectfont
  \boldmath
}
\makeatother

\newcommand{\ie}{\emph{i.e.}\ }
\newcommand{\eg}{\emph{e.g.}\ }

\def\beq{\begin{equation}}
\def\eeq{\end{equation}}
\def\({\left(}
\def\){\right)}
\def\[{\left[}
\def\]{\right]}

\newcommand{\GeV}{\ensuremath{\,\text{GeV}}\xspace}

\newcommand{\abs}[1]{\left|#1\right|}

\newcommand{\ptsup}[1]{\ensuremath{p_{\text{T}}^{#1}}\xspace}

\newcommand{\sherpa}{S\protect\scalebox{0.8}{HERPA}\xspace}

\newcommand{\pythia}{P\protect\scalebox{0.8}{YTHIA}\xspace}

\newcommand{\amegic}{A\protect\scalebox{0.8}{MEGIC}\xspace}
\newcommand{\comix}{C\protect\scalebox{0.8}{OMIX}\xspace}
\newcommand{\CSS}{CSS\protect\scalebox{0.8}{HOWER}\xspace}
\newcommand{\alaric}{A\protect\scalebox{0.8}{LARIC}\xspace}

\newcommand{\Centauro}{C\protect\scalebox{0.8}{ENTAURO}\xspace}

\newcommand{\rivet}{R\protect\scalebox{0.8}{IVET}\xspace}

\newcommand{\powheg}{P\protect\scalebox{0.8}{OWHEG}\xspace}

\newcommand{\LHAPDF}{L\protect\scalebox{0.8}{HA}P\protect\scalebox{0.8}{DF}\xspace}

\newcommand{\muR}{\ensuremath{\mu_{\text{R}}}}
\newcommand{\muF}{\ensuremath{\mu_{\text{F}}}}

\newcommand{\zcut}{\ensuremath{z_{\text{cut}}}}
\newcommand{\alphaS}{\alpha_\text{s}\xspace}
\newcommand{\xBj}{\ensuremath{x}}

\newcommand{\GrTau}{\ensuremath{\tau_{\text{Gr}}}}
\newcommand{\LO}{\text{LO}\xspace}

\newcommand{\MEPSatLO}{\text{\textsc{MEPS@LO}}\xspace}
\newcommand{\MEPSatNLO}{\text{\textsc{MEPS@NLO}}\xspace}

\newcommand{\MCatNLO}{\text{\textsc{MC@NLO}}\xspace}

\newcommand{\EIC}{EIC\xspace }
\newcommand{\HERA}{HERA\xspace}
\newcommand{\ZEUS}{ZEUS\xspace}

\def\draftdate{\relax}
\def\mda{\relax}
\def\mua{\relax}
\def\mla{\relax}
\def\draft{
\def\thtystars{******************************}
\def\sixtystars{\thtystars\thtystars}
\typeout{}
\typeout{\sixtystars**}
\typeout{* Draft mode!
         For final version remove \protect\draft\space in source file *}
\typeout{\sixtystars**}
\typeout{}
\def\draftdate{\today}
\def\mua{\marginpar[\boldmath\hfil$\uparrow$]%
                   {\boldmath$\uparrow$\hfil}\color{black}%
                    \typeout{marginpar: $\uparrow$}\ignorespaces}
\def\mda{\color{red}\marginpar[\boldmath\hfil$\downarrow$]%
                   {\boldmath$\downarrow$\hfil}%
                    \typeout{marginpar: $\downarrow$}\ignorespaces}
\def\mla{\marginpar[\boldmath\hfil$\rightarrow$]%
                   {\boldmath$\leftarrow $\hfil}%
                    \typeout{marginpar: $\leftrightarrow$}\ignorespaces}
\def\Mua{\marginpar[\boldmath\hfil$\Uparrow$]%
                   {\boldmath$\Uparrow$\hfil}\color{black}%
                    \typeout{marginpar: $\uparrow$}\ignorespaces}
\def\Mda{\color{red}\marginpar[\boldmath\hfil$\Downarrow$]%
                   {\boldmath$\Downarrow$\hfil}%
                    \typeout{marginpar: $\downarrow$}\ignorespaces}
\def\Mla{\marginpar[\boldmath\hfil\textcolor{red}{$\Rightarrow$}]%
                   {\boldmath\textcolor{red}{$\Leftarrow $}\hfil}%
                    \typeout{marginpar: $\leftrightarrow$}\ignorespaces}
\overfullrule 5pt
\oddsidemargin 15mm
\marginparwidth 29mm
}

\usepackage{color}
\usepackage{footnote}
\definecolor{darkblue}{rgb}{0,0,0.5}
\definecolor{darkred}{rgb}{0.5,0,0}
\definecolor{darkgreen}{rgb}{0,0.5,0}

\hypersetup{
  pdfauthor={Peter Meinzinger, Daniel Reichelt, Federico Silvetti},
  pdftitle={Event generation at \MEPSatNLO accuracy for jet physics in neutral and
  charged current DIS at the \EIC}
}

\preprint{ZU-TH 43/25\\CERN-TH-2025-110\\IPPP/25/31\\MCNET-25-13}

\usepackage[symbol]{footmisc}

\author[1]{Peter~Meinzinger\footnote[1]{Email: \texttt{peter.meinzinger@uzh.ch}}}
\author[2]{Daniel~Reichelt\footnote[2]{Email: \texttt{d.reichelt@cern.ch}}}
\author[3]{Federico~Silvetti\footnote[3]{Email: \texttt{federico.silvetti@durham.ac.uk}}}

\affil[1]{Physik-Institut, Universität Zürich, Winterthurerstrasse 190, CH-8057 Zürich, Switzerland}
\affil[2]{CERN, Theoretical Physics Department, CH-1211 Geneva 23, Switzerland}
\affil[3]{Institute for Particle Physics Phenomenology, Department of Physics, Durham University,
  Durham DH1 3LE, United Kingdom}

\title{Event generation at \protect\MEPSatNLO accuracy in neutral and charged current DIS
  at the \protect\EIC}
\begin{document}
\maketitle

\begin{abstract}
  We present state-of-the-art hadron-level predictions for the deep-inelastic
  scattering process at next-to-leading-order precision for several
  multiplicities, consistently merged in one sample. For the first time at this
  level of accuracy, we consider both
  neutral and charged current deep-inelastic scattering at the
  Electron-Ion Collider, and present the first application of consistent
  next-to-leading-order merging to charged current deep-inelastic scattering in
  general. We critically examine inclusive predictions using multileg merging techniques,
  contrasting perturbative and nonperturbative uncertainties. Further, we study
  typical kinematic deep-inelastic scattering observables as well as jet
  measurements and 1-jettiness with realistic cuts implied by expected and past
  detector resolution.
  On the perturbative side, we see large corrections toward small virtualities
  and Bjorken-$x$, which can be captured by higher-multiplicity matrix elements
  and the merging procedure. Nonperturbative effects, while negligible in most
  jet observables, can reach similar size as the perturbative uncertainties
  around the peak of the 1-jettiness distributions especially at low values of $Q^2$.
\end{abstract}

\clearpage
\vspace{10pt}
\noindent\rule{\textwidth}{1pt}
\tableofcontents
\noindent\rule{\textwidth}{1pt}
\vspace{10pt}

\section{Introduction}

Lepton-hadron collision experiments, as conducted for example at \HERA, are crucial for probing hadron
structure~\cite{Klein:2008di,Abramowicz:1998ii,Newman:2013ada}. These
interactions occur via the exchange of at least one electroweak boson. When the
virtuality of this mediator is sufficiently large, i.e., $Q^2 \gg
m_P^2 \sim 1\GeV^2$, the
interaction enters the deeply inelastic Scattering (DIS) regime. In this limit,
the factorization of the cross section into matrix elements and parton
distribution functions (PDFs) has been proven explicitly~\cite{Ellis:1978ty,
  Collins:1989gx}, hence allowing the application of perturbative
techniques. Specifically, in the case of neutral-current (NC) DIS mediated by photons
and Z-bosons, in contrast to charged-current (CC) DIS mediated by W-bosons,
measurements can be performed fully inclusively with respect to the hadronic
final state, providing a high-precision probe of the proton’s partonic
structure.

In recent years, lepton-hadron scattering has garnered more attention again due to the
upcoming  planned for the Electron-Ion Collider (\EIC) at Brookhaven National Laboratory.
The \EIC aims for unprecedented precision in determining the spatial and
transverse-momentum distributions of partons in the
proton~\cite{Accardi:2012qut, AbdulKhalek:2021gbh}. Maximizing the available
statistics for experimental analyses requires lowering the virtuality cutoff,
as the cross section scales as $1 / Q^4$. However, at lower $Q^2$, additional
scales -- such as jet transverse momenta -- can become dominant, leading to
increased theoretical uncertainties. This challenge is addressed by the
computation of higher-order corrections in fixed-order (FO) perturbation theory,
particularly via real radiation corrections in QCD. While current FO
calculations extend up to next-to-next-to-next-to-leading order (N$^3$LO)~\cite{Vermaseren:2005qc, Moch:2008fj,
  Currie:2018fgr}, further improvements are computationally prohibitive in the
foreseeable future. Additionally, the lower center-of-mass
energy than, \eg, at \HERA shifts the attention toward smaller values of $Q^2$
as well as nonperturbative corrections which should approximately be of the
order $\Lambda_\mathrm{QCD}/Q$ or $\Lambda^2_\mathrm{QCD}/Q^2$.

An alternative approach is multijet merging, which combines tree-level matrix
elements of varying final-state multiplicities and parton shower evolution. This
method was initially developed at leading order (LO)~\cite{Hoeche:2009rj,Carli:2010cg} and later
extended to next-to-leading-order (NLO)~\cite{Hoeche:2012yf,Gehrmann:2012yg}. Multijet merging for DIS
was recently also studied at LO within the \pythia event
generator~\cite{Helenius:2024wjg}, too. Although not a complete higher-order
correction, this approach effectively captures the dominant effects at small
virtualities, in particular in the range $Q^2 \simeq \mathcal{O}(1\text{-}10\GeV^2)$.

Most previous studies of inclusive DIS phenomenology for the \EIC have relied on
LO calculations interfaced to parton showers (LO+PS), primarily using
\pythia~\cite{Page:2019gbf,Chien:2021yol,Arratia:2019vju,Arratia:2020nxw,Zheng:2018ssm,Arratia:2020azl,Arratia:2022oxd}.
Notably, Ref.~\cite{Page:2019gbf} explored the full-$Q^2$ phase space, including the low-$Q^2$ photoproduction regime at LO.
DIS has also been recently studied at matched NLO
accuracy in the \powheg formalism in Refs.~\cite{Banfi:2023mhz,Borsa:2024rmh,Buonocore:2024pdv}.
Several other tools~\cite{Karlberg:2024hnl,NNLOJET:2025rno} 
are available to compute DIS observables at next-to-next-to-leading-order (NNLO) accuracy for the \EIC. 
Furthermore, jet predictions in the photoproduction regime were computed at NLO
in Ref.~\cite{Meinzinger:2023xuf}; see also Ref.~\cite{Andersen:2024czj}.

LO+PS calculations will provide only a leading-order description of the
inclusive DIS kinematics (i.e. $Q^2$ and $x_{Bj}$ spectra, see Sec.~\ref{sec:DISkin} for
details) and describe additional jets within the parton shower approximation\footnote{Since
we are mainly focused on the inclusion of higher order matrix elements here, we
choose to largely avoid a full discussion of the formal accuracy achieved by the
parton shower. For example the discussion in \cite{Dasgupta:2018nvj} (see
also \cite{vanBeekveld:2023chs} for a DIS specific discussion) indicates that the
default \sherpa dipole shower used here asymptotically resums leading logarithms
(LL) of event shapes at leading color, while breaking next-to-leading logarithms due to
its recoil scheme. On the other hand, parton showers include important physical
constrains such as exact momentum conservation and unitarity
\cite{Hoche:2017kst}. Within the \sherpa framework we use here, a shower, dubbed
\alaric, satisfying the recoil safety requirement for NLL resummation has been implemented
\cite{Herren:2022jej,Assi:2023rbu} and equipped with the same merging techniques at LO
\cite{Hoche:2024dee} for initial- and final state evolution and matching and
merging at NLO for final state emissions \cite{Hoche:2025gsb}. In practical
terms it has shown similar performance to the shower used here for the cases
where \alaric is available so far.}. NLO calculations matched to parton showers
will naturally describe the inclusive 
kinematics at NLO accuracy, and observables related to one additional emission,
such as the $p_T$ spectrum of an additional jet, at leading order. Observables sensitive
to more emissions are again described in the parton shower
approximation. Fixed-order calculations are available at higher perturbative
orders; however, there 
are no automated matching schemes available for these so far. Without matching
to a parton shower, and subsequent hadronization via a dedicated model, there
are no methods to relate such a calculation to a fully differential hadronic
final state. Merging NLO (for higher multiplicities LO) matrix elements, as
described in this manuscript, allows us to describe observables sensitive to
further emissions, for example the $p_T$-spectrum of the $n$th jet, at NLO (LO)
accuracy, while maintaining the ability to include a parton shower and hence a
fully differential distribution for arbitrary observables.

In this work, we present a study of precision DIS phenomenology at the \EIC using the \sherpa event generator.
Specifically, we employ its automated matching and merging capabilities to produce the most precise, fully exclusive hadron-level predictions to date.
We analyze both traditional DIS observables and hadronic event shapes, including 1-jettiness, jet multiplicities, and leading-jet transverse momentum.
To assess theoretical uncertainties, we perform independent evaluations of
perturbative and nonperturbative effects, identifying limitations in
theoretical predictions relevant to \EIC analyses.

We structure our paper as follows. In Sec.~\ref{sec:DISkin}, we introduce our
notation for DIS kinematics and define the observables to be
studied. Section~\ref{sec:event-generation} details our simulation framework
within \sherpa, and Sec.~\ref{sec:cc-dis-validation} presents a validation of
the CC DIS simulation against \ZEUS data. We then analyze NC
and CC DIS events at the \EIC in Sec.~\ref{sec:eic-predictions},
before concluding in Sec.~\ref{sec:conclusions}.

\section{Overview of DIS kinematics and analyzed observables} \label{sec:DISkin}

We consider DIS events between an electron and a
proton with momentum $p$ and $P$ respectively,
\begin{equation}
  e^-\(p\) + p \(P\) \rightarrow l\(p'\) + X\(p_X\) \label{eq:DISreaction}
\end{equation}
This results in collection of hadrons with momentum $p_X$ plus a recoiling
fermion momentum $p'$ which can be either another electron, $e^-$, or a
neutrino, $\nu_e$, if the interaction is driven by neutral or charged current,
respectively\footnote{While we focus on the case of electron beams here, the
described methods are of course likewise available for positron beams}. In both
cases the virtual boson carries momentum $q = p - p^\prime$. When the momentum
transfer is large enough, the virtual boson acts as a probe of the internal
structure of the proton, which is dominated by color-confined strong
interactions. Thus, measurements of this family of scattering process are an
excellent experimental handle on the behavior of strong nuclear forces and the
nature of the elementary degrees of freedom of the hadronic side of the Standard Model.

It is customary to parametrize the scattering using two out of three of the following invariants
\begin{subequations}
  \begin{align}
    \text{virtuality}\;\;\;Q^2 &= -q^2 \ , \label{eq:virtuality} \\
    \text{Bjorken~x}\;\;\; \xBj &= \frac{Q^2}{2 P\cdot q} \ , \label{eq:Bjorkenx}\\
    \text{and inelasticity}\;\;\; y   &= \frac{P\cdot q}{P\cdot p} \  \label{eq:inelasticity} \ .
  \end{align}\label{eq:DISvars}
\end{subequations}
The observable $Q^2$ defines the energy scale of the boson probing the proton structure, while Eq.~\eqref{eq:Bjorkenx}
defines the lowest fraction of momentum that a struck quark or gluon inside the
proton can carry.
Finally, Eq.~\eqref{eq:inelasticity} quantifies the energy loss of the electron.
This choice of variables is especially relevant in the neutral current case as
it allows to fully characterize the cross section from measurements of the final-state lepton only.
Indeed, one of the main goals of the \EIC~program is a high-precision determination of the
proton structure and its dependence on $Q^2$~\cite{Accardi:2012qut}, this motivates
theoretical effort in improving the prediction for the same observables.

Beside the inclusive cross section, we can learn more about the features of the strong interactions,
 \textit{e.g.}, the value of strong coupling $\alphaS$ or the details of the QCD radiation patterns,
 by studying directly properties of the ensemble of hadrons emerging in the final state.
Event shapes observables are one way to approach this task as they evaluate the geometrical distribution of particles in the final state.
Specifically for this work, we consider the 1-jettiness distribution,
\begin{equation}
  \tau = \frac{2}{Q^2}\sum_{i \in X} \min(p_i\cdot \(\xBj P\),\, p_i \cdot \( \xBj P + q \))\,. \label{eq:1-jettiness}
\end{equation}
This is a measure the collimation of hadrons around the direction of the incoming or produced parton and in DIS is 
equivalent to the thrust event shape~\cite{Kang:2013nha}.

A complementary approach to the above study of the global hadronic final state
is the clustering of hadrons into jets. In modern experimental analyses, jets are usually defined in terms of
sequential clustering algorithms \cite{JADE:1988xlj}. For
this purpose, we consider three variants from Refs.~\cite{Catani:1993hr, Cacciari:2008gp,
  Arratia:2020ssx}, defined by the clustering distance measures for two hadrons $p_i$, $p_j$
\begin{subequations}
  \begin{align}
    d_{ij}^{k_{t}}             = & \min\left(k_{ti}^2, \ k_{tj}^2\right) \frac{\Delta^2_{ij}}{R^2}, \quad d_{iB} = k_{ti}^{2}, \\
    d_{ij}^{{\rm anti}-k_{t}}  = & \min\left(\frac{1}{k_{ti}^2}, \ \frac{1}{k_{tj}^2}\right) \frac{\Delta^2_{ij}}{R^2}, \quad d_{iB} = \frac{1}{k_{ti}^2}, \\
    d_{ij}^{{\rm Centauro}} = & \frac{1}{R^2}\[ \left( f_i - f_j \right)^2 + 2 f_i f_j \(1- \cos \Delta\phi_{ij}\)\] \quad \mathrm{with} \, f_i = 2 \sqrt{1 + \frac{q\cdot p_i}{\xBj P \cdot p_i}} , \quad d_{iB} = 1,
  \end{align}
\end{subequations}
where $\Delta_{ij}^2 = \(y_i-y_j\)^2 + (\phi_i^2 - \phi_j^2)$ is the distance in
the azimuthal plane between the two hadrons and $y$, $\phi$ and $k_t$ are the
rapidity, azimuthal angle and transverse momentum relative to the beam axis of
each hadron, respectively. All quantities here are defined in the Breit
frame for our purposes\footnote{The Breit frame, or brick wall frame, is defined as the frame in
which the momentum transfer is $$q = Q \left( 0,\,0,\,0,\,-1\right),$$
this provides a natural separation hemispheres between the leptonic and hadronic side of DIS at Born level.
}. The parameter $R$ controls the expected average size of the jet.
All these tools allow for the characterization of events in terms of leading jet
transverse momentum and pseudorapidity, as well as the overall jet multiplicity
and invariant masses of di-/trijet systems.

Finally, given a definition of jets, one more handle on the behavior of QCD
radiation is offered by the internal structure of the jet themselves. Analyses
of jet substructure are a tool extensively used at the LHC
\cite{Larkoski:2017jix,Kogler:2018hem,Marzani:2019hun,Marzani:2024mtt}. A common
goal is to remove soft radiation at wide angles from the jet axis, which is typically dominated by
contamination from sources outside the jet and hadronization
corrections. This can be achieved by the so called soft-drop grooming technique
\cite{Larkoski:2014wba}. While originally introduced in the context of jets at hadron
colliders, and the LHC specifically, it has since been explored in many
other contexts such as event shapes at hadron \cite{Baron:2020xoi} as well as
lepton colliders \cite{Baron:2018nfz, Marzani:2019evv}, in Higgs decays
\cite{Gehrmann-DeRidder:2024avt} and for jets measured at lower-energy
experiments such as RHIC \cite{Chien:2024uax}. Here, we consider the application
in DIS described in Ref.~\cite{Makris:2021drz} to soft drop on top of the \Centauro
clustering (see above). Related observables have recently been measured in DIS
for the first time \cite{H1:2024pvu}. Briefly, this entails first clustering the
hadronic ensemble $X$ with the standard algorithm and then stepping backwards
through the jet clustering history. At each branching, if
\begin{equation}
  \frac{\min(z_i,z_j)}{z_i+z_j} < \zcut, \quad z_i = \frac{P\cdot p_i}{P \cdot q}~, \label{eq:soft-drop}
\end{equation}
where $\zcut$ defines the cutoff parameter for the procedure, the softer
subjet is discarded and the procedure continues along the harder branch. The algorithm
terminates if either Eq.~\eqref{eq:soft-drop} is not met or there is only one
particle left in the jet.
After this grooming procedure is completed the surviving particles can be used to compute any hadronic observable.
For this work, we will consider the groomed 1-jettiness:
\begin{subequations}
  \begin{align}
    \GrTau &= \frac{2}{Q^2}\sum_{i \in X_\mathrm{Groomed}} \min(p_i\cdot \(\xBj P\),\, p_i \cdot \( \xBj P + q \))\,. \label{eq:Groomed1-jettiness}
  \end{align}
\end{subequations}

\section{Event generation for DIS in \sherpa}\label{sec:event-generation}

For our hadron-level predictions, we use the \sherpa 3~\cite{Sherpa:2024mfk}
event generator. It provides matrix elements at tree level through the internal
generators \amegic \cite{Krauss:2001iv} and \comix \cite{Gleisberg:2008fv}.
Virtual corrections for the DIS process are likewise included internally. We use
\sherpa's default \CSS \cite{Schumann:2007mg} based on Catani-Seymour dipoles
\cite{Catani:1996vz}. It can be matched to NLO matrix elements using the
\MCatNLO \cite{Frixione:2002ik} method. Matrix elements with different
multiplicities are combined using the CKKW merging method \cite{Catani:2001cc}.
Events can be hadronized using the cluster hadronization
model \cite{Webber:1983if} as implemented in \sherpa
\cite{Winter:2003tt,Chahal:2022rid} or alternatively the Lund model using an
interface to \pythia 8 \cite{Bierlich:2022pfr}.  The events are analyzed using
\sherpa's interface to \rivet~\cite{Bierlich:2019rhm}.
For the parton densities, we use the NNPDF30\_nlo\_as\_0118 set~\cite{NNPDF:2014otw}
using \LHAPDF~\cite{Buckley:2014ana}. \LHAPDF is also used for the $\alphaS$
evolution, ensuring consistency with the PDF set and in particular implying
$\alphaS(M_Z)=0.118$.

\subsection{\protect\MEPSatLO, \protect\MCatNLO and \protect\MEPSatNLO} \label{ssec:MC-modes}

The automation of matching and merging for DIS in \sherpa has been presented
in~\cite{Carli:2010cg}. Matching to higher-order matrix elements is achieved
through \sherpa's implementation of the \MCatNLO prescription~\cite{Hoeche:2011fd}.
Merging in the CKKW method is achieved by applying a jet criterion at the
matrix-element level and restricting the available phase space for the parton
shower. To maintain the logarithmic accuracy, a backward clustering has to be
performed to dress matrix elements with appropriate Sudakov factors and evaluate
$\alphaS$ at the correct scale. The techniques used in this work are known as
\MEPSatLO~\cite{Hoeche:2009rj} and
\MEPSatNLO~\cite{Hoeche:2012yf,Gehrmann:2012yg}, and we refer to
the respective publications for the details of the procedures and subtleties
related to the use of NLO matrix elements.

In our calculations, we compute single- and dijet production at NLO and
three- and four-jet production at LO, \ie,
\begin{equation}\label{eq:MEPS_MEs_NC}
  e^-p\to e^- + 1,2\,j\,@\,\mathrm{NLO}+3,4\,j\,@\,\mathrm{LO}
\end{equation}
for neutral-current DIS and
\begin{equation}\label{eq:MEPS_MEs_CC}
  e^-p\to \nu +1,2\,j\,@\,\mathrm{NLO}+3,4\,j\,@\,\mathrm{LO}
\end{equation}
for CC DIS at \MEPSatNLO, and analogous simplified processes for
\MEPSatLO, \MCatNLO and LO predictions. In the neutral-current interactions, we
treat the light quarks as massless and additionally compute and merge in processes
with massive $c$ and $b$ quarks at LO. {In either case we use the five-flavor
NNPDF30\_nlo\_as\_0118 PDF set.} In the parton shower, $c$ and $b$ quarks are
likewise treated using massive splitting functions~\cite{Catani:2002hc}. For the
charged-current simulation, we work with four massless quarks. The $b$ does not
contribute to jet production since we assume a diagonal CKM matrix, and is
anyway heavily suppressed by the PDF.

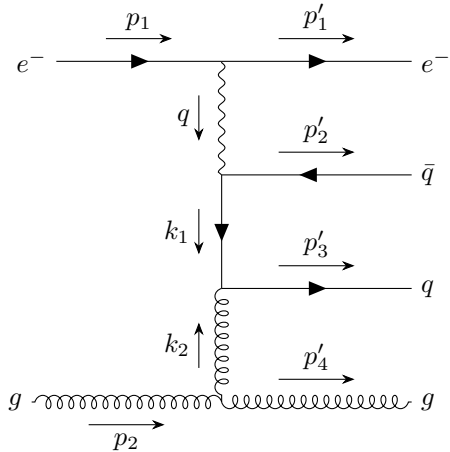
\begin{figure}[htpb]
    \centering
    \begin{tikzpicture}
    \centering
    \begin{feynman}
        \vertex (a) {$e^-$};
        \vertex[right=2.5cm of a] (b);
        \vertex[right=2.5cm of b] (c) {$e^-$};
        \vertex[below=1.5cm of b] (d);
        \vertex[below=1.5cm of d] (e);
        \vertex[right=2.5cm of d] (h) {$\bar{q}$};
        \vertex[right=2.5cm of e] (f) {$q$};
        \vertex[below=1.5cm of e] (k);
        \vertex[left=2.5cm of k] (g) {$g$};
        \vertex[right=2.5cm of k] (i) {$g$};

        \diagram*{
            (a) --[fermion,momentum={[arrow shorten=0.3]$p_1$}] (b),
            (b) --[fermion,momentum={[arrow shorten=0.3]$p'_1$}] (c),
            (b) --[boson,momentum'={[arrow shorten=0.3]$q$}] (d),
            (d) --[fermion,momentum'={[arrow shorten=0.3]$k_1$}] (e),
            (d) --[anti fermion,momentum={[arrow shorten=0.3]$p'_2$}] (h),
            (e) --[fermion,momentum={[arrow shorten=0.3]$p'_3$}] (f),
            (e) --[gluon,reversed momentum'={[arrow shorten=0.3]$k_2$}] (k),
            (k) --[gluon,reversed momentum={[arrow shorten=0.3]$p_2$}] (g),
            (k) --[gluon, momentum={[arrow shorten=0.3]$p'_4$}] (i)
        };
    \end{feynman}
    \end{tikzpicture}
    \caption{Sketch of an example Feynman diagram with three final-state partons in DIS. Depending on the hierarchy of the internal momenta, the event is clustered back to an underlying $2\to2$ core process. Only if $q^2 \gg k_1^2,\, k_2^2$ holds is the core process DIS-like. }\label{fig:trijet-dis-sketch}
\end{figure}

When merging higher-multiplicity matrix elements, each process has to be
clustered back to a $2 \to 2$ configuration. An exemplary diagram which involves
all possible clustering steps is shown in
Fig.~\ref{fig:trijet-dis-sketch}. Depending on the hierarchy of the propagators'
momenta, $q$ and $k_i$, for any given event one can distinguish three
different core processes and their corresponding scale $\mu_{\mathrm{core}}$:

\begin{enumerate}
  \item[(i)] virtual photon exchange, \ie, $ej\to ej$, where $\mu_{\mathrm{core}}^2=Q^2$;
  \item[(ii)] interaction of the virtual photon with a QCD parton,
    \ie, $\gamma^*j\to j_1j_2$, with $\mu^2_{\mathrm{core}}=m_{\perp,1}
    m_{\perp,2}$ defined as the product of the two jet transverse masses
    $m_{\perp,i} = \sqrt{m^2_i + p_{\perp,i}^2}$ relative to the beam axis,
  \item[(iii)] pure QCD channels, \ie, $jj\to jj$, where
    $\mu^2_{\mathrm{core}}=-\frac{1}{\sqrt{2}}\left(s^{-1}+t^{-1}+u^{-1}\right)^{-1}$
    is a scaled harmonic mean of the Mandelstam variables $s,t,u$.
\end{enumerate}

The factorization and shower starting scales are then set to the core
scale, \ie, $\mu_{\mathrm{F}}=\mu_{\mathrm{R}}=\mu_\mathrm{Q}=\mu_{\mathrm{core}}$.
For higher multiplicities, the renormalization scale is determined according to
the clustering algorithm~\cite{Hoeche:2009rj} with $\mu_\mathrm{core}$ as the scale
of the core $2\to2$ process.

The merging scale $Q_\mathrm{cut}$, \ie, the separation of the phase space into a
region filled by additional legs in the matrix elements instead of the parton
shower, is dynamically given by

\begin{equation}
  Q_{\mathrm{cut}} = \frac{\bar{Q}_{\mathrm{cut}}}{\sqrt{1+\frac{\bar{Q}^2_{\mathrm{cut}}}{S_{\mathrm{DIS}} Q^2}}}
  \label{eq:Qcut}
\end{equation}

using $\bar{Q}_{\mathrm{cut}} = 25\GeV$ and $S_\mathrm{DIS} = 0.4$.
The former avoids unnecessary multileg matrix elements, while the latter
ensures a decrease of the merging scale towards smaller virtualities, therefore
improving the description in this region.

\subsection{Systematic uncertainties of event generation in DIS} \label{ssec:uncertainties}

A systematic assessment of theoretical uncertainties is crucial in high-energy
phenomenology due to increasing precision in calculations, simulations, and
experiments. For a Monte Carlo event generator like \sherpa, two primary sources
of uncertainties exist: theoretical uncertainty from truncation of perturbative
series expansion, and uncertainty in modeling and tuning parton fragmentation
and hadronization due to the lack of a first-principle solution for long-range
QCD interactions.

The first source encompasses several perturbative ingredients of the
calculation, i.e. the evaluation of hard matrix elements at various
multiplicities in merged samples, as well as parton shower evolution.
We follow the established approach of estimating the size of missing higher
perturbative orders by examining the residual dependence of resulting
observables on the renormalization and factorization scales. We note, though, that
this approach has severe limitations and especially in the context of even
higher-order calculations it might be appropriate to investigate other sources
of uncertainty
\cite{Cacciari:2011ze,Duhr:2021mfd,Ghosh:2022lrf,Tackmann:2024kci,Lim:2024nsk}. Here,
we perform seven-point variations around a central choice $(\mu_{F,0},\mu_{R,0})$,
considering all ratios
\begin{equation}
  \centering
  (\mu_F/\mu_{F,0},\mu_R/\mu_{R,0}) \in \left\lbrace (1,1),\,(1,2),\,(1,0.5),\,(2,1),\,(0.5,1),\,(0.5,0.5),\,(2,2)\right\rbrace .
\end{equation}
We apply these variations in the matrix-element calculation and the
parton-shower evolution simultaneously. Ultimately, we build our estimate of the perturbative
uncertainty from the envelope of all scale choices. It is worth stressing that,
in this procedure, we are varying scales appearing in the parton shower, which
affects the evaluation of $\alphaS$ and the PDF ratios. Providing more extensive
parton shower uncertainties is an open topic of research, complicated by the
fact that parton showers include significant subleading contributions
\cite{Hoche:2017kst}. We stay restricted here to the traditional scale
variations.

The particular choice of the merging scale, while arbitrary in principle, also affects the perturbative side of a merged simulation.
In our approach, see
Eq.~\eqref{eq:Qcut}, it is dynamically determined by the two parameters
$\bar{Q}_\text{cut}$ and $S_\text{DIS}$. To asses the related uncertainty, we
 only vary the resulting scale $Q_\text{cut}$ here by a factors $1/2$ and $2$ and
form an envelope of those predictions. Note this procedure is equivalent to
varying only $\bar{Q}_\text{cut}$ at large $Q^2$, where $\bar{Q}_\text{cut}$
effectively acts as a fixed scale,  or only $S_\text{DIS}$ at small $Q^2$ where
the effective scale is just proportional to that.

\sherpa performs this operation on the fly~\cite{Bothmann:2016nao},
which significantly reduces computing costs compared to running separate
simulations for each scale setting.

The second source of uncertainty relates to the nonperturbative modeling of
parton fragmentation into hadrons. These uncertainties arise from
phenomenological models rather than first-principle calculations, leading to
a significant model dependence. To evaluate fragmentation corrections, we use \sherpa's
internal cluster hadronization model~\cite{Chahal:2022rid} and the Lund String model via an interface
to \pythia 8 \cite{Bierlich:2022pfr}. The symmetrized difference between these
two models is taken as the uncertainty\footnote{Note this procedure can in principle result in uncertainties
covering negative (differential) cross sections. These are clearly unphysical
and we cut off the envelope at zero.}, using the cluster
model as the central value.

Another aspect of nonperturbative modeling involves tuning model parameters to
data. We use the replica tunes from Ref.~\cite{Knobbe:2023njd} to estimate
uncertainties related to the tuning of final-state fragmentation parameters and
the variations from Ref.~\cite{Knobbe:2023ehi} for parameters governing cluster
decays involving beam remnants. This introduces two sets of variations within
the cluster model itself, for which we take the respective envelopes for each
histogram as the related uncertainty. As a result, three sources of
nonperturbative uncertainty arise: model choice, final-state fragmentation
tune, and beam fragmentation tune. Arguably the different sources of uncertainty
should be combined. We, however, find that usually only one, in most cases the
model choice, dominates the overall uncertainty. We hence only show the three
sources independently, without a combination.

Since estimating nonperturbative uncertainties requires independent Monte Carlo
runs for each parameter set, and for both models, performing this study for the
full \MEPSatNLO, \MEPSatLO and \MCatNLO setups is infeasible. Instead, we derive
the relative uncertainty based on parton showered leading-order matrix
elements merged together in the \MEPSatLO approach. A merged approach is still
beneficial to ensure sufficient phase-space coverage. In the following we only
treat the results of two tunes or models as different if we can statistically
distinguish them at a level of 3 standard deviations, and otherwise indicate the
statistical error that serves as an upper bound on the uncertainty.

We make no attempt to directly apply this uncertainty to
higher- or lower-precision simulations, though one could envision a procedure based on
transfer matrices introduced in Ref.~\cite{Reichelt:2021svh}. Additionally, we do not
combine nonperturbative and perturbative uncertainty estimates but rather focus
on comparing their relative sizes in the following analysis.

\section{Event generation for \protect\HERA and \protect\MEPSatNLO in charged current DIS}\label{sec:cc-dis-validation}
\begin{figure}[ht]
  \centering
      \includegraphics[width=0.4\linewidth]{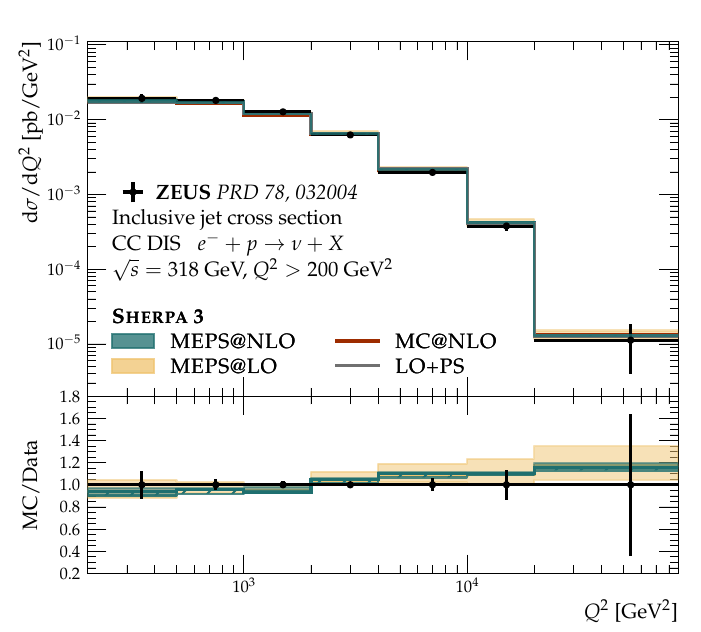}
      \includegraphics[width=0.4\linewidth]{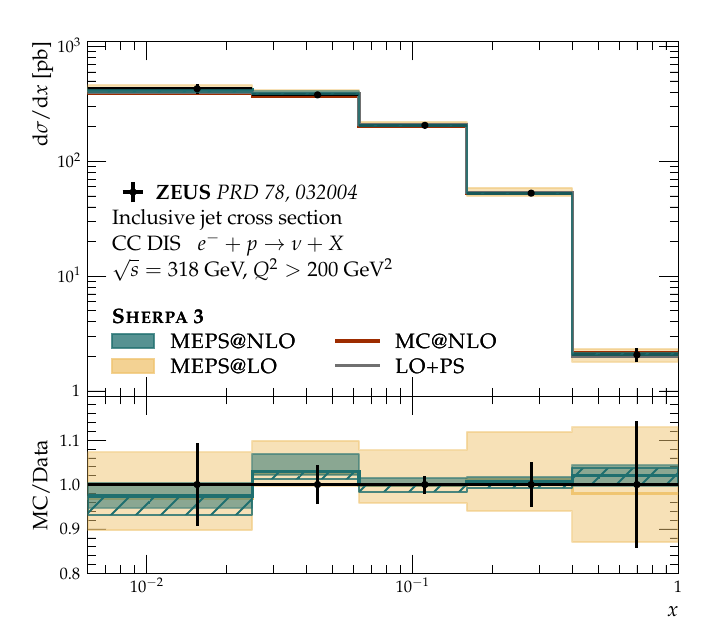}
  \caption{Differential distributions for $Q^2$ and $x$
    at LO, \protect\MCatNLO, \protect\MEPSatLO and \protect\MEPSatNLO in
    inclusive CC DIS at the \protect\HERA. The lower panels show
    the ratio between data and the merged \protect\MEPSatLO and
    \protect\MEPSatNLO predictions, while we omit the ratio for the \protect\LO
    and \protect\MCatNLO simulations for clarity. The full (hatched) band corresponds to variations of $\muF,\muR$ ($Q_{\mathrm{cut}}$). Experimental data points are from Ref.~\cite{ZEUS:2008arl}.
    \label{fig:cc-hera2:dis-obs-1}}
\end{figure}
The \sherpa framework has been extensively used in the past to describe neutral
current DIS at \HERA, see for example Ref.~\cite{Carli:2010cg}. Early versions of
\sherpa 3 were also used in a variety of recent event shape and jet analyses by
the H1 Collaboration\cite{H1:2023fzk, H1:2024nde, H1:2024aze, H1:2024pvu,
  H1:2024mox}, usually finding satisfactory agreement with the data. A
\MEPSatNLO setup similar to the one used here was compared to resummed
calculations for event shapes in Refs.~\cite{Knobbe:2023ehi,Knobbe:2024rci}. The
latter also determined part of the current tune of nonperturbative
parameters. With this, we consider the framework and \MEPSatNLO to be
sufficiently validated in neutral current DIS, and we focus in the following on
the first detailed comparison to CC data from \HERA.

We compare our Monte Carlo setups against multijet cross sections in CC DIS
measured by the \ZEUS Collaboration~\cite{ZEUS:2008arl}. We consider beams of
electrons at $27.5 \GeV$ and protons at $920 \GeV$. To model the proton
contents, we use the NNPDF30\_nlo\_as\_0118 set as discussed in
Sec.~\ref{ssec:MC-modes}.

In principle, a caveat for CC DIS is that, in contrast to NC DIS,
it is impossible to use the final-state neutrino to reconstruct $q$. Therefore, the measurements of
Eqs.~\eqref{eq:DISvars} involve the reconstruction of the full hadronic final
state under the condition of large missing $p_t$ from the invisible high-energy
neutrino. The details of this procedure are discussed in
Refs.~\cite{ZEUS:2008arl,Amaldi:1979qp}. Since the data we compare to have been corrected for detector effects,
we can
ignore this complication and consider the recoil neutrino as known for the determination of $Q^2$, $x$ and $y$.
In the analysis, events are selected following the experimental phase space cuts:

\begin{equation}
Q^2 > 200 \GeV^2,\qquad y < 0.9, \qquad N_{\mathrm{jets}} \geq 1, \qquad E_{T,\text{leading jet}} > 14 \GeV, \qquad E_{T, \text{any jet}} > 5 \GeV~.
\end{equation}

\begin{figure}[htpb]
  \centering
      \includegraphics[width=0.4\linewidth]{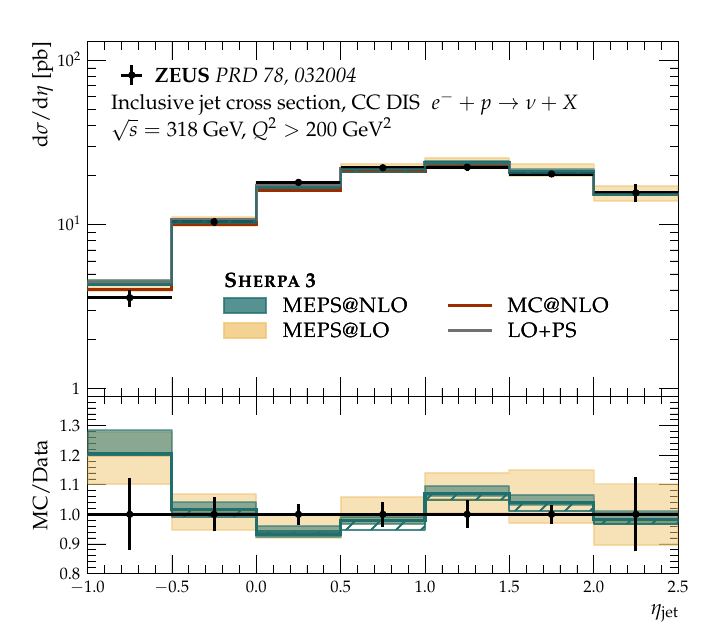}
      \includegraphics[width=0.4\linewidth]{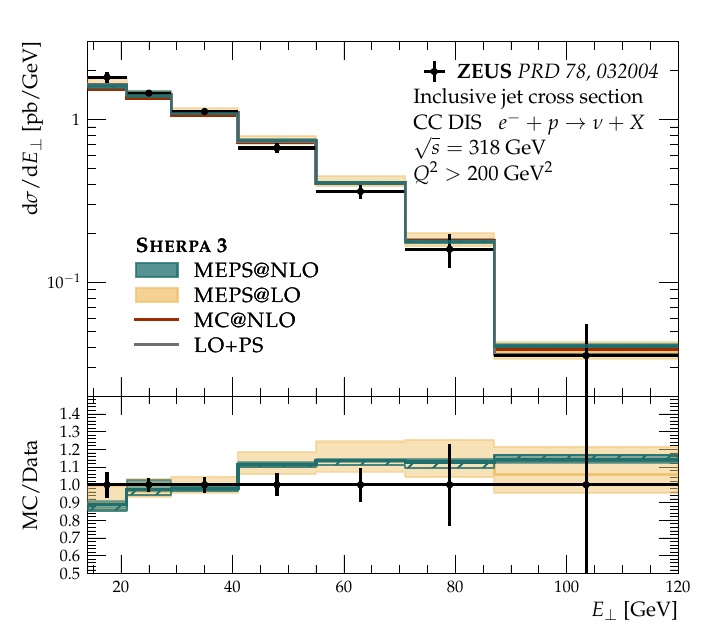}
  \caption{Differential distributions for $\eta_{\mathrm{jet}}$ and $E_{T,\mathrm{jet}}$
    at LO, \protect\MCatNLO, \protect\MEPSatLO and \protect\MEPSatNLO in
    inclusive CC DIS at the \protect\HERA. The lower panels show
    the ratio between data and the merged \protect\MEPSatLO and
    \protect\MEPSatNLO predictions, while we omit the ratio for the \protect\LO
    and \protect\MCatNLO simulations for clarity. The full (hatched) band corresponds to variations of $\muF,\muR$ ($Q_{\mathrm{cut}}$). Experimental data points are from Ref.~\cite{ZEUS:2008arl}.
    \label{fig:cc-hera2:dis-obs-2}}
\end{figure}

\begin{figure}[htpb]
  \centering
  \begin{tabular}{cc}
      \includegraphics[width=0.4\linewidth]{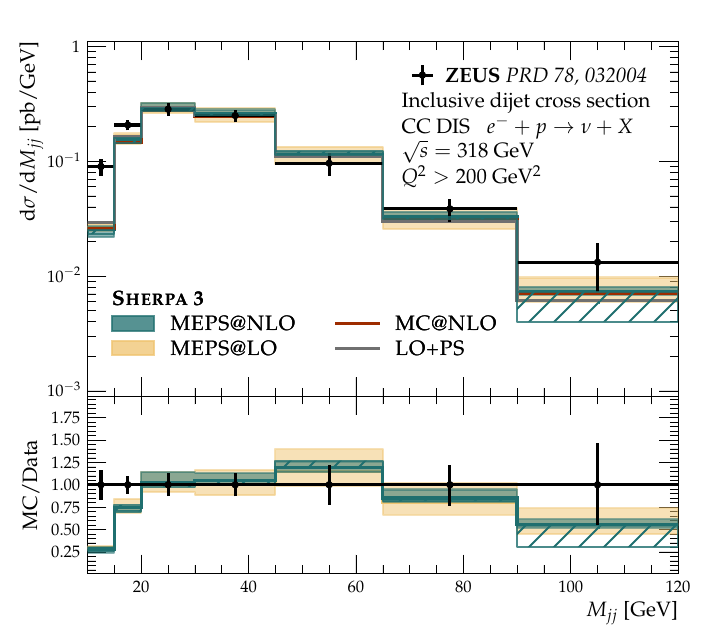} &
      \includegraphics[width=0.4\linewidth]{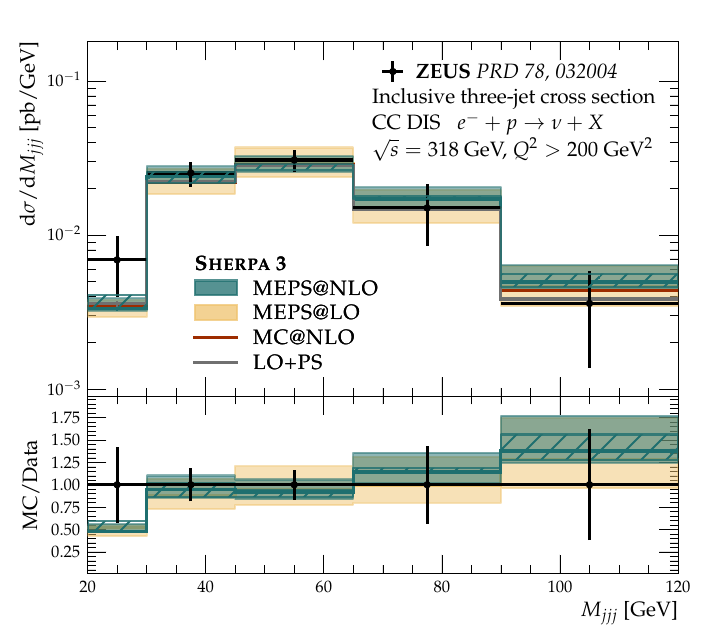}
  \end{tabular}
  \caption{Multijet distributions differential in the mass of the two (left) or
    three (right) leading jets at LO, \protect\MCatNLO, \protect\MEPSatLO and
    \protect\MEPSatNLO in inclusive CC DIS at the \protect\HERA. The lower panels show
    the ratio between data and the merged \protect\MEPSatLO and
    \protect\MEPSatNLO predictions, while we omit the ratio for the \protect\LO
    and \protect\MCatNLO simulations for clarity. The full (hatched) band corresponds to variations of $\muF,\muR$ ($Q_{\mathrm{cut}}$).
    Experimental data points are from Ref.~\cite{ZEUS:2008arl}.
    \label{fig:cc-hera2:dis-multi-jet}}
\end{figure}

In Figs.~\ref{fig:cc-hera2:dis-obs-1} and \ref{fig:cc-hera2:dis-obs-2}, we reproduce the differential cross sections for inclusive-jet production as functions of $Q^2$, $\xBj$, and
leading-jet $\eta$ and $E_{\perp}$. Due to the relatively large $Q^2$ requirement,
higher-order corrections are rather small, and we find the simulations at all
accuracy levels to be consistent with each other and with the data. In
particular, the lack of a sizable shift between the \protect\LO and
\protect\MEPSatNLO predictions can be attributed to the high $Q^2$ cut.
This requirement restricts the phase space to regions where merging effects remain small.
In the lower panels, we show the ratio of the merged simulations to data.
Since the central values for all Monte Carlo accuracies are very consistent,
we do not show ratios for the \LO and \MCatNLO runs to increase legibility.
For the two remaining runs, we also plot the band obtained from scale variations.
Notably, while the central value is almost unchanged when going from \MEPSatLO to \MEPSatNLO,
the scale dependence is reduced dramatically. This leads to some visible systematic
trend of the \MEPSatNLO simulation against data, especially in the transverse
energy distribution, but the data are not precise enough there for a definite
conclusion.
Finally, we include an uncertainty band corresponding to the
  variations of the merging scale. In order to keep the plots legible, we only
  show the band for the \MEPSatNLO calculation. The size is typically a few
  percent, though at the level of uncertainty indicated by the scale variations
  it can be the dominant effect for  some bins.
In Fig.~\ref{fig:cc-hera2:dis-multi-jet}, the dijet and three-jet invariant
mass cross sections are presented. The different Monte Carlo samples are
again very consistent with each other. For the dijet mass, some deviations
appear for very small and very large masses. However only the deviations at very small masses
have any statistical significance, and are driven by hadronization
  effects as we have confirmed by comparing parton and hadron level in our
  simulations. In the range $20 \GeV < M_{jj} < 120 \GeV$, the 
Monte Carlo predictions are consistent with the data. In the three-jet mass
case, prediction and data are consistent over the full measured range, albeit
admittedly with significantly larger uncertainties on the data. The scale
variations are reduced in both cases, although they remain sizable in the
three-jet case. The merging scale uncertainties are of similar relative
  size as in the inclusive jet case, and hence much smaller than the scale
  variations even for the \MEPSatNLO calculation.

Overall, we find the agreement between Monte Carlo simulation and data
satisfactory for CC DIS as well. In the future it would be
interesting to study further analyses performed by the HERA collaborations in
this process. Unfortunately, the availability of automated analyses in \rivet is
limited in this case.

\section{Event generation for the \protect\EIC and its systematic uncertainties}\label{sec:eic-predictions}
In this section, we present predictions for DIS at the \EIC for both neutral and
charged current interactions. As outlined in Sec.~\ref{sec:DISkin},
we analyze a combination of global observables, including the $Q^2$, $y$, and
$\xBj$ distributions. For neutral current interactions, we additionally provide
results for the transverse momentum of the recoil lepton, while for charged
current interactions, we examine the missing transverse momentum.

Furthermore, we compare jet properties
obtained using the $k_t$, anti-$k_t$, and \Centauro clustering algorithms,
highlighting differences in jet multiplicities and leading jet transverse
momentum. Finally, to probe the hadronic final-state, we study the 1-jettiness
distribution, presented in slices of $Q^2$ and $\xBj$.

These observables are evaluated using all Monte Carlo configurations discussed
in Sec.~\ref{ssec:MC-modes}, allowing for a detailed comparison of the effects
of matching, merging, and higher-order corrections.
Our main goal is to evaluate the role of \MEPSatNLO and in what phase space it
is to be preferred to the other configurations.
We also estimate hadronization uncertainties as explained in
Sec.~\ref{ssec:uncertainties}.

For both neutral and charged current DIS, we assume energies of $18$ and
$275\GeV$ for the lepton and proton beams respectively; this corresponds to
the high-luminosity configuration of the \EIC. We do not consider any
polarization of the beams.

To arrive at realistic predictions, we discard particles outside an assumed
detector acceptance of $\abs{\eta} > 4$ during the analysis and apply a cut on
the inelasticity, keeping the interval $0.2 < y < 0.9$. Jets are clustered with
a minimal transverse momentum of $5\GeV$ and a jet radius of $R=1$.
While we follow the setup of tradition studies like Ref.~\cite{Page:2019gbf},
 the \Centauro algorithm presents new advantageous options that should be explored 
 in dedicated studies~\cite{H1:2025}.

\subsection{Neutral current}

\begin{figure}[htpb]
  \centering
  \begin{tabular}{cc}
      \includegraphics[width=0.4\linewidth]{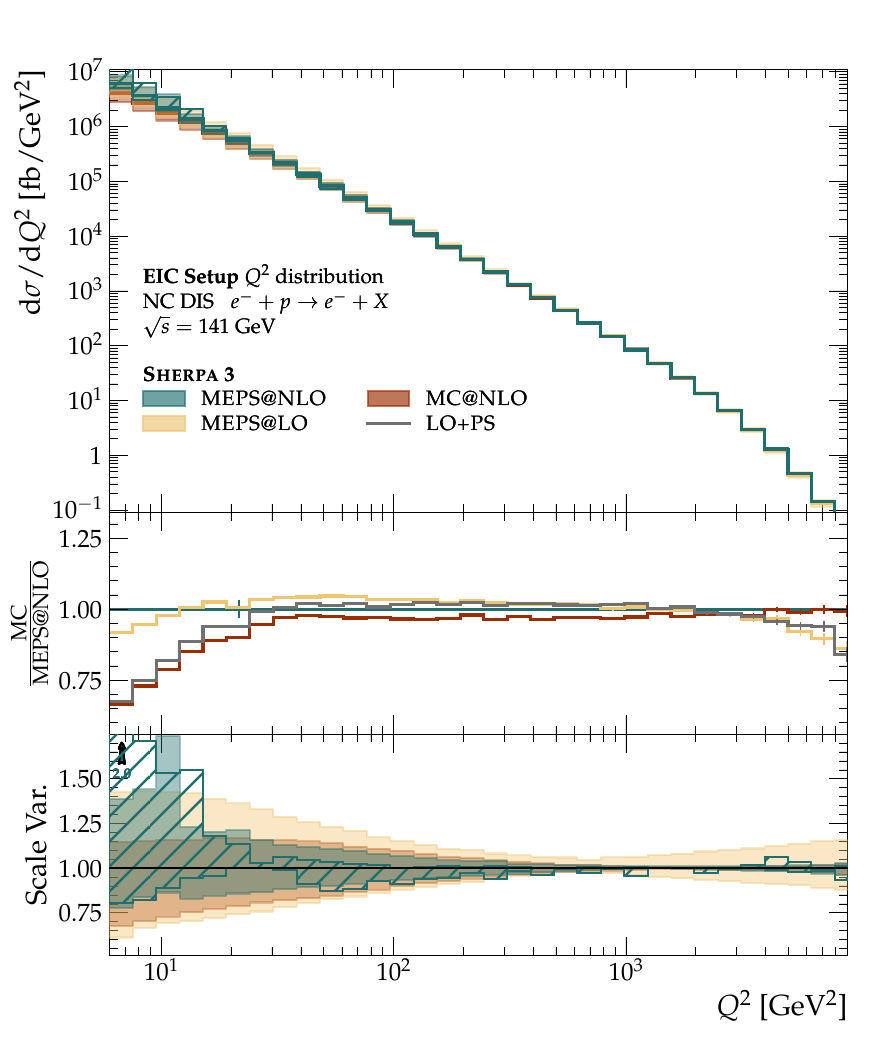} &
      \includegraphics[width=0.4\linewidth]{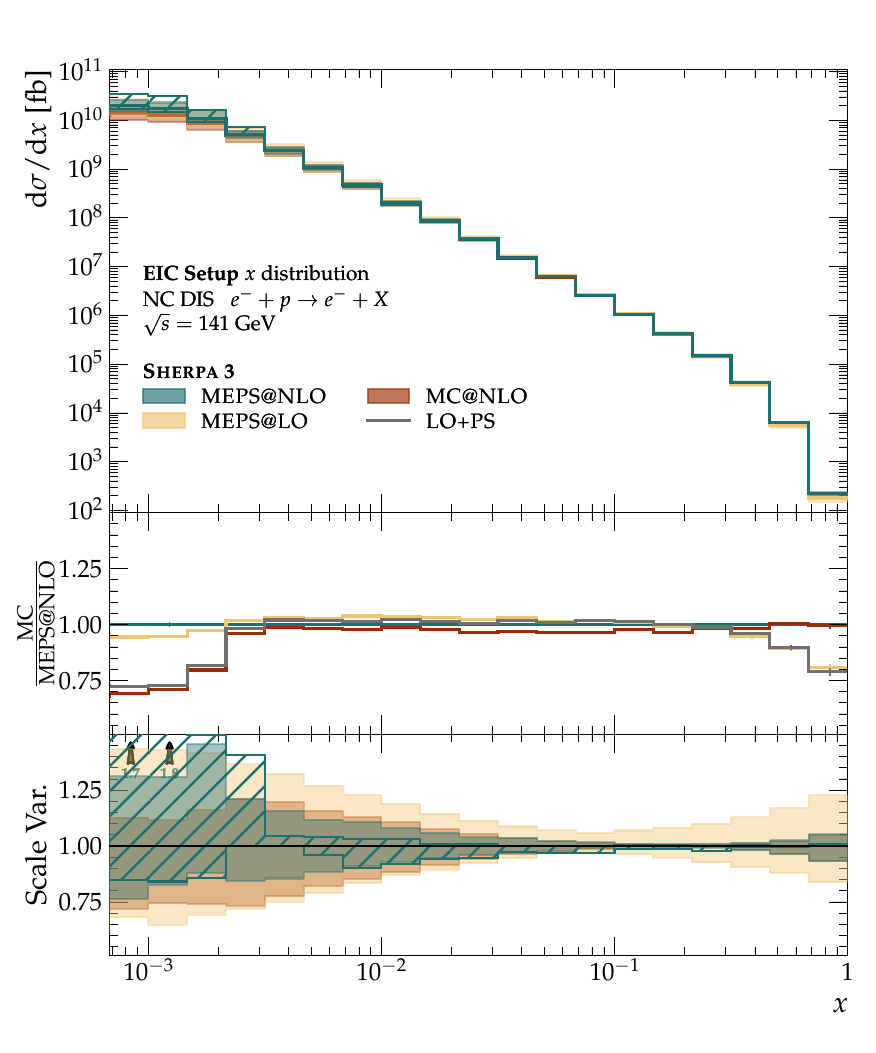} \\
      \includegraphics[width=0.4\linewidth]{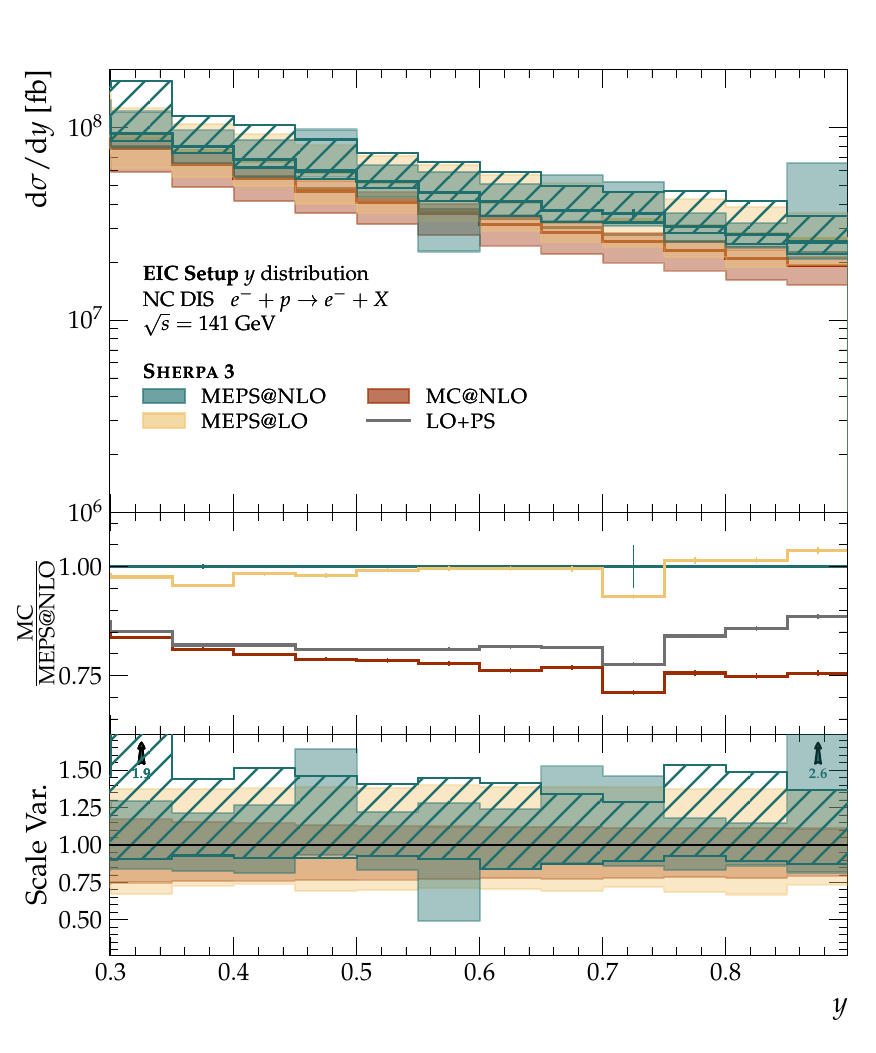} &
      \includegraphics[width=0.4\linewidth]{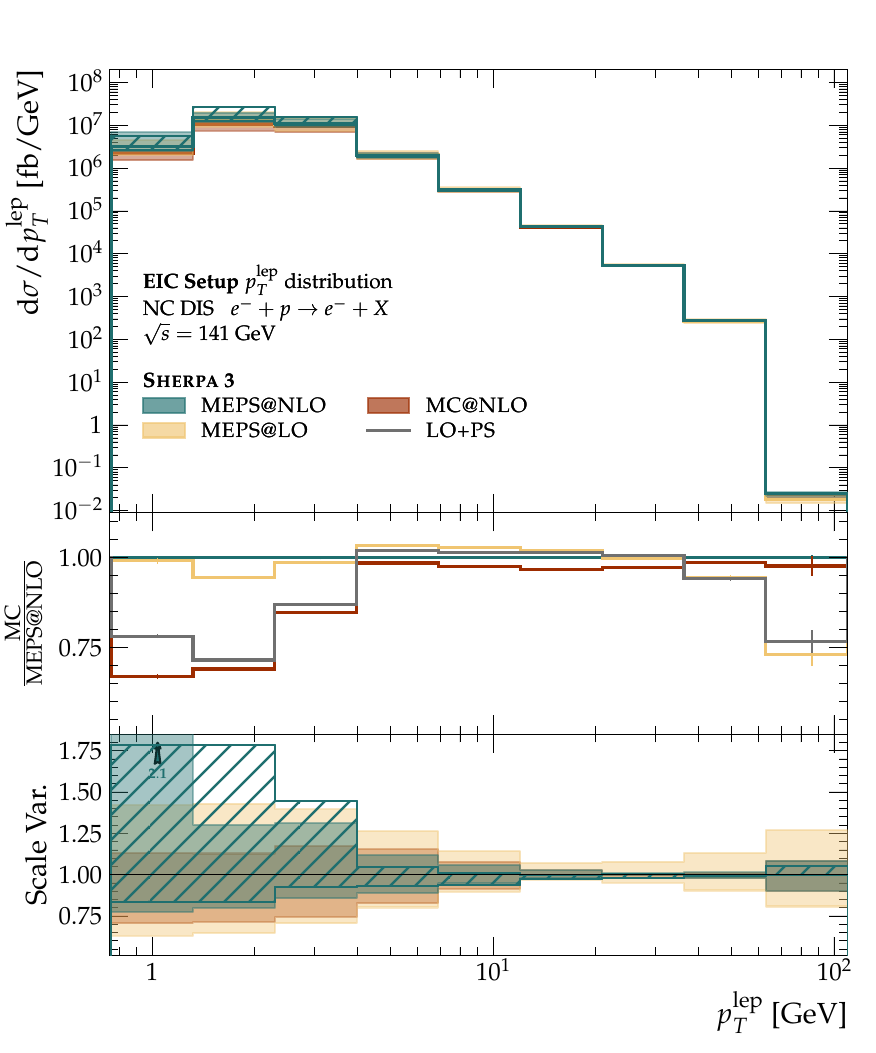}
  \end{tabular}
  \caption{Differential distributions for $Q^2$ (top left), $x$ (top right), $y$
    (bottom left) and lepton transverse momentum $\ptsup{\mathrm{lep}}$ (bottom
    right) in NC DIS at the \protect\EIC, comparing LO+PS,
    \protect\MCatNLO, \protect\MEPSatLO and \protect\MEPSatNLO predictions. The
    main panels show results with scale variation uncertainties for all but the
    LO+PS calculation. The first ratio panel displays the relative deviation
    from \MEPSatNLO to highlight merging effects. The second ratio shows the
    relative scale uncertainties. The hatched band represents merging
     scale variation $Q_{\mathrm{cut}}$ in the \MEPSatNLO sample. \label{fig:nc-eic:dis-obs}}
\end{figure}

We consider the neutral current setup first. In this case, we place a small
virtuality cut, $ Q^2 > 6 \GeV^2$. We present the global DIS observables
$Q^2$, $\xBj$, $y$, and $\ptsup{\mathrm{lep}}$ in
Fig.~\ref{fig:nc-eic:dis-obs}. For $Q^2$ and $\xBj$, the merged calculations
exhibit distinct behavior from the matched \MCatNLO approach. In the $Q^2$
distribution (top left), we observe that for $Q^2 < 20 \GeV^2$, the \MEPSatNLO
prediction exceeds LO and \MCatNLO of up to 30\%, while showing a more modest
10\% enhancement over \MEPSatLO. This enhancement exceeds the scale uncertainty
bands between \MEPSatNLO and \MCatNLO but remains within the \MEPSatLO
uncertainties. The predictions converge in the intermediate region $20\GeV^2
\leq Q^2 \leq 5\cdot10^{3}\GeV^2$. At higher virtualities ($Q^2 >
5\cdot10^{3}\GeV^2$), we observe a systematic 10\% deficit in the LO and
\MEPSatLO predictions relative to the NLO-accurate calculations, showing the
importance of the virtual $\mathcal{O}(\alphaS)$ corrections in this
regime.

Similar patterns can be seen in the $\xBj$ and $\ptsup{\mathrm{lep}}$
distributions (top and bottom right). Multijet merging effects dominate at
$\xBj < 1.5\cdot10^{-3}$ ($\ptsup{\mathrm{lep}} < 5 \GeV$), while NLO matching
becomes crucial at $\xBj > 0.7$ ($\ptsup{\mathrm{lep}} > 65 \GeV$). The
\MEPSatNLO calculation shows reduced scale uncertainties compared to \MEPSatLO
in the low-$\xBj$ and low-$\ptsup{\mathrm{lep}}$ regions. It also shows smaller
uncertainties than \MCatNLO in the intermediate ranges of around $Q^2 \approx
10^2 \GeV^2$ and $\xBj \approx 10^{-2}$ as a consequence of calculating the
$2\to 3$ process to NLO. The inelasticity distribution (bottom left) reveals
good agreement between \MEPSatNLO and \MEPSatLO across the full range, while LO
and \MCatNLO predictions are uniformly suppressed by approximately 25\%, though
this deviation remains within the combined scale uncertainty bands.

The merging cut variations naturally are largest in the regions where
  merging has the largest impact, i.e. for small $Q^2$ and $x$, indicating a
  significant uncertainty on the way the DIS regime is
  effectively  interpolated to the photoproduction region. The inelasticity
  distribution is affected everywhere by a flat shift from the scale
  variations. This is simply due to the fact that we integrate over the full
  $Q^2$ range, if we were to cut more aggressively to ensure a measurement
  cleanly in the DIS regime, we would expect a much smaller uncertainty.

The observations above are in line with the general trends for example at
HERA energies. We do not display any hadronic uncertainty as these observables
are not affected within the models we consider.

\begin{figure}[htpb]
  \centering
  \begin{tabular}{ccc}
    \includegraphics[width=0.3\linewidth]{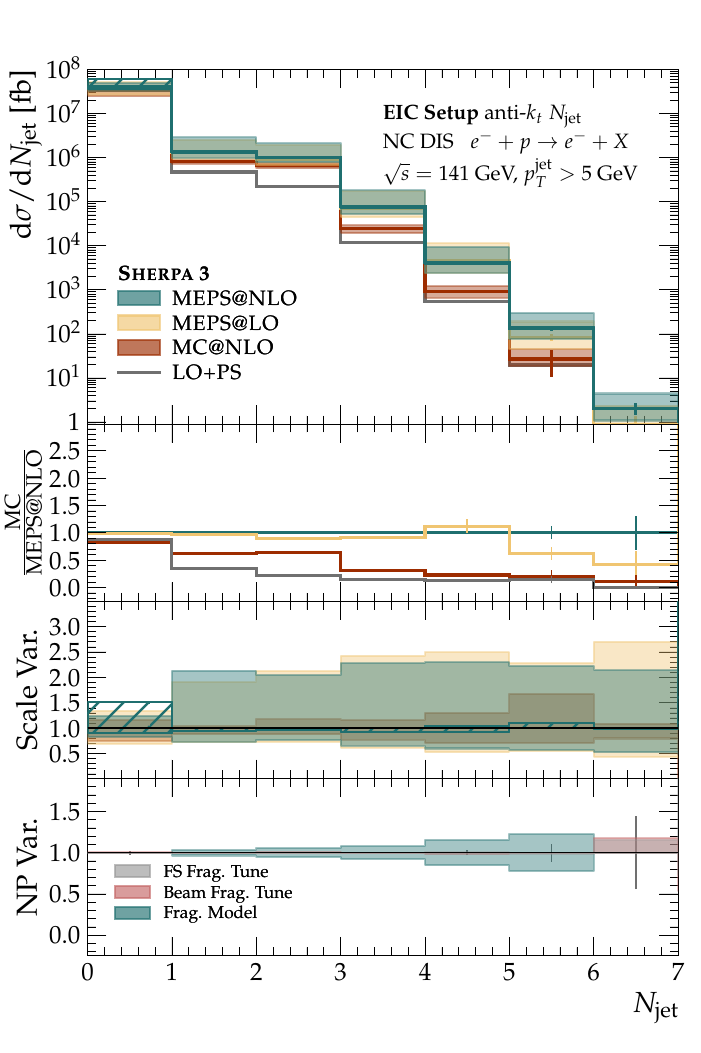} &
    \includegraphics[width=0.3\linewidth]{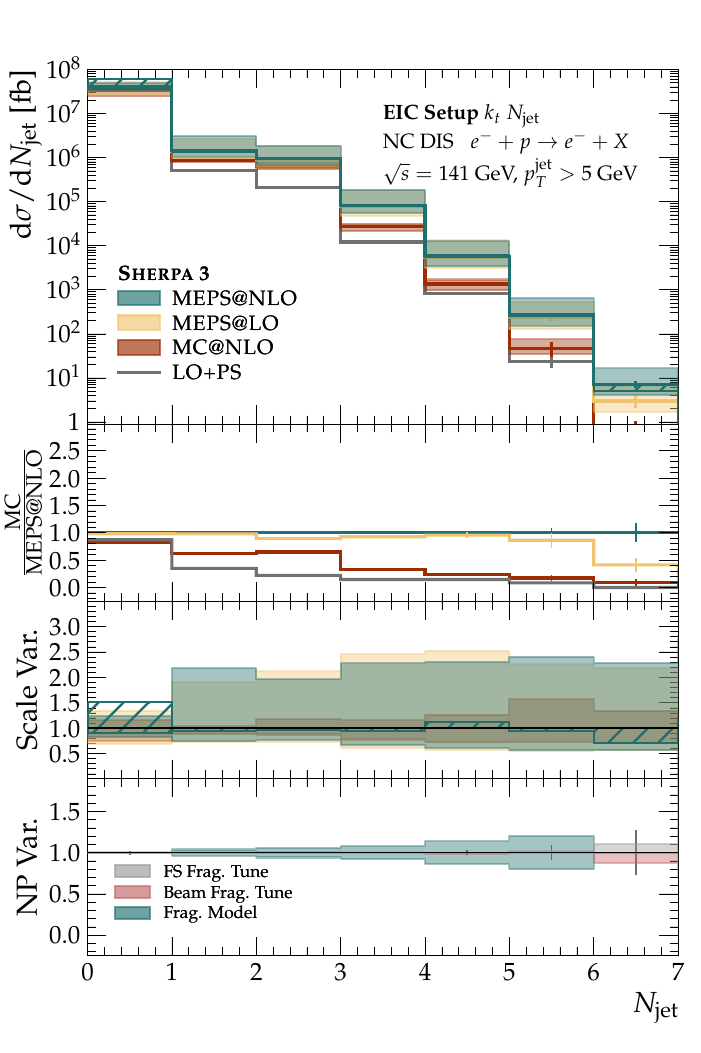} &
    \includegraphics[width=0.3\linewidth]{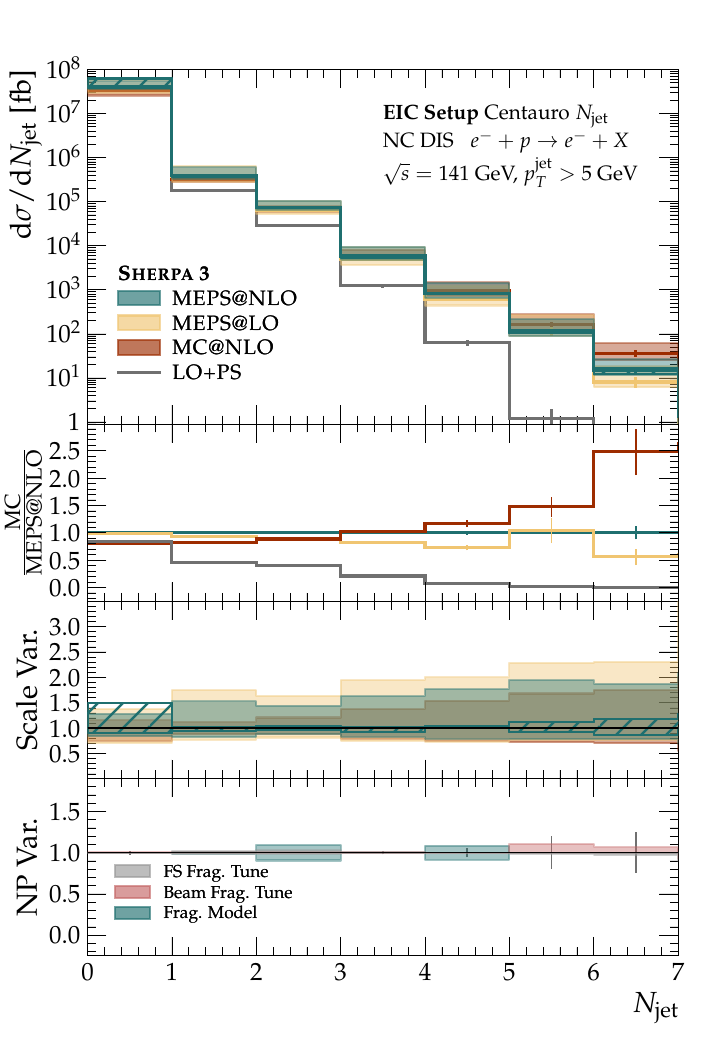} \\
    \includegraphics[width=0.3\linewidth]{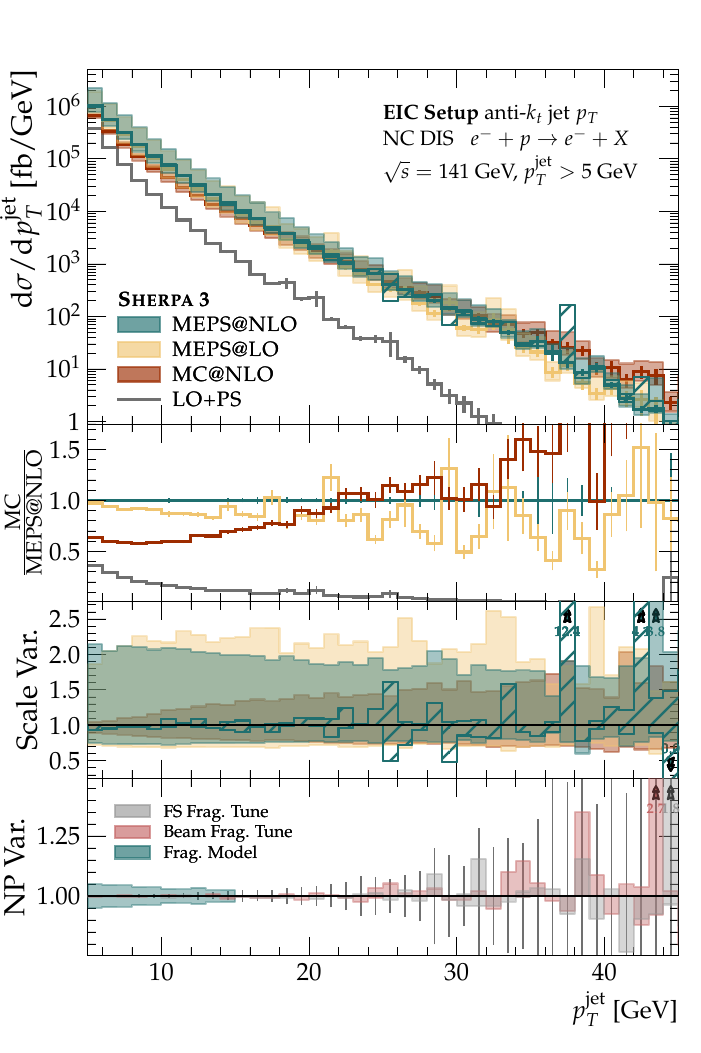} &
    \includegraphics[width=0.3\linewidth]{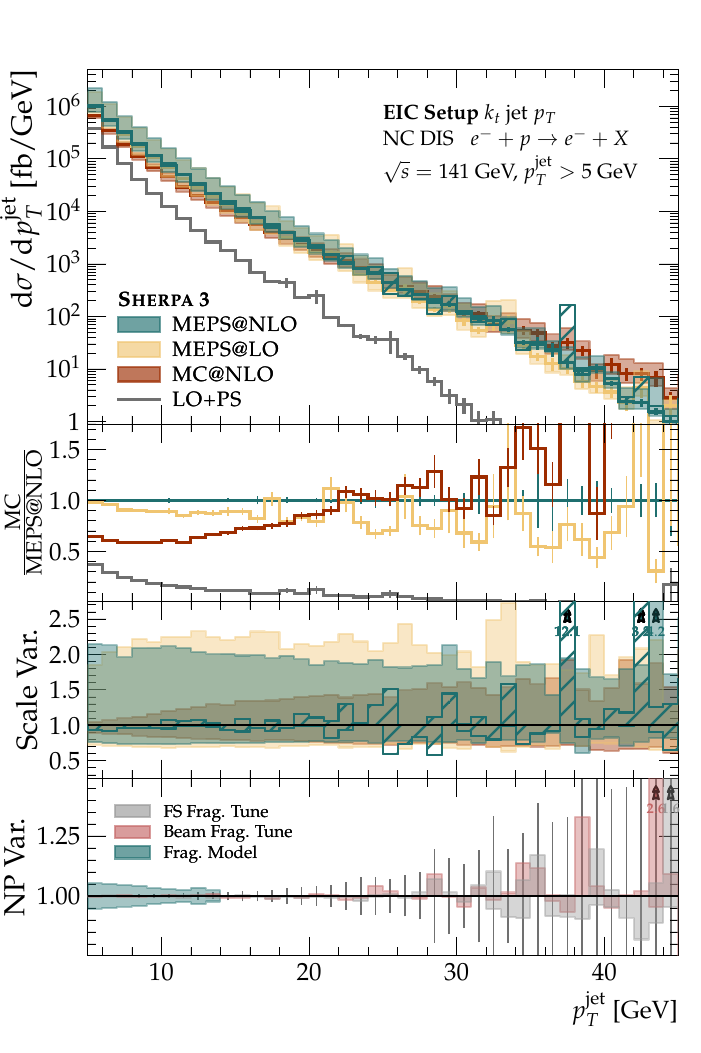} &
    \includegraphics[width=0.3\linewidth]{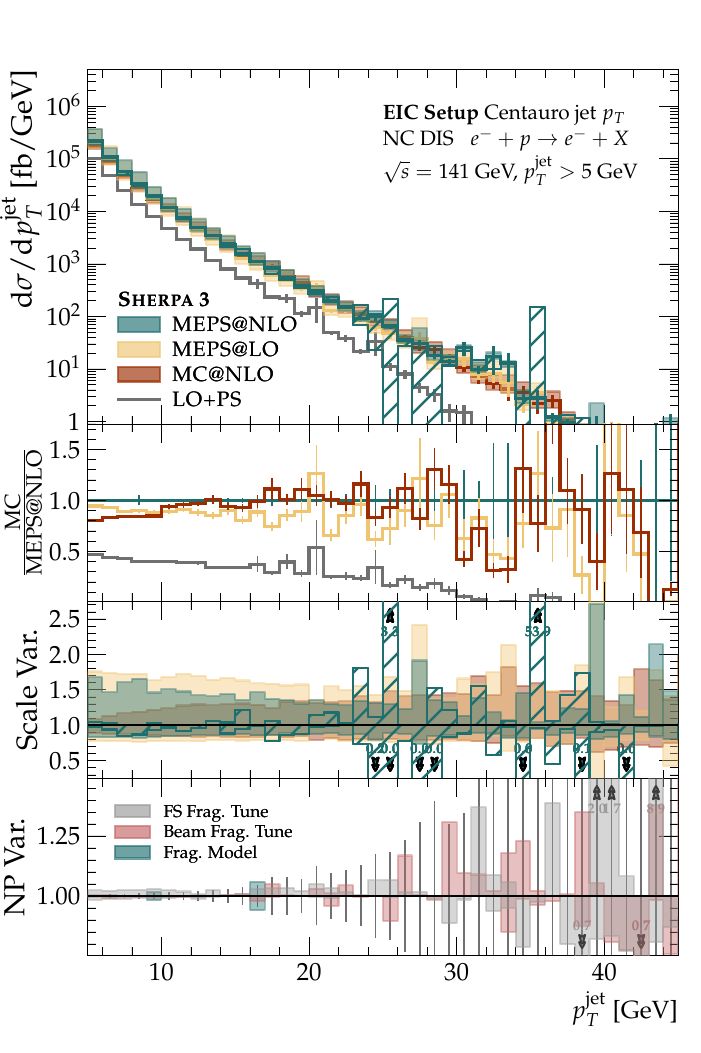}
  \end{tabular}
  \caption{Jet multiplicity $N_\mathrm{jet}$ (top row) and leading jet
    transverse momentum $p_T^\mathrm{jet}$ (bottom row) distributions for
    anti-$k_t$ (left), $k_t$ (middle) and \Centauro (right) jet algorithms in
    NC DIS at the \protect\EIC. The bottom ratios in each row
    indicate, from top to bottom, the ratio of each prediction to the
    \MEPSatNLO one, the uncertainty from scale variations and the uncertainty
    from nonperturbative model and tuning variations.
     The hatched band represents merging scale variation
      $Q_{\mathrm{cut}}$ in the \MEPSatNLO sample. The error bars in the
    lowest panel indicate the statistical uncertainty of the difference between
    cluster and string model.\label{fig:nc-eic:jet-obs}}
\end{figure}

Figure~\ref{fig:nc-eic:jet-obs} presents the jet multiplicity and leading jet
transverse momentum distributions for three different clustering
algorithms. While \MEPSatNLO and \MEPSatLO predictions show good concordance,
both \MCatNLO and LO calculations yield systematically lower rates. Despite
reduced scale uncertainties in \MEPSatNLO compared to \MEPSatLO, both merged
calculations exhibit significant scale variations, reflecting the LO accuracy of
higher-multiplicity configurations that dominate the low-$Q^2$ regime and that
contribute significantly in these observables. Merging cut variations
play a subleading role for these observables.
It is also worth noting that the \Centauro clustering algorithm leads to
smaller higher-order corrections in this analysis. Indeed, the LO simulation
is unable to fill the whole phase space. For the $k_t$ and anti-$k_t$, the \MCatNLO
sample is still off, compared to the merged ones, by a $K$-factor of about 1.5 for the low-$p_T$
region. Conversely, the \Centauro algorithm shows much smaller $K$-factors, in fact only a
factor of 1.2 for the low-$p_T$ region with respect to the \MEPSatNLO
prediction.

Hadronization uncertainties are generally smaller than the perturbative uncertainties and,
as one might have expected, are largest for small $p_T$ (up to $5\%$) and for large jet multiplicities $(1-2\%)$.
In the latter, we find that only a variation of the fragmentation model induces a significant effect,
 while replica tunes within the cluster model are negligible.
Instead, in the $p_T$ spectrum, fragmentation model uncertainty is still the largest nonperturbative contribution
for the anti-$k_t$ and $k_t$ jets, while variations from replica tunes are larger than model ones for \Centauro jets, 
although both are especially small.
Beyond small values of $p_T$, the difference between hadronization models becomes unresolvable within the accumulated statistics.
Instead, the combined statistical uncertainty from the two models is taken as an overestimate.
These values are still comparable or subdominant to the perturbative uncertainties in the same part of the spectrum.

\begin{figure}[ht]
  \centering
  \includegraphics[width=\linewidth]{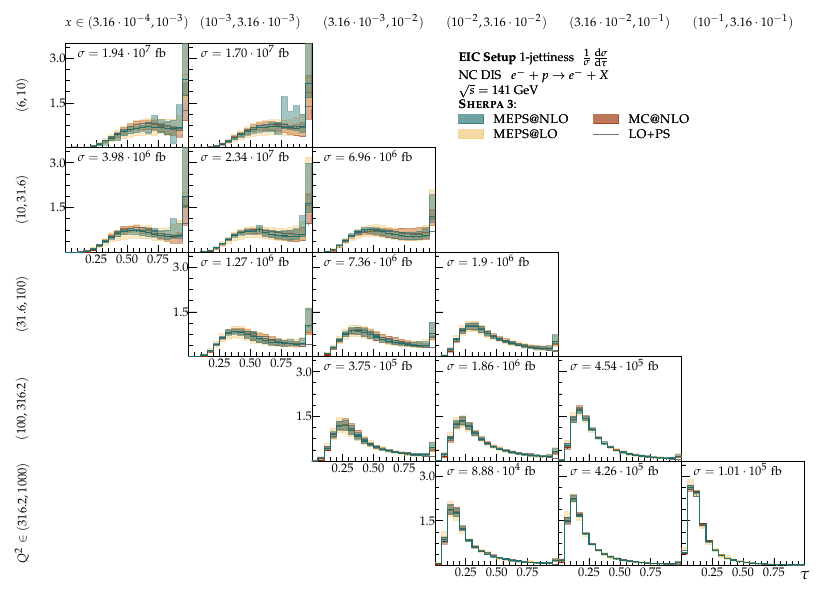}
  \caption{Differential distributions of 1-jettiness $\tau$ in different bins of
    $Q^2$ and $x$ in NC DIS at the \protect\EIC. The total \MEPSatNLO cross section for each $x$-$Q^2$ bin is 
    reported on top of the corresponding plot.}
    \label{fig:nc-eic:thrust-in-bins}
\end{figure}
We finally consider 1-jettiness, an event shape variable that is sensitive to the
underlying parton dynamics and can be used, e.g., for global fits of the strong
coupling constant $\alphaS$ \cite{Dasgupta:2002dc}. In Ref.~\cite{Ee:2025scz}, predictions of 1-jettiness at
the \EIC were computed with NNLO+N$^3$LL accuracy.
Here, we present in Fig.~\ref{fig:nc-eic:thrust-in-bins} an overview of the
1-jettiness $\tau$ across bins of $\xBj$ (left to right) and $Q^2$ (top to
bottom). For the sake of presentation we omit the ratio panels here, and the full
plots can be found in the Appendix.
Overall, the shape of the distributions agrees well between the different
configurations. Similar to the behavior at the recent H1
measurement~\cite{H1:2024aze}, we observe that the high-$Q^2$ has a pronounced
peak at small $\tau$ values, while the low-$Q^2$ distributions tend to higher
values, which underlines the importance of high-multiplicity matrix elements in
the latter regime.

\begin{figure}
  \centering
  \resizebox{\textwidth}{!}{
  \includegraphics[width=0.3333\textwidth]{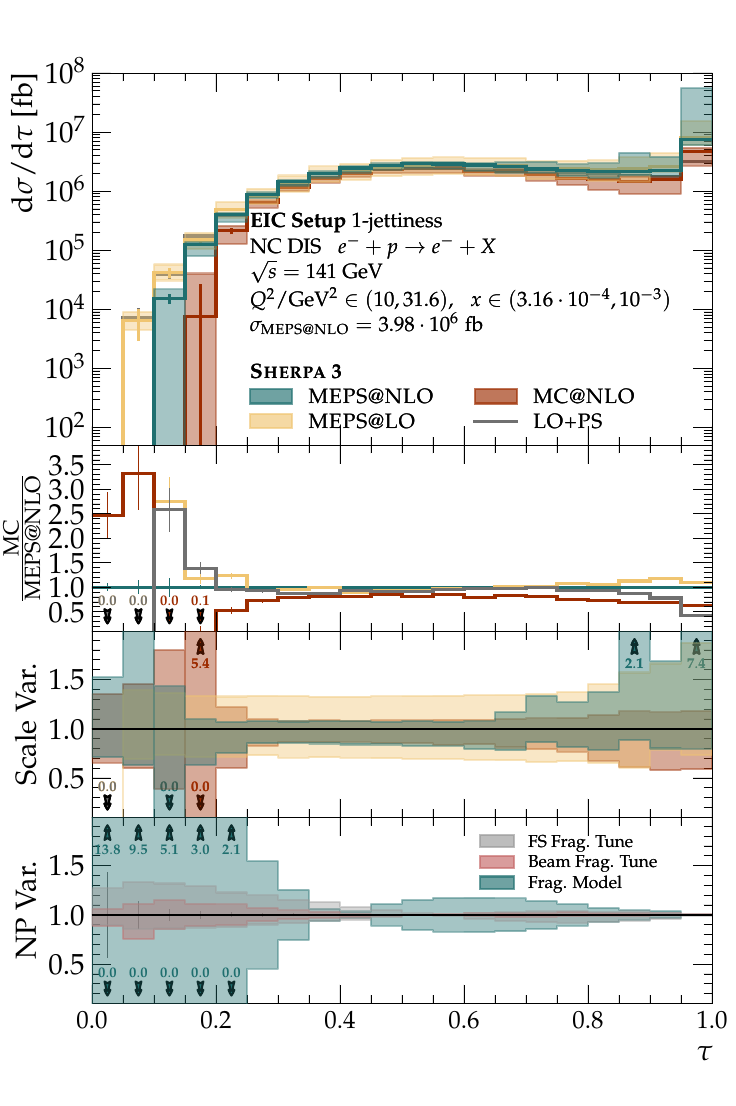}
  \includegraphics[width=0.3333\textwidth]{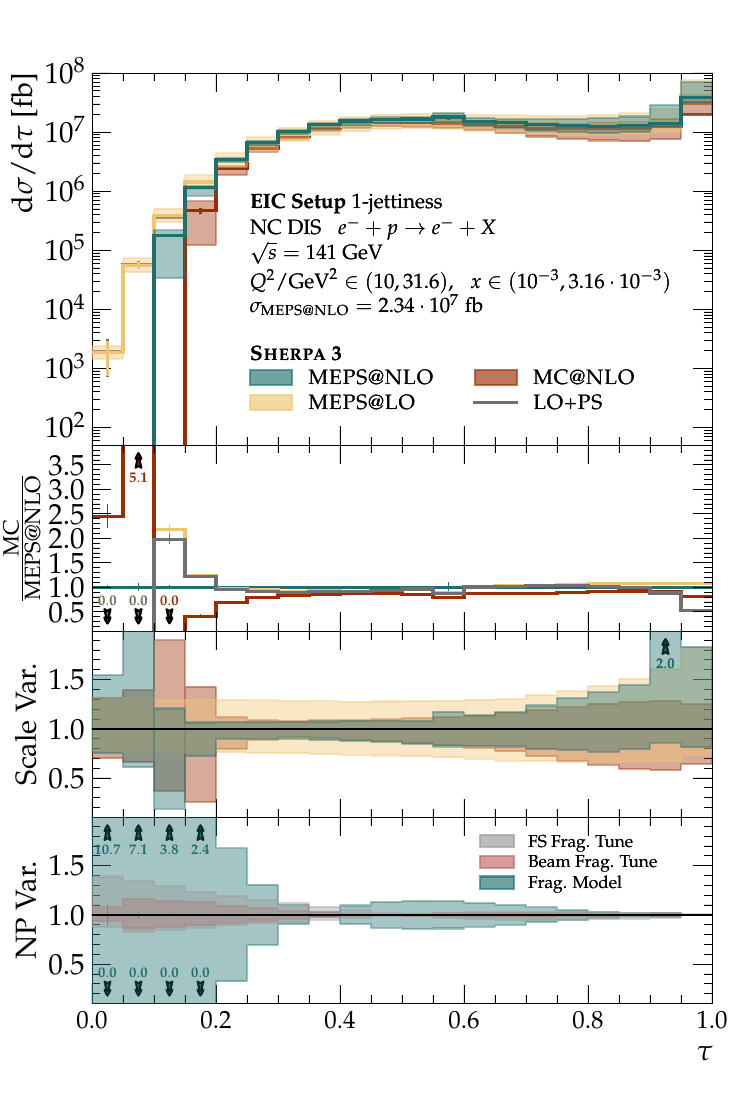}
  \includegraphics[width=0.3333\textwidth]{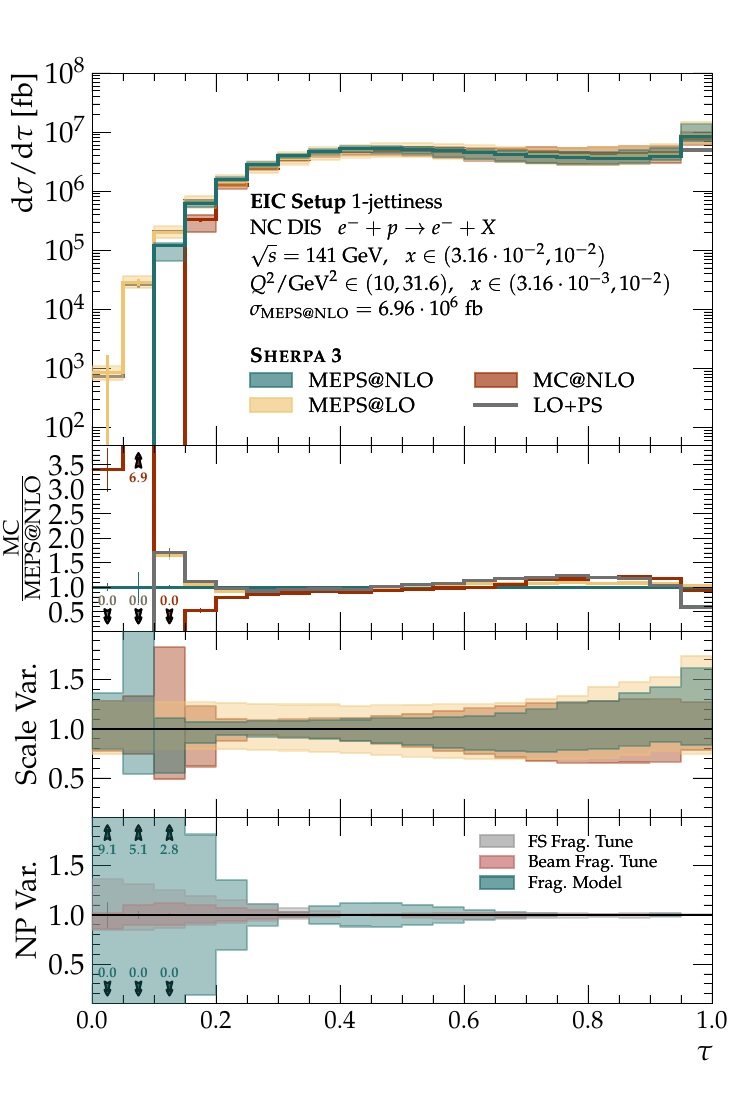}
  }
  \caption{Differential distributions of 1-jettiness $\tau$ in a fixed slice of
    $Q^2$ and different bins of $x$ in NC DIS at the \protect\EIC. The bottom ratios in each row
    indicate, from top to bottom, the ratio of each prediction to the
    \MEPSatNLO one, the uncertainty from scale variations and the uncertainty
    from nonperturbative model and tuning variations. The error bars in the
    lowest panel indicate the statistical uncertainty of the difference between
    cluster and string model. \label{fig:nc-eic-tau-Q2-slice}}
\end{figure}

For a more thorough discussion, we show the full breakdown of ratios and
uncertainties in Fig.~\ref{fig:nc-eic-tau-Q2-slice} for three different $\xBj$
regions in the $10 < Q^2/\GeV^2 < 31.6$ range. Plots for the other $\xBj$-$Q^2$
ranges in Fig.~\ref{fig:nc-eic:thrust-in-bins} can be found in
Appendix~\ref{app:nc-tau}.  We observe that the
nonperturbative uncertainties are non-negligible for small $\tau$ values and of
similar size as the \MCatNLO and \MEPSatNLO scale
uncertainties in that region. The same qualitative behavior is visible in
  both the tuning variations and the difference between string and cluster
  models, while the latter are significantly larger. This is the case for both
tuning uncertainties and variations of the hadronization model, where the latter
are dominating. For larger values of 1-jettiness, either of the considered
nonperturbative uncertainties are negligible, consistent with the observations
in Ref.~\cite{Knobbe:2023ehi}.
Furthermore, the nonperturbative uncertainties are---unsurprisingly---largest
for small values of $Q^2$ and \xBj. The shape of the distributions at small
$Q^2$ underlines the importance of multiparton topologies in this region.
  We find that variations in $Q_{\text{cut}}$ primarily affect the overall cross section
  and not the distribution itself, apart from regions where NP effect are completely dominant.
  Thus, we choose to omit them and show the 1-jettiness distributions in each $Q^2$--$x$ bin with no normalizing factor to emphasize the
effect of the different MC setups. 

\begin{figure}[htpb]
  \centering
  \resizebox{\textwidth}{!}{
  \begin{tabular}{ccc}
      \includegraphics[width=0.333\linewidth]{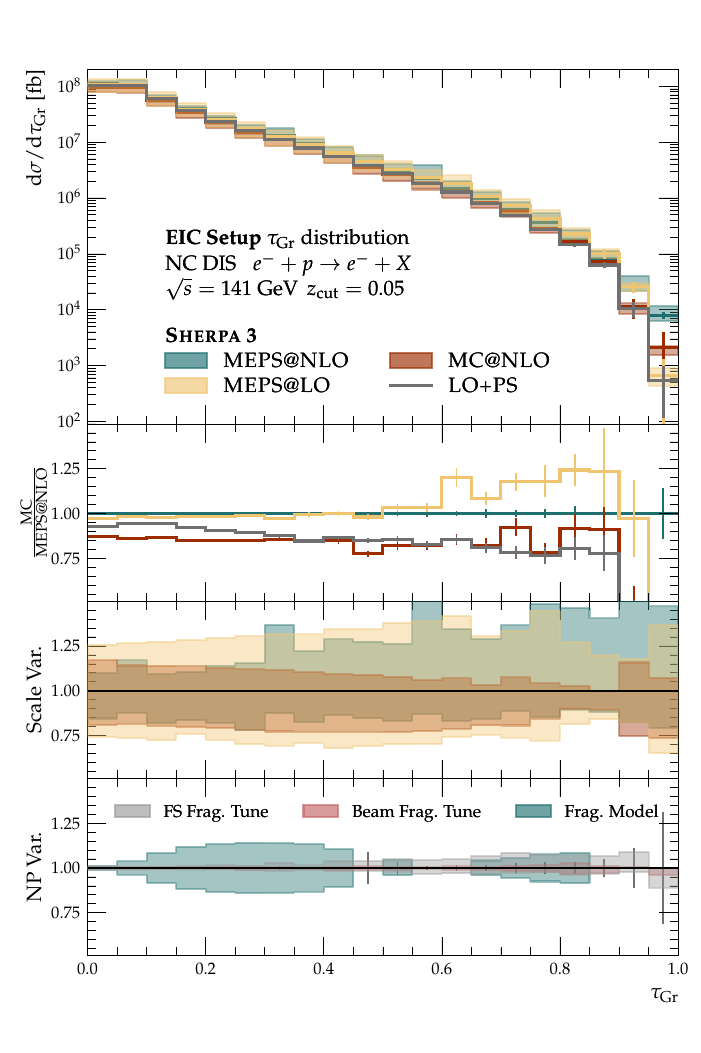} &
      \includegraphics[width=0.333\linewidth]{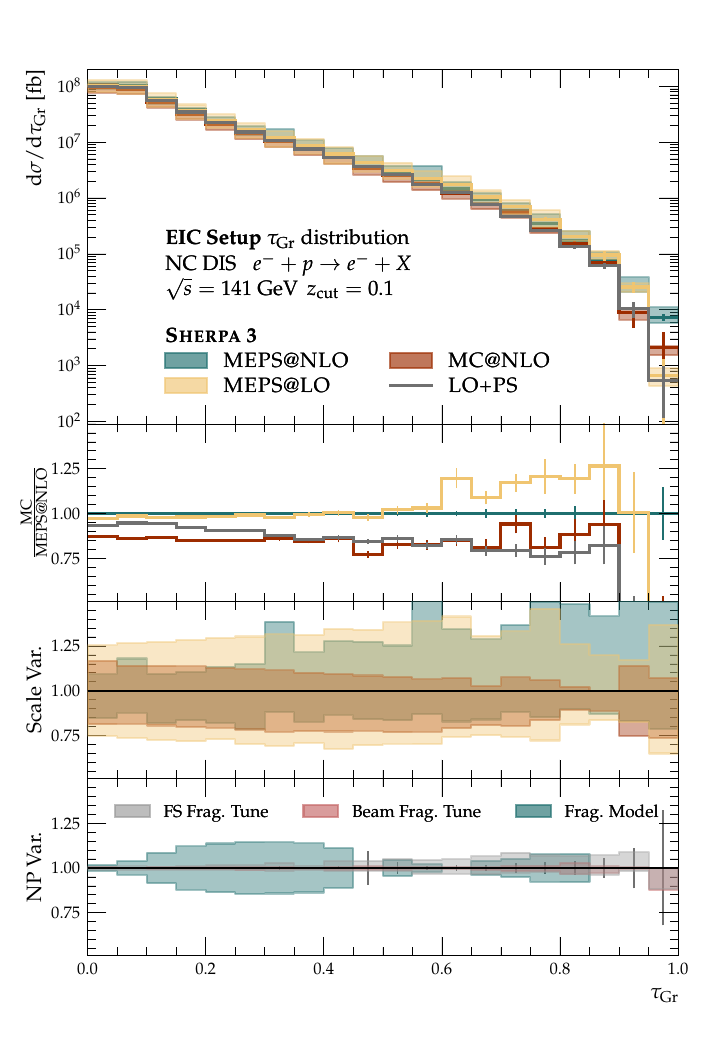} &
      \includegraphics[width=0.333\linewidth]{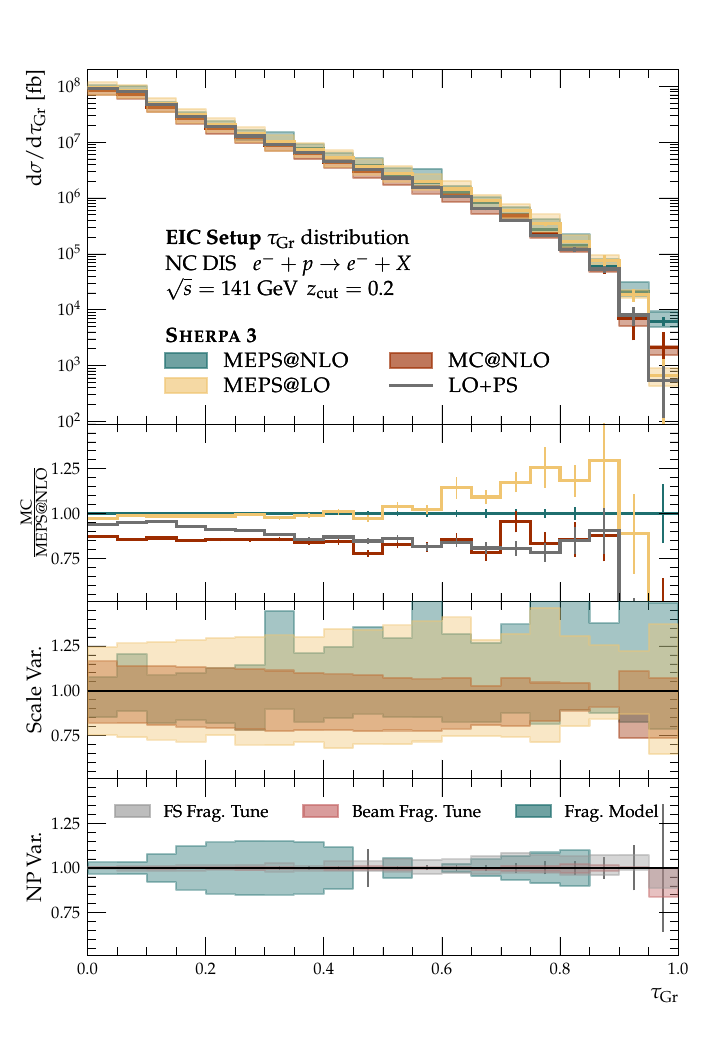}
  \end{tabular}
  }
  \caption{Groomed 1-jettiness in NC DIS at the \protect\EIC. From
    left to right the soft drop cut $z_{\text{cut}}$ is $0.05$, $0.1$ and $0.2$. The bottom ratios in each row
    indicate, from top to bottom, the ratio of each prediction to the
    \MEPSatNLO one, the uncertainty from scale variations and the uncertainty
    from nonperturbative model and tuning variations. The error bars in the
    lowest panel indicate the statistical uncertainty of the difference between
    cluster and string model.
    \label{fig:nc-eic:groomed-tau}}
\end{figure}

At last, in Fig.~\ref{fig:nc-eic:groomed-tau} we report the 1-jettiness event
shape after applying the soft drop to \Centauro jets. We consider three values
of the grooming cut from $z_{\text{cut}} \in (0.05,0.1,0.2)$, while remaining
inclusive in $Q^2$ and $x$.
As expected from the removal of soft and low scale radiation, the size of
nonperturbative uncertainties is significantly reduced compared to the ungroomed
1-jettiness and does not exhibit strong dependence on $z_{\text{cut}}$. With
respect to the \MEPSatNLO, the \MEPSatLO configuration behaves consistently up
to $\tau_{\text{Gr}} < 0.5$. Conversely, the \MCatNLO and LO+PS ones are
suppressed by a factor $~10\%$. These deviations are still small compared to
scale variations, which are reduced switching from \MEPSatLO to \MEPSatNLO up to
$\tau_{\text{Gr}}< 0.8$.
Moreover, \MEPSatNLO scale uncertainties are generally larger than \MCatNLO ones,
reinforcing the idea in the low-$Q^2$ regime LO contributions from
higher-multiplicity events provide a large correction.

\subsection{Charged current}

For the analysis of CC events, we use the same setup as for neutral
current events, but we increase the virtuality cut to $Q^2 > 10\GeV^2$. This
is to account for the experimentally difficult extraction of the correct event
kinematics.

\begin{figure}[htpb]
  \centering
      \includegraphics[width=0.49\linewidth]{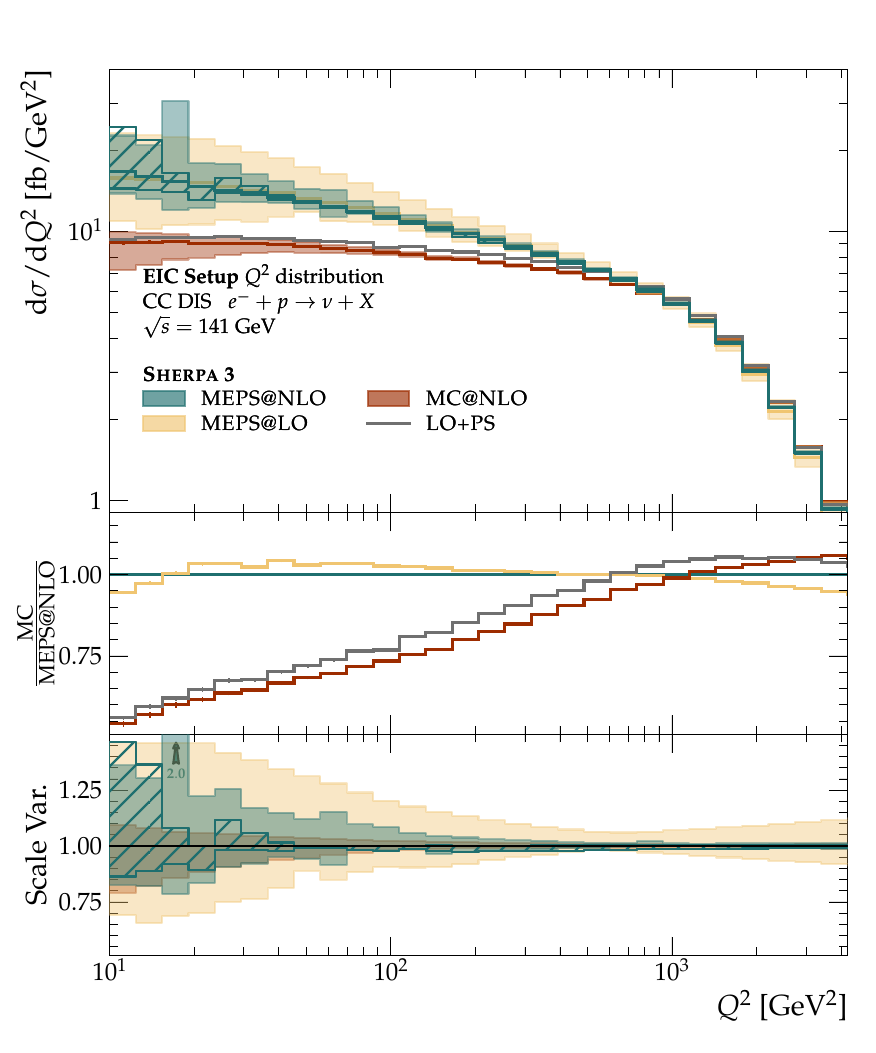}
      \includegraphics[width=0.49\linewidth]{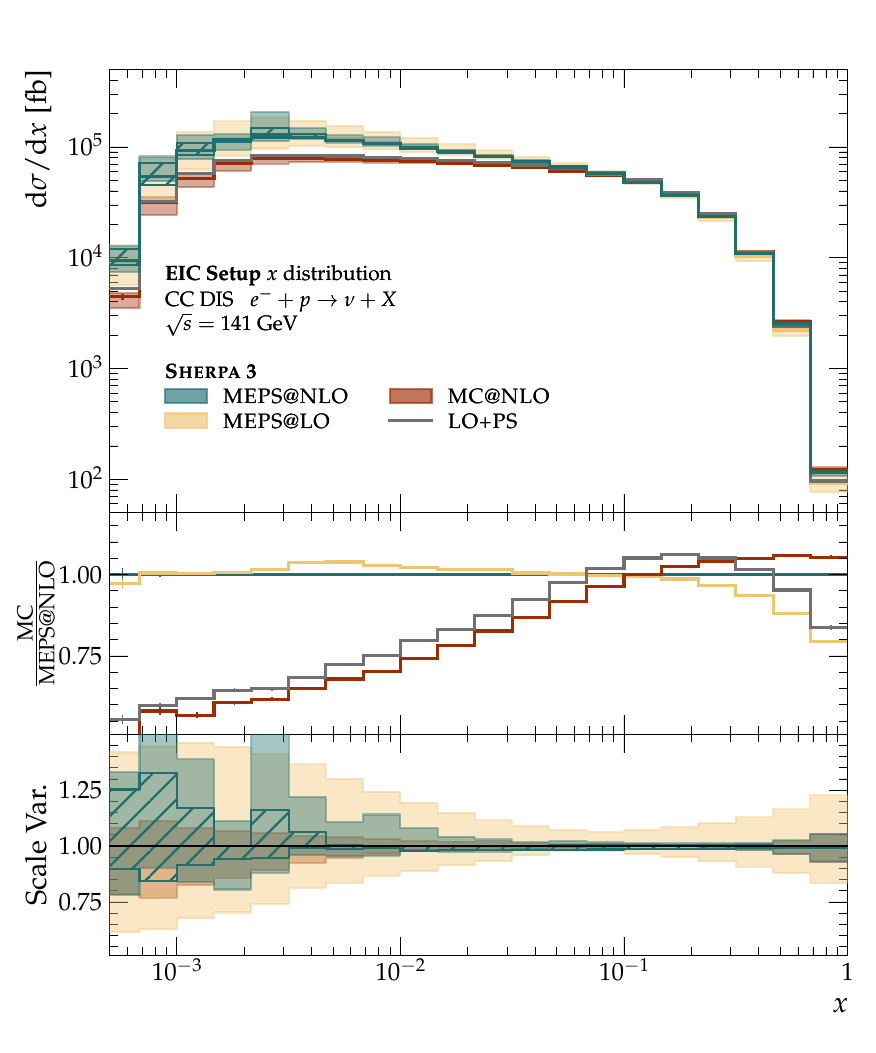}
  \caption{Differential distributions for $Q^2$ and $x$ in charged current DIS at the \protect\EIC, comparing LO+PS, \protect\MCatNLO, \protect\MEPSatLO and \protect\MEPSatNLO predictions. The main panels show results with scale variation uncertainties for all but the LO+PS calculation. The first ratio panel displays the relative deviation from \MEPSatNLO to highlight merging effects.
   The second ratio shows the relative scale uncertainties,  as detailed Sec.~\ref{ssec:uncertainties}. Full (hatched) band for $\muF,\muR$ ($Q_{\mathrm{cut}}$) respectively.
   }
  \label{fig:cc-eic:dis-obs-1}
\end{figure}

\begin{figure}[ht]
  \centering
  \includegraphics[width=0.49\linewidth]{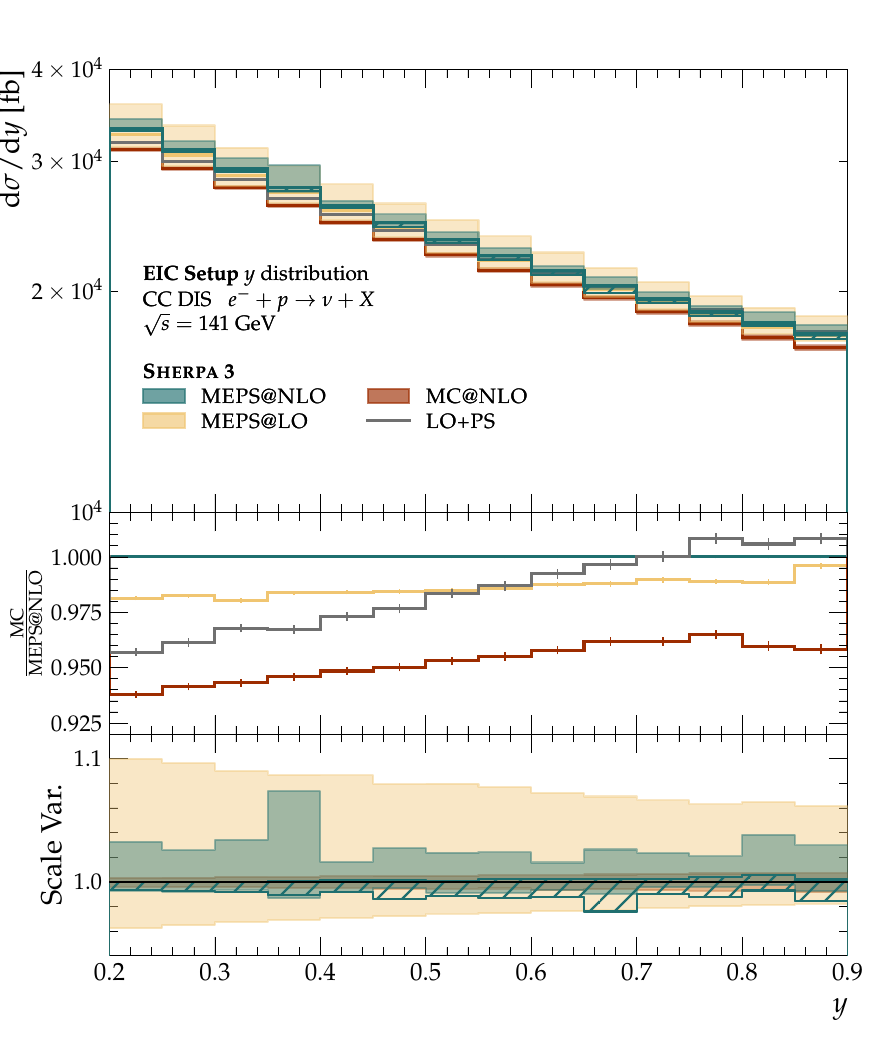}
  \includegraphics[width=0.49\linewidth]{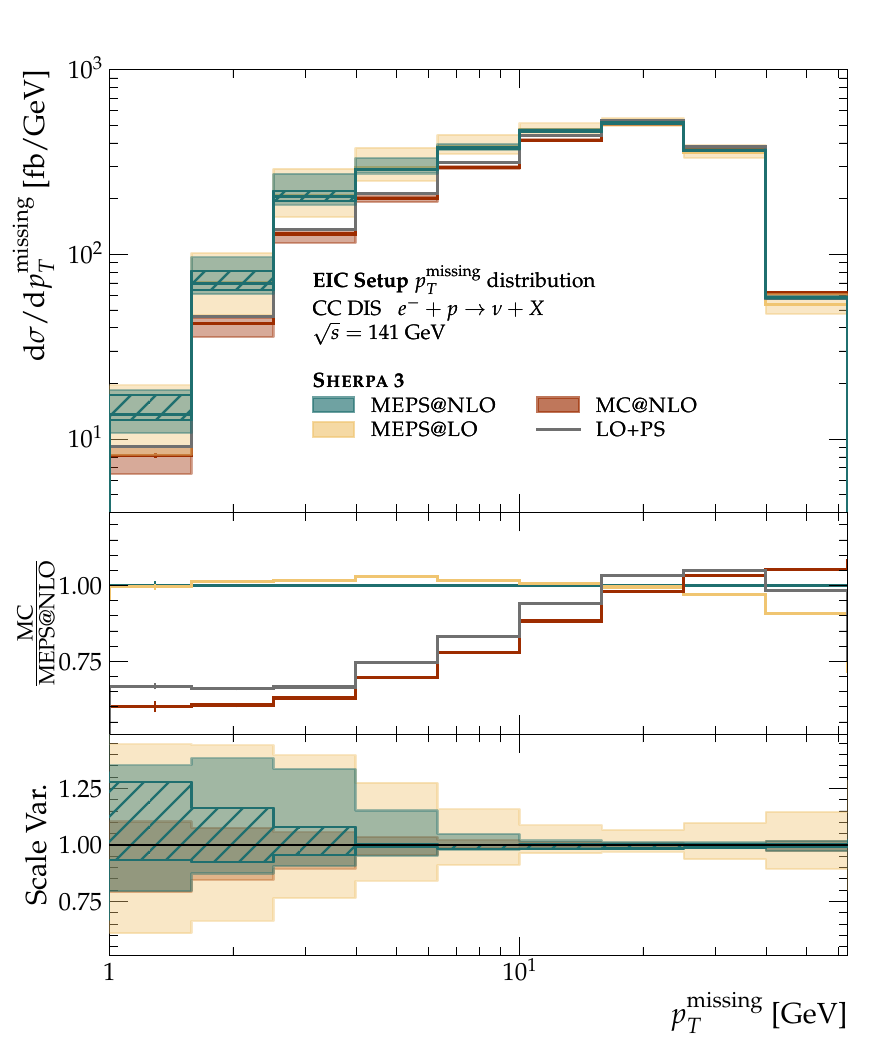}
  \caption{Differential distributions for $y$ and $\ptsup{\mathrm{missing}}$ in charged current DIS at the \protect\EIC, comparing LO+PS, \protect\MCatNLO, \protect\MEPSatLO, and \protect\MEPSatNLO.
   Main panels show scale variation uncertainties (excluding LO+PS). Ratio panels display merging effects (hatched band) and scale uncertainties (full bands).}
  \label{fig:cc-eic:dis-obs-2}
\end{figure}

We again start our discussion with the global observables, $Q^2$, $\xBj$, $y$ and missing $p_T$ in Figs.~\ref{fig:cc-eic:dis-obs-1} and~\ref{fig:cc-eic:dis-obs-2}.
While the inelasticity does not exhibit large differences between the different calculations, we observe rather large corrections for the virtuality and the Bjorken-$x$ distributions.
The LO and \MCatNLO predictions agree well with each other within the uncertainties; however, the merged setups receive large corrections toward decreasing $Q^2$, $\xBj$ and missing $p_T$.
This trend starts at values of $10^3 \GeV^2$ and $10^{-1}$, respectively, and
reach up to almost a factor of 2 for the lowest $Q^2$ and $\xBj$ values. With
respect to each other the \MEPSatLO and \MEPSatNLO agree well, and the uncertainties are significantly reduced in the latter, as already observed before.
Naturally, the \MEPSatNLO shows smaller uncertainties than the \MEPSatLO and
comparable ones to the \MCatNLO at high values for the three observables. As
multijet events dominate at low $Q^2$ and these events are only computed at LO,
the \MEPSatNLO shows larger uncertainties than \MCatNLO
there. Likewise, merging scale variations are considerable in this
  range, analogous to our discussion for the NC case. nonperturbative uncertainties were negligible for these observable as they are measured by means of the lepton kinematics and are hence not affected by hadronization.

\begin{figure}[htpb]
  \centering
  \begin{tabular}{ccc}
    \includegraphics[width=0.3\linewidth]{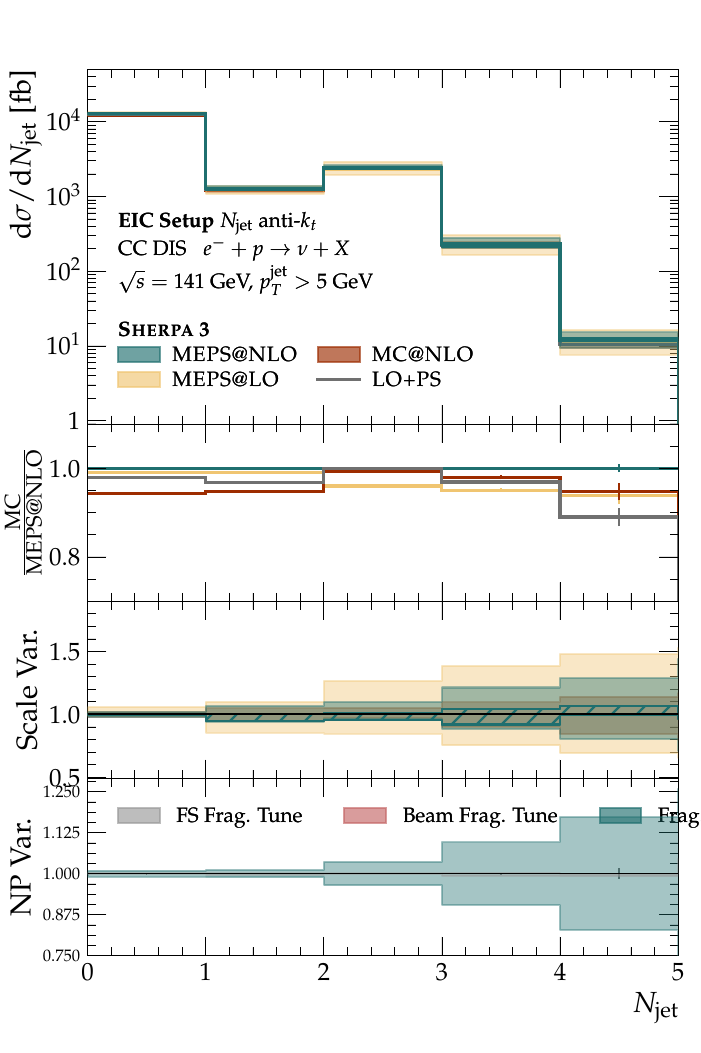} &
    \includegraphics[width=0.3\linewidth]{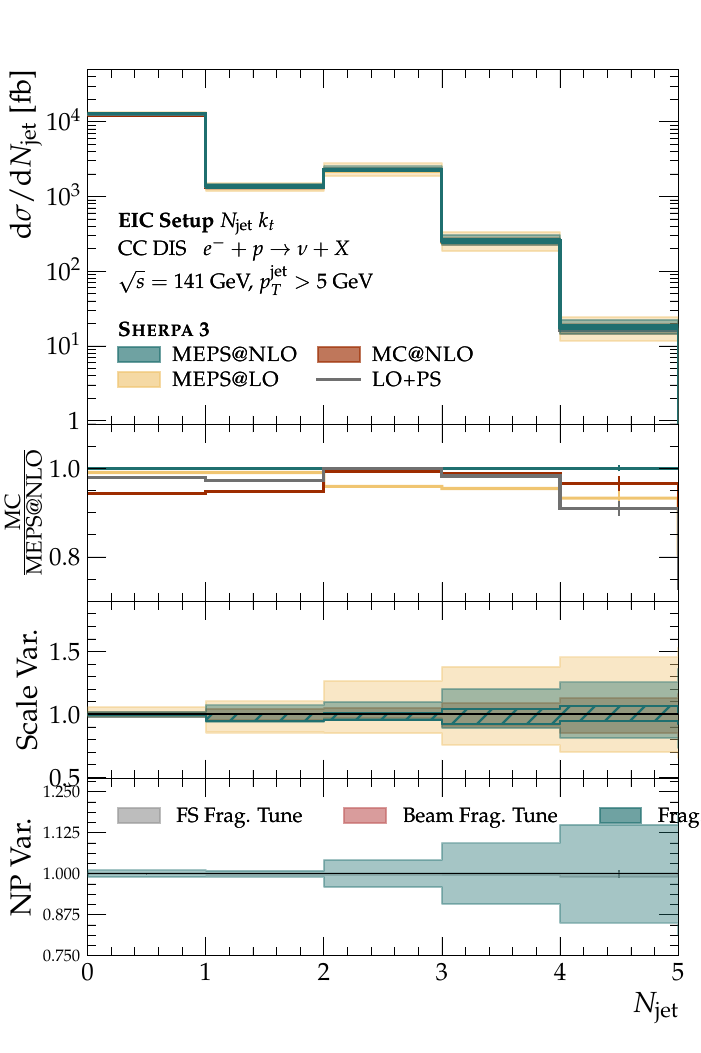} &
    \includegraphics[width=0.3\linewidth]{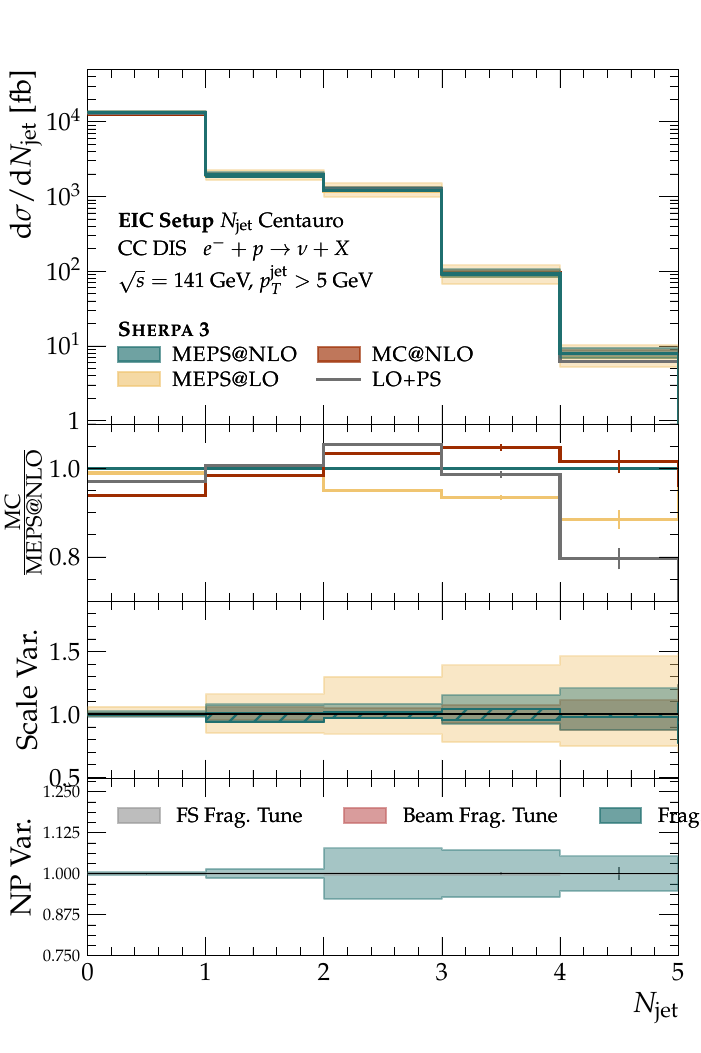} \\
    \includegraphics[width=0.3\linewidth]{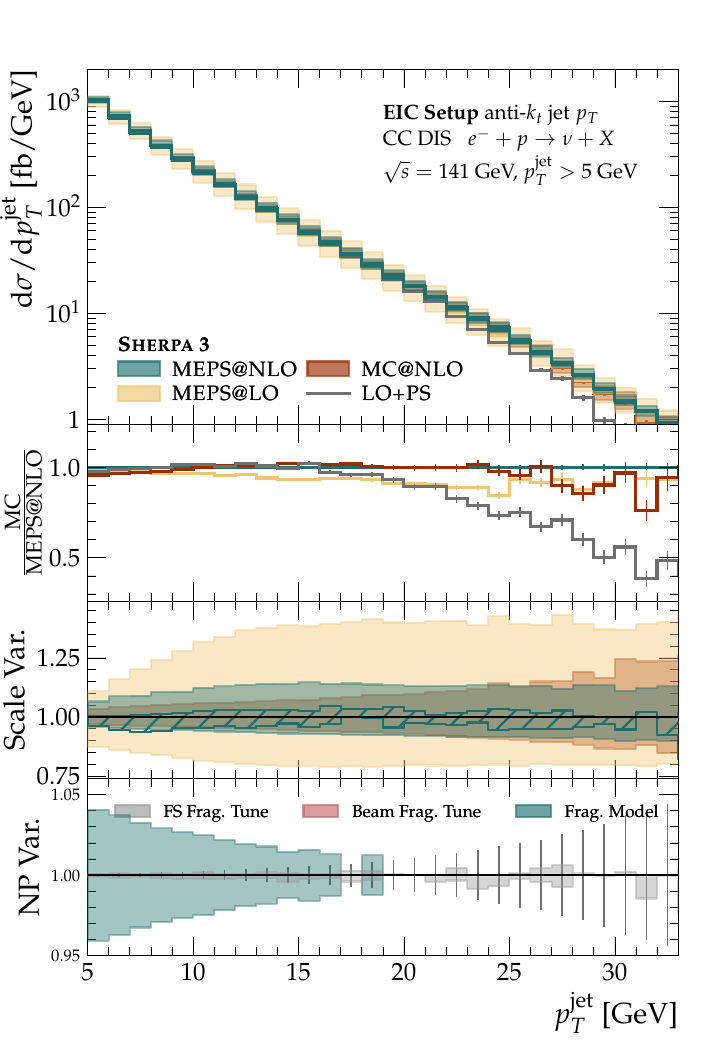} &
    \includegraphics[width=0.3\linewidth]{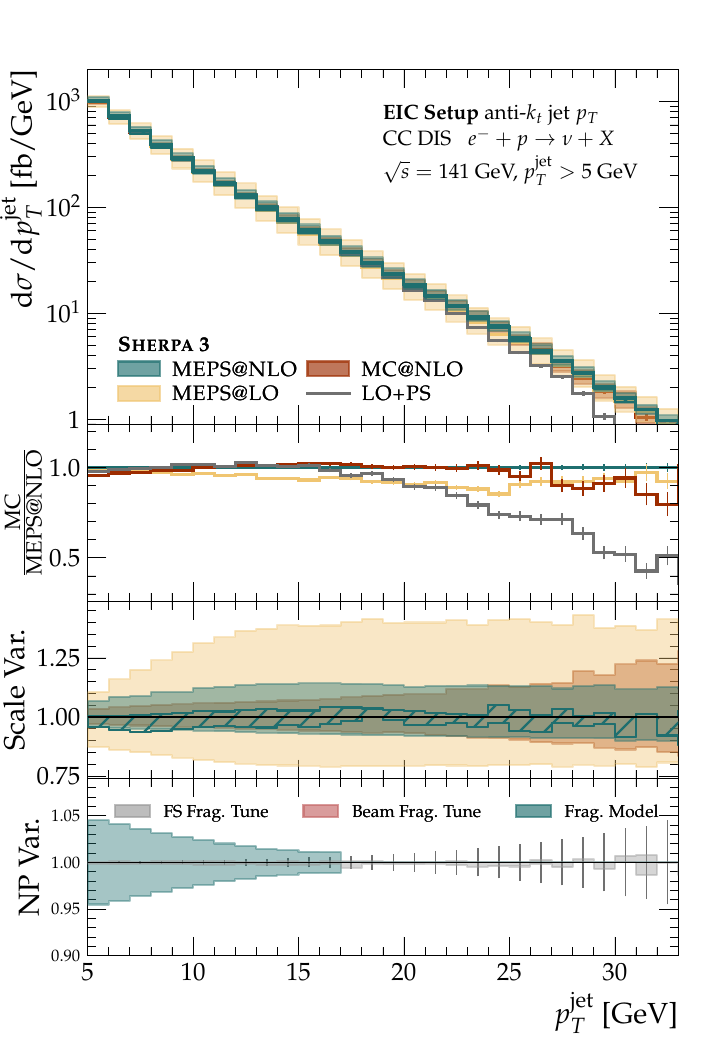} &
    \includegraphics[width=0.3\linewidth]{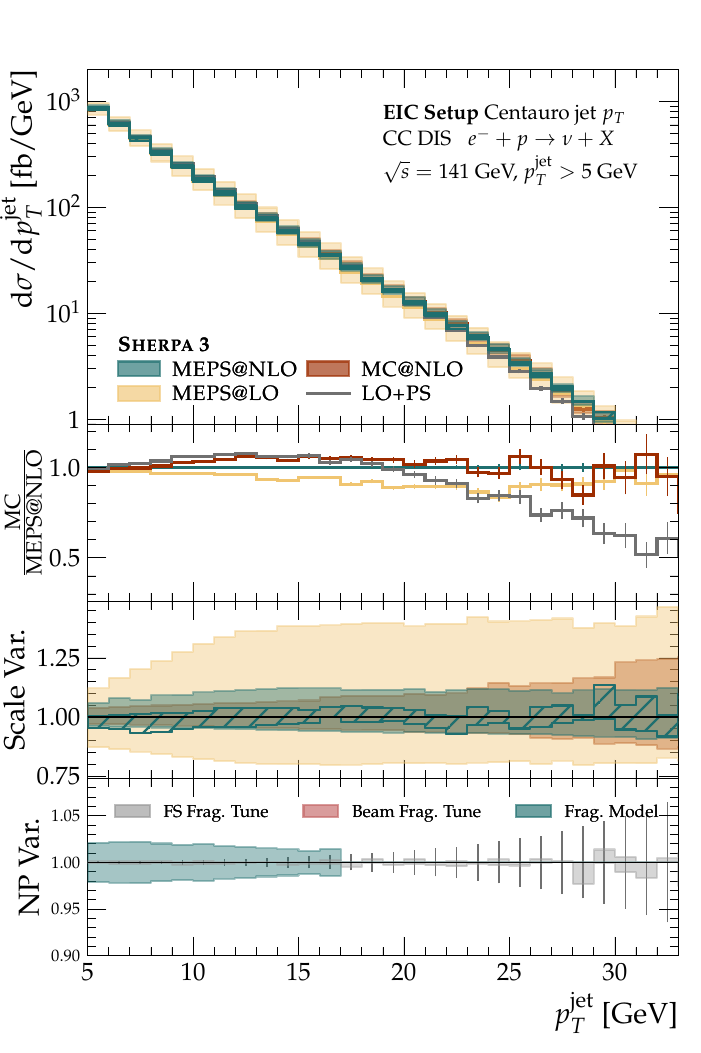}
  \end{tabular}
  \caption{Jet multiplicity $N_\mathrm{jet}$ (top row) and leading jet
    transverse momentum $p_T^\mathrm{jet}$ (bottom row) distributions for
    anti-$k_t$ (left), $k_t$ (middle) and \Centauro (right) jet algorithms in
    CC DIS at the \protect\EIC. The bottom ratios in each row
    indicate, from top to bottom, the ratio of each prediction to the
    \MEPSatNLO one, the uncertainty from scale variations (full band for $\muF\, \mathrm{and}\,\muR$, hatched for $Q_{\mathrm{cut}}$) and the uncertainty
    from nonperturbative model and tuning variations. The error bars in the
    lowest panel indicate the statistical uncertainty of the difference between
    cluster and string model. \label{fig:cc-eic:jet-obs}}
\end{figure}

In Fig.~\ref{fig:cc-eic:jet-obs}, we study the three different jet algorithms in
terms of jet multiplicity and leading jet transverse momentum. In the latter, and
unlike to the neutral current case, there are no corrections with respect to the
LO calculation in the low-$p_T$ region for neither of the clustering
algorithms. At high $p_T$ the simulation at LO accuracy yields a slightly lower
cross section by about 50\% compared to the MEPS@NLO result while the other calculations agree with each other.
Hence, we can assume that the large corrections
that can be seen in $Q^2$ and $\xBj$ are driven by small-$p_T$ jets, which
are removed by the $p_T$ cut. Seeing that already \MCatNLO does not show large
differences to the merged setups, we conclude that for jet observables with the
used $p_T$ cut of $5 \GeV$, already the real correction of the Born process
captures any corrections in this part of the phase space. As a consequence, the
scale uncertainties of the \MCatNLO and the \MEPSatNLO do not differ as much as
in the neutral current case before. Merging scale variations are
  again less important than scale variations.
As in the NC case, nonperturbative uncertainties are generally smaller than
perturbative ones and affect significantly only the high-multiplicity and low-$p_T$
regions. At most, they account for a  relative $\sim5\%$ correction, which matches
the one induced by scale variations only at very low $p_T$ for all jet algorithms considered.
Fragmentation model uncertainty is the dominant contribution, while
replica tunes within the cluster model do not generate an observable effect within
statistical fluctuations.

\begin{figure}[ht]
  \centering
  \includegraphics[width=\linewidth]{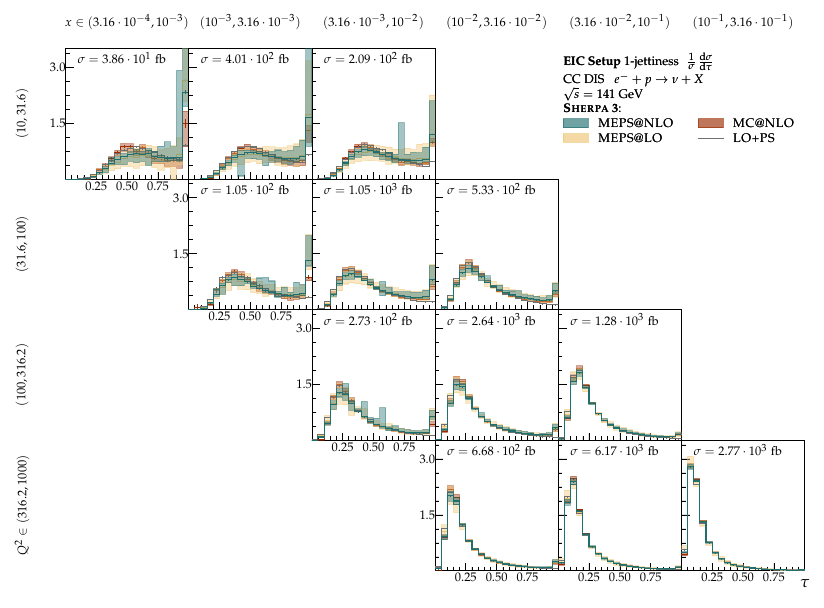}
  \caption{Differential distributions of 1-jettiness $\tau$ in different bins of $Q^2$ and $x$ in CC DIS at the \protect\EIC.
  The total \MEPSatNLO cross section for each $x$-$Q^2$ bin is reported on top of the corresponding plot.}
  \label{fig:cc-eic:thrust-in-bins}
\end{figure}

\begin{figure}[ht]
  \centering
  \includegraphics[width=0.32\linewidth]{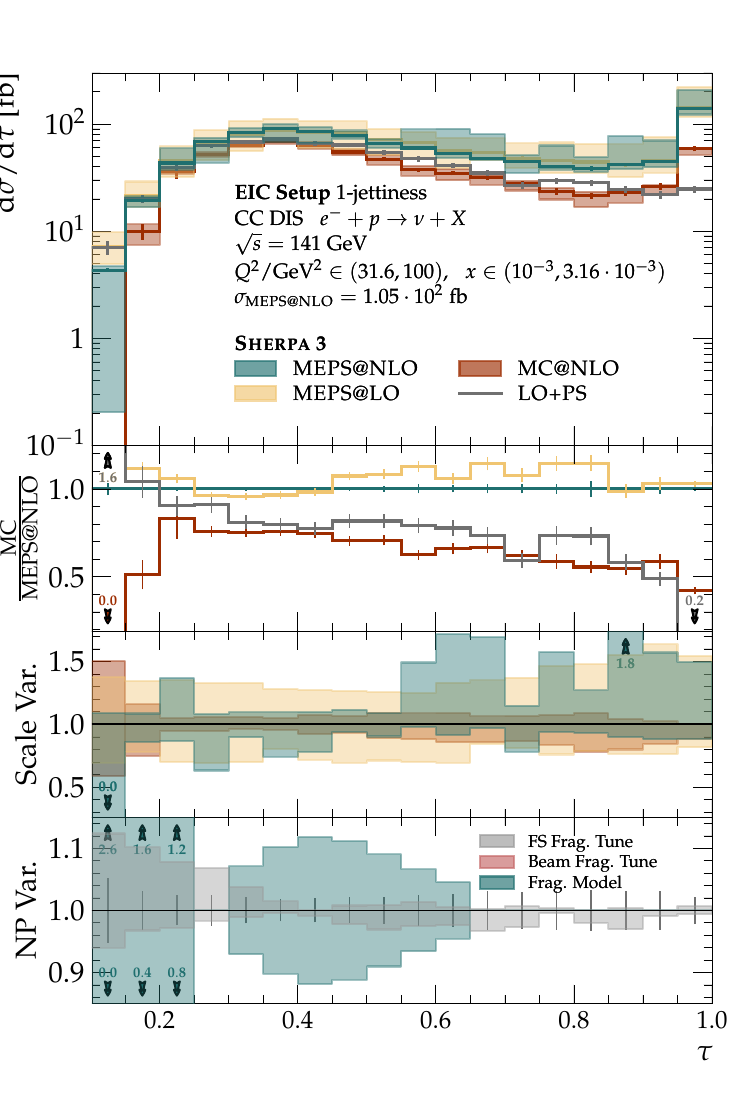}
  \includegraphics[width=0.32\linewidth]{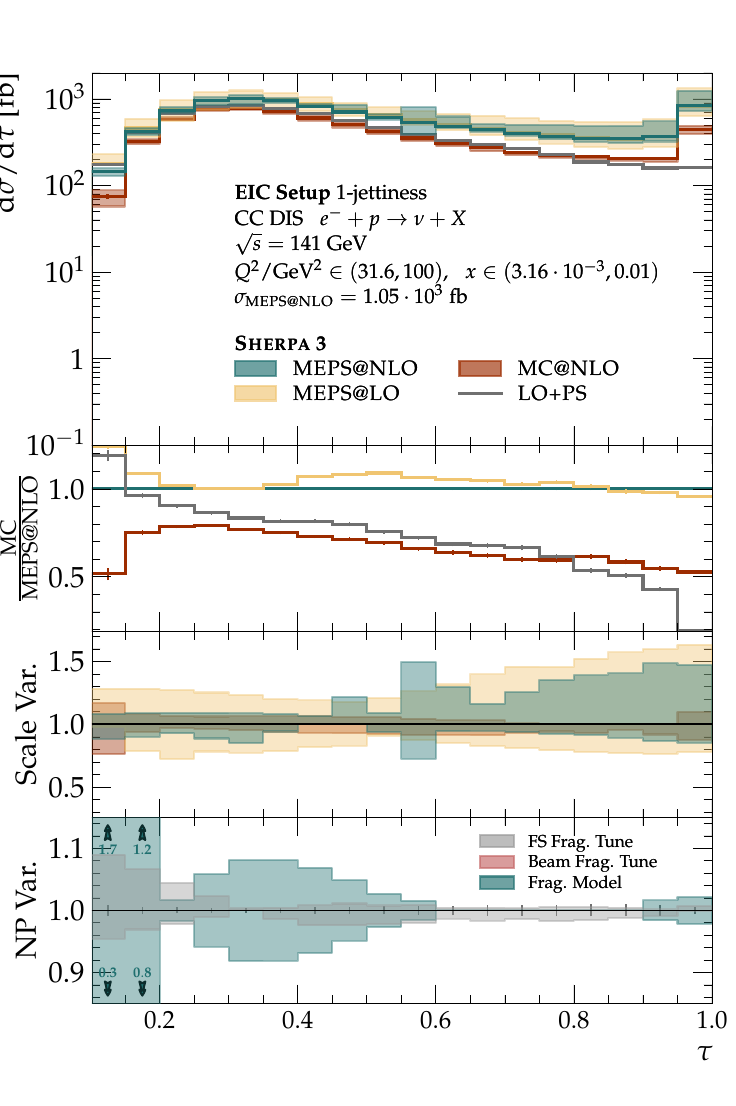}
  \includegraphics[width=0.32\linewidth]{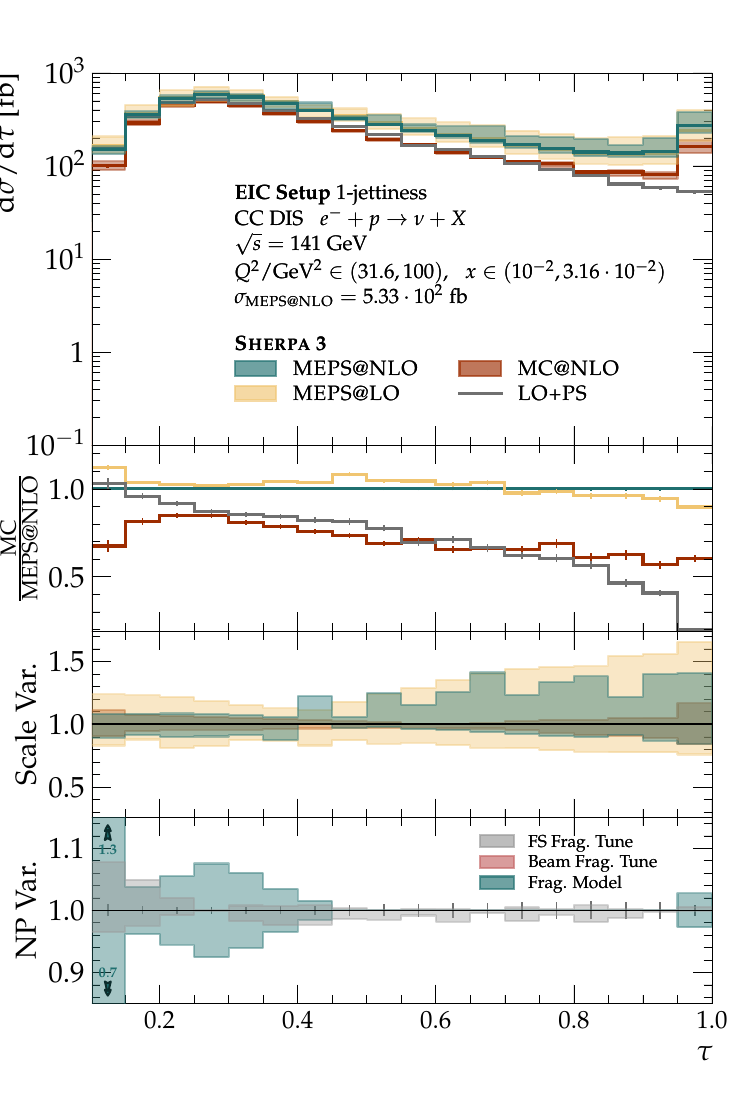}
  \caption{Differential distributions of 1-jettiness $\tau$ in a fixed slice of
    $Q^2$ and different bins of $x$ in charged current DIS at the \protect\EIC. The bottom ratios in each row
    indicate, from top to bottom, the ratio of each prediction to the
    \MEPSatNLO one, the uncertainty from scale variations and the uncertainty
    from nonperturbative model and tuning variations. The error bars in the
    lowest panel indicate the statistical uncertainty of the difference between
    cluster and string model. \label{fig:cc-eic-tau-Q2-slice}}
\end{figure}

Analogously to before, we show the 1-jettiness in bins of $Q^2$ and $x$ in
Fig.~\ref{fig:cc-eic:thrust-in-bins} without ratio plots for clarity and for
the range $31.6 < Q^2/\GeV^2 < 100$ in three $\xBj$ ranges including ratio plots in
Fig.~\ref{fig:cc-eic-tau-Q2-slice}. Note that we pick a different $Q^2$ slice with respect to the NC case,
as the total cross section in the lower slice is smaller. 
Plots for the remaining $\xBj$-$Q^2$ ranges
can again be found in Appendix~\ref{app:cc-tau}.
Overall, we see similar changes in the shape when going from small to large virtualities.
As the calculation of the 1-jettiness operates inclusively, \ie, without a jet
cut, we can observe large corrections on the cross section for low $Q^2$ by
merging additional legs. With regard to the hadronization, we also see the same
pattern as before; \ie, only the peak is affected and the phase space region of
small values of \xBj{} and $Q^2$ has larger uncertainties.
It is only in these regions that the perturbative and the nonperturbative
uncertainties are of comparable size. While 1-jettiness is only an example, this
suggests that global analyses of charged-current events at the \EIC need higher-order corrections for the small virtualities.
Depending on the observable, a thorough understanding of nonperturbative
effects will also be necessary.

\FloatBarrier
\section{Conclusion}\label{sec:conclusions}

In this work we presented the first predictions at \MEPSatNLO precision for both
neutral and charged current DIS at the \EIC. Producing full hadron-level
predictions, we  contrasted perturbative and
nonperturbative uncertainties and identified regions where higher-order
corrections will be necessary.
To this end, we used the multijet merging capabilities of the \sherpa event
generator, which lead to non-negligible corrections in the cross sections
compared to LO.
In the analysis we focused on global DIS observables, jet multiplicities, and transverse momenta in the Breit frame and 1-jettiness distributions.
While estimating the perturbative higher-order uncertainties with seven-point scale
variations, we examined the nonperturbative uncertainties associated with the
fragmentation modeling by means of replica tunes of the hadronization and the
uncertainties associated with the choice of nonperturbative model.

We found that the merging of additional legs in the matrix element in the
\MEPSatLO and \MEPSatNLO schemes, respectively, captures large corrections of up
to factor 2 in the low $Q^2$ and \xBj{} phase space.
The virtual corrections lead to an observable effect on the cross section, especially at high $Q^2$ and \xBj.
Comparing the $k_t$, anti-$k_t$ and \Centauro jet algorithms, we observed very
large $K$-factors for the two former, but only a small $K$-factor for the
latter. We assume this to be a consequence of the \Centauro algorithm being
better at clustering back to a Born kinematics, whereas only higher-order
corrections contribute for $k_t$-type algorithms in the Breit frame.

We then studied 1-jettiness as an example for a global event-shape observable, enabling
among others the extraction of the strong coupling in measurements at the \EIC.
We analyzed it in slices of $Q^2$ and \xBj{} and observed a change
in event shapes from single-jet topologies at high virtualities toward more
uniform shapes at small virtualities ones.
nonperturbative uncertainties can reach comparable size as the scale variations
in limited region of the phase space and especially toward smaller
virtualities. Those can to an extent be mitigated by modern grooming techniques.
However, especially with the availability of NNLO FO calculations, and the advent of general NNLO
matching techniques in the foreseeable future, this implies a strong necessity of understanding better nonperturbative and
higher-twist corrections due to the lower center-of-mass energy at the \EIC than
at \HERA, in order to fully exploit the potential of the data.
Tuning cluster and string model in comparable setups, using
simultaneously LEP and DIS data, should be a pragmatic first step in this direction.
Ideally, this exercise would also produce tuning uncertainties for
both model tunes.

Our study showcases the state-of-the-art of precision hadron-level predictions
for the \EIC and examines the leading uncertainties.
At virtualities of $Q^2 \approx 1\text{-}10 \GeV^2$ in NC DIS, the
photoproduction region comes into play, and the merging of additional legs in
the matrix elements are the tool to interpolate toward this region.

While many exclusive measurements are planned during the operation of the \EIC, an understanding of the inclusive background is crucial for any analysis.
Here, we showed a global analysis of the DIS region, future studies will need to
understand better the photoproduction region and the interpolation between the
two to fully allow inclusive jet measurements at the \EIC. This would allow
performing jet studies at the \EIC in a fully inclusive way over all virtualities,
which has so far been formulated as a goal in the experimental program, but
has not seen much attention from the theoretical side.
Calculations at even higher order, at NNLO and beyond, and ideally combined
with either resummation or matched to the parton shower, should be within reach
in the upcoming years and in time for the data taking at the \EIC.

\section*{Acknowledgments}
We are grateful to Frank Krauss for encouraging us to investigate this topic and
for his comments on our manuscript.
D.R.\ is supported by the European Union under the HORIZON program in Marie
Sk{\l}odowska-Curie Project No. 101153541.
F.S.\ is supported by the STFC under Grant No. ST/P006744/1.
P.M.\ is supported by the Swiss National Science Foundation (SNF) under Contract
No. 200020-204200.

\FloatBarrier

\appendix
\section{Further observables}
In this Appendix we present further observables and versions of plots that we
omitted in the main text for brevity.
\subsection{Neutral current 1-jettiness}\label{app:nc-tau}
\begin{figure}[ht]
  \centering
  \resizebox{\textwidth}{!}{
    \includegraphics[width=0.333\linewidth]{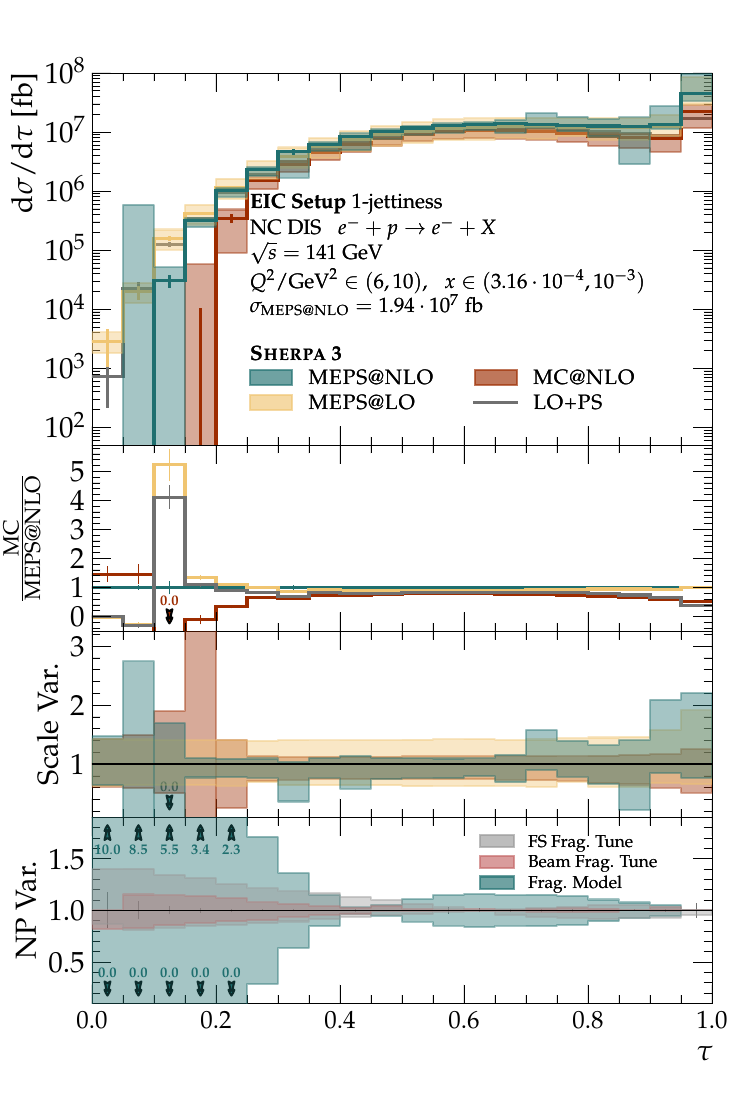}
    \includegraphics[width=0.333\linewidth]{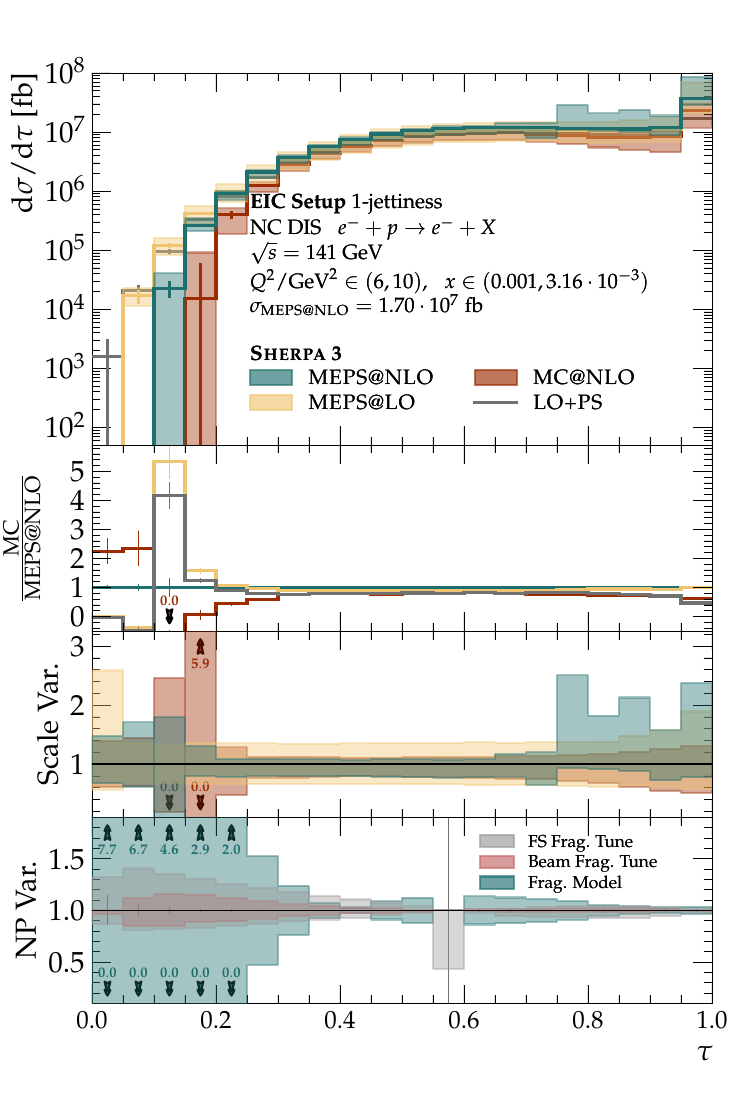}
    \hspace{0.333\linewidth}
  }
  \resizebox{\textwidth}{!}{
    \includegraphics[width=0.333\linewidth]{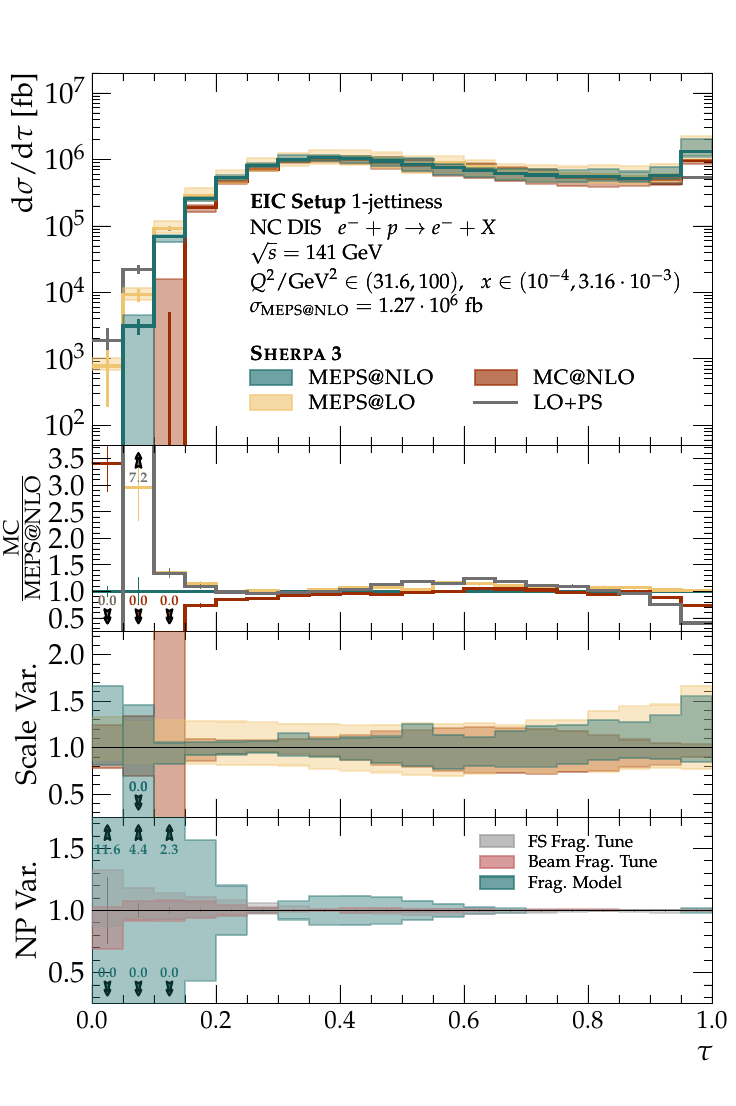}
    \includegraphics[width=0.333\linewidth]{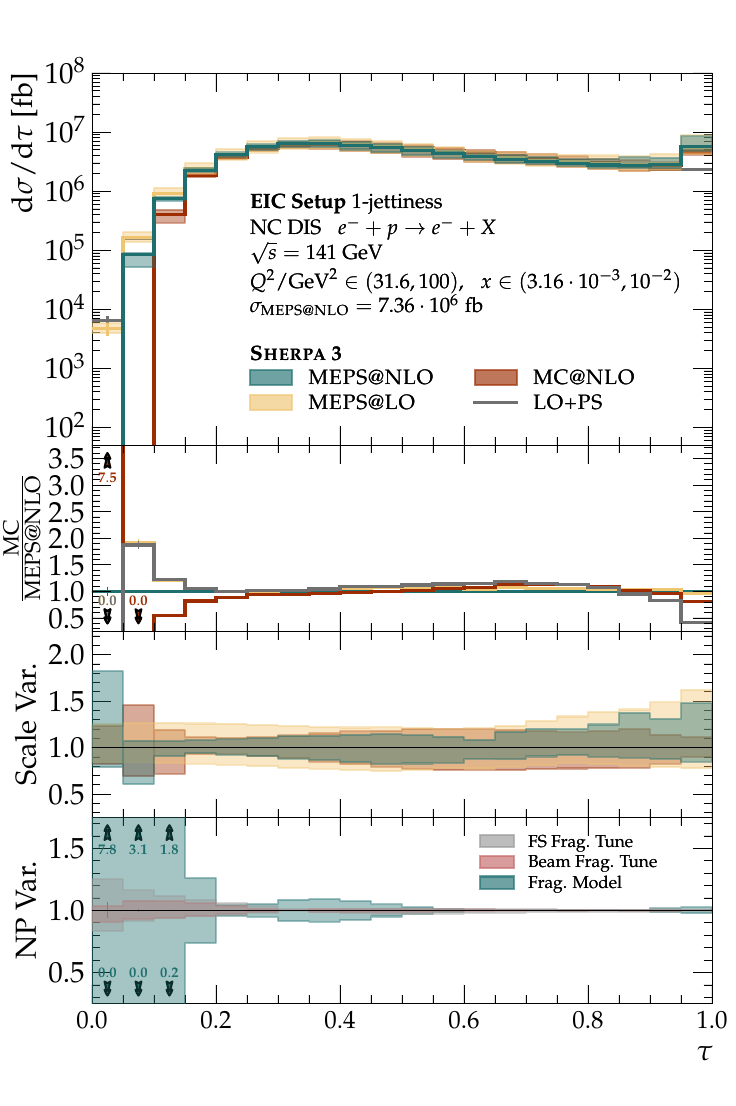}
    \includegraphics[width=0.333\linewidth]{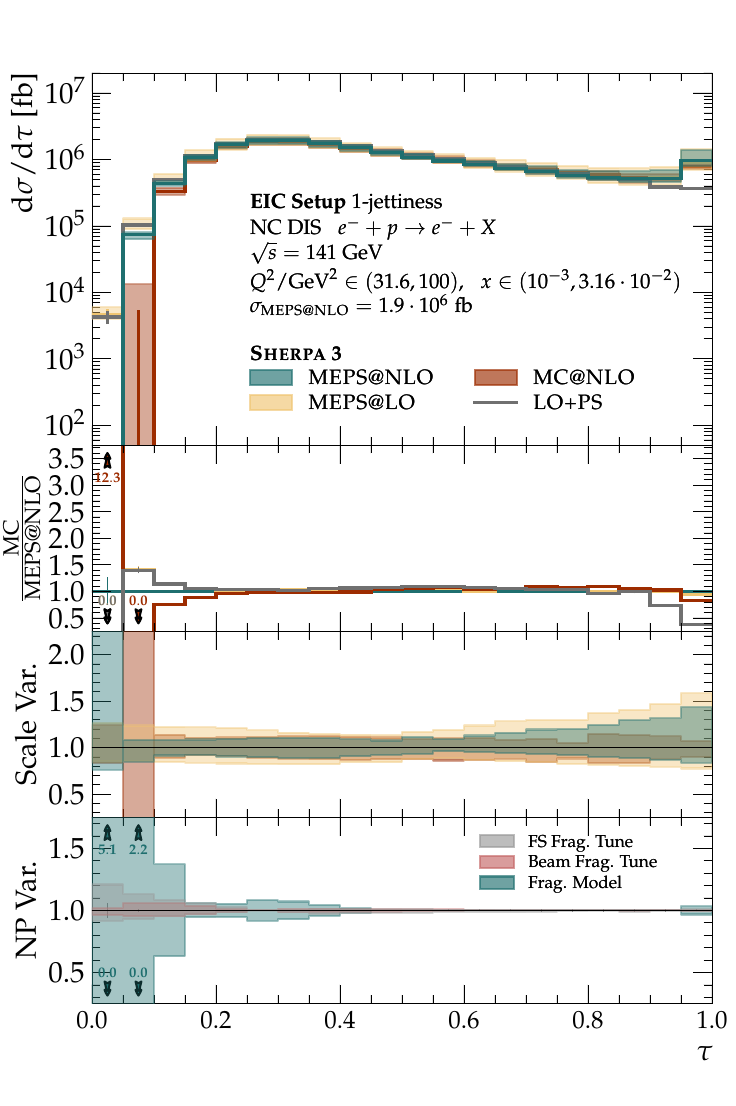}
  }
  \caption{Differential distributions of 1-jettiness $\tau$ in a fixed slice of
    $Q^2$ and different bins of $x$ in NC DIS at the \protect\EIC. The bottom ratios in each row
    indicate, from top to bottom, the ratio of each prediction to the
    \MEPSatNLO one, the uncertainty from scale variations and the uncertainty
    from nonperturbative model and tuning variations. The error bars in the
    lowest panel indicate the statistical uncertainty of the difference between
    cluster and string model.}\label{fig:app-1-1}
  \label{app:A-tau-nc}
\end{figure}

\begin{figure}[ht]
  \centering
  \resizebox{\textwidth}{!}{
    \includegraphics[width=0.333\linewidth]{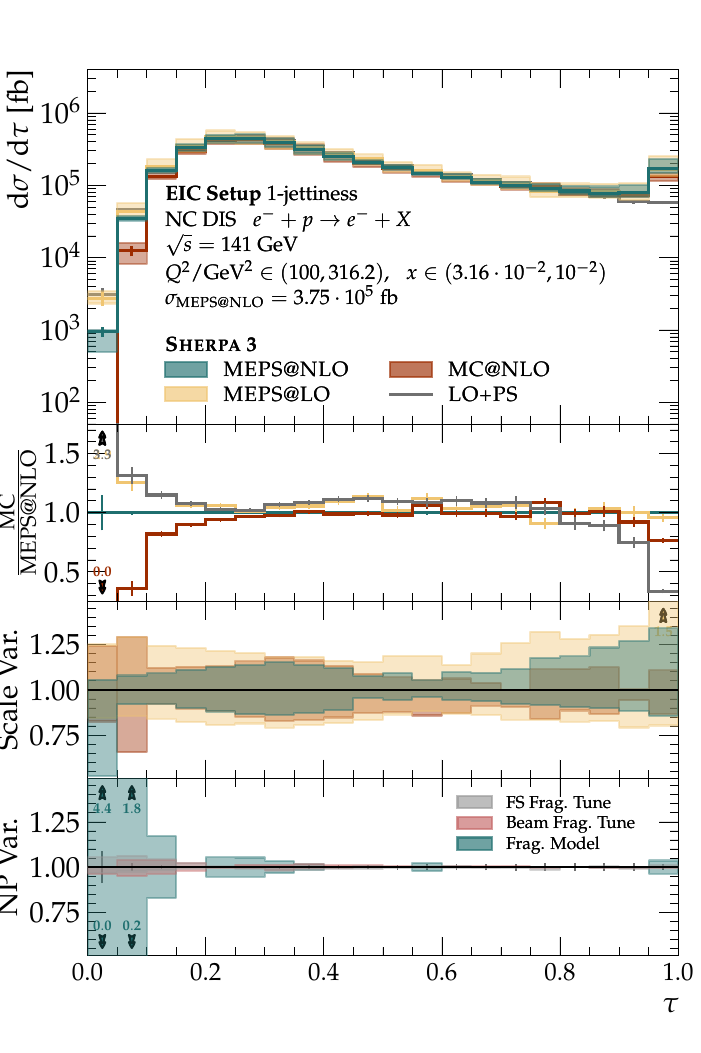}
    \includegraphics[width=0.333\linewidth]{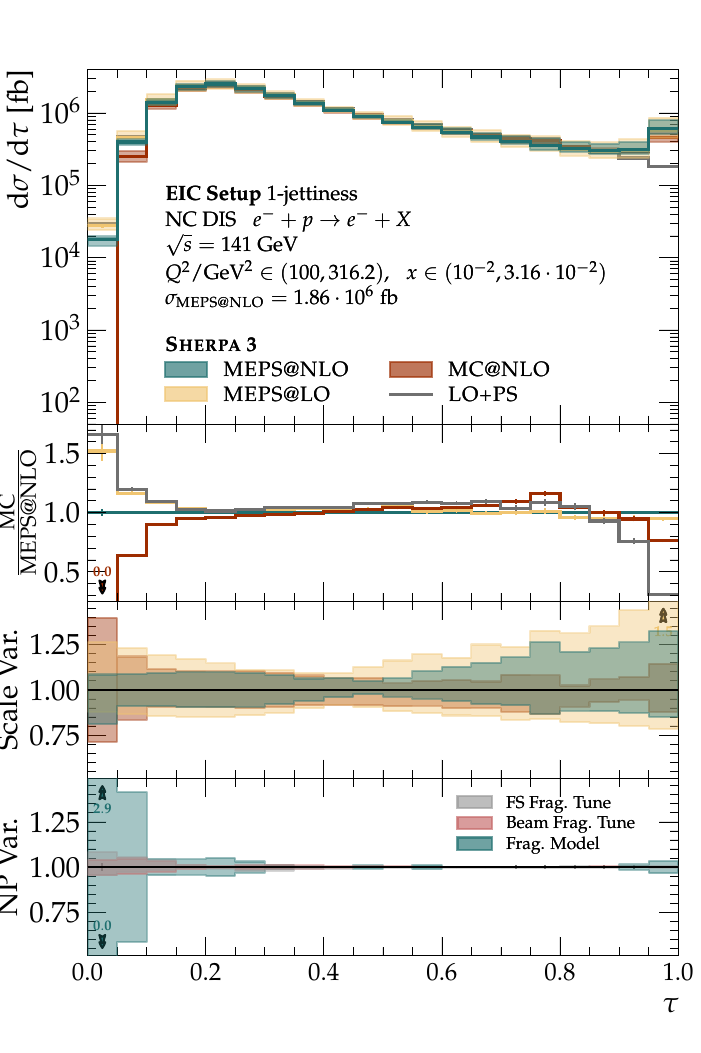}
    \includegraphics[width=0.333\linewidth]{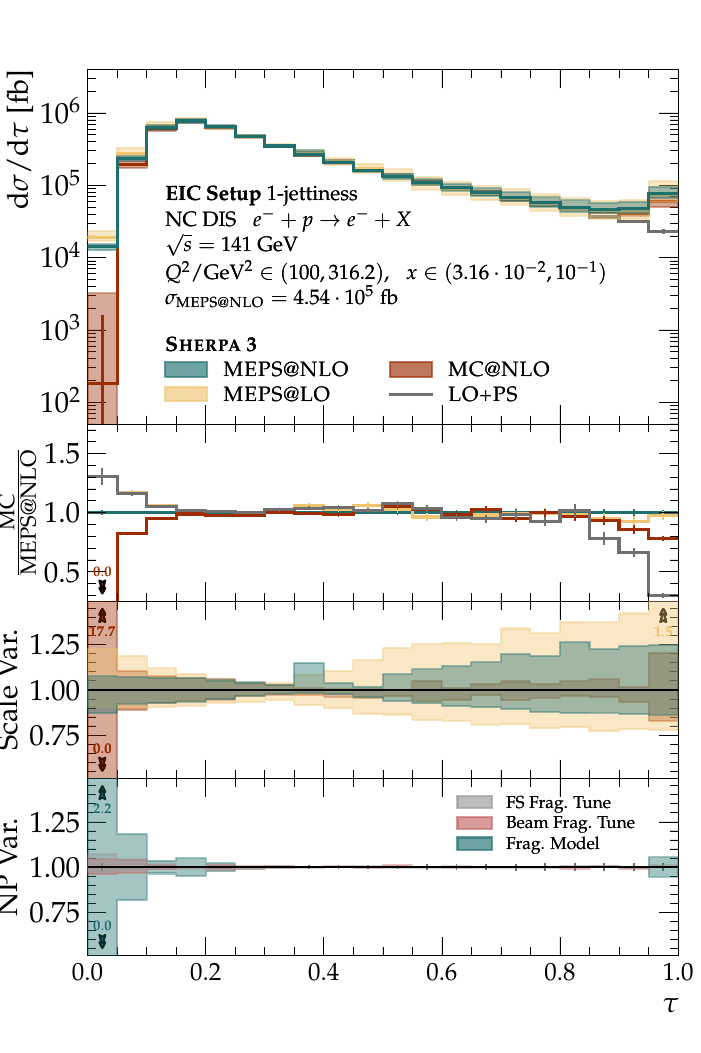}
  }
  \resizebox{\textwidth}{!}{
    \includegraphics[width=0.333\linewidth]{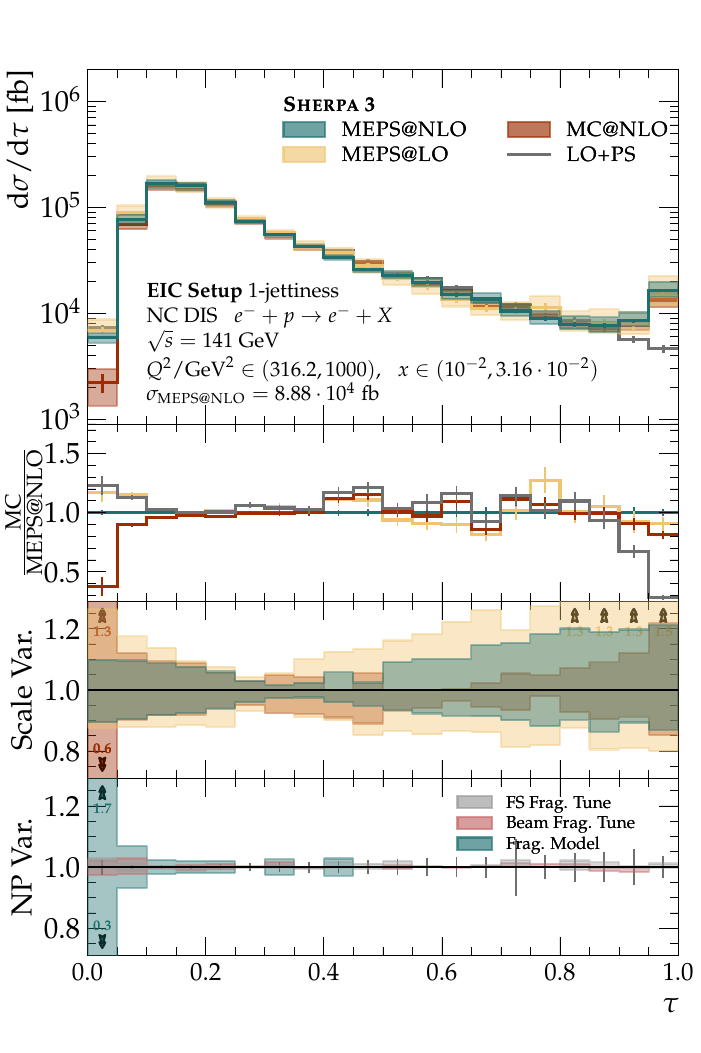}
    \includegraphics[width=0.333\linewidth]{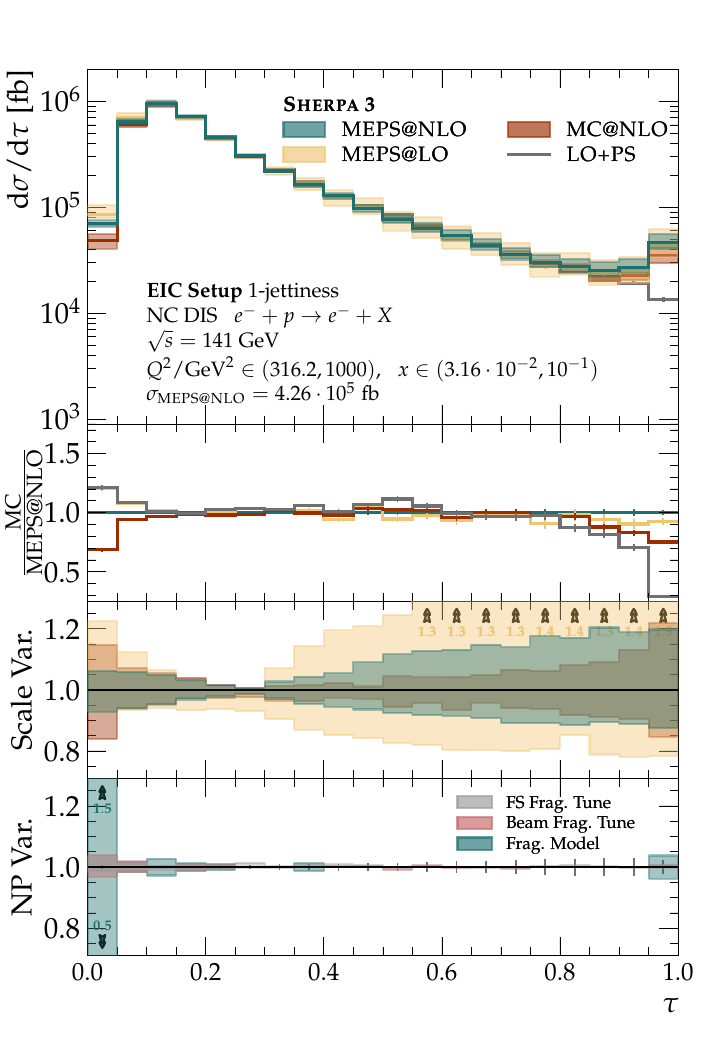}
    \includegraphics[width=0.333\linewidth]{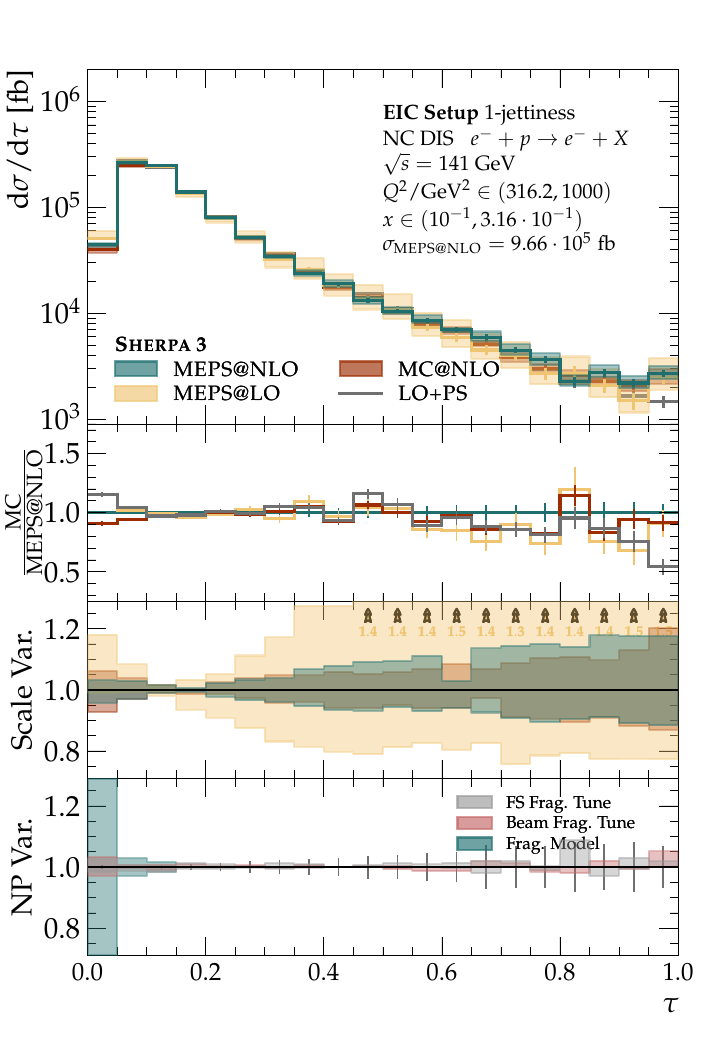}
  }
    \caption{Differential distributions of 1-jettiness $\tau$ in a fixed slice of
    $Q^2$ and different bins of $x$ in NC DIS at the \protect\EIC. The bottom ratios in each row
    indicate, from top to bottom, the ratio of each prediction to the
    \MEPSatNLO one, the uncertainty from scale variations and the uncertainty
    from nonperturbative model and tuning variations. The error bars in the
    lowest panel indicate the statistical uncertainty of the difference between
    cluster and string model.}\label{fig:app-1-2}
\end{figure}

Here, we show the remaining 1-jettiness histograms in NC DIS in the $\xBj$ and
$Q^2$ ranges from Fig.~\ref{fig:nc-eic:thrust-in-bins} with the same ratio
panels as in Fig.~\ref{fig:nc-eic-tau-Q2-slice}. These plots are shown in
Figs.~\ref{fig:app-1-1} and \ref{fig:app-1-2}.
\FloatBarrier
\clearpage
\subsection{Charged current 1-jettiness}\label{app:cc-tau}
\begin{figure}[ht]
  \centering
  \resizebox{\textwidth}{!}{
    \includegraphics[width=0.333\linewidth]{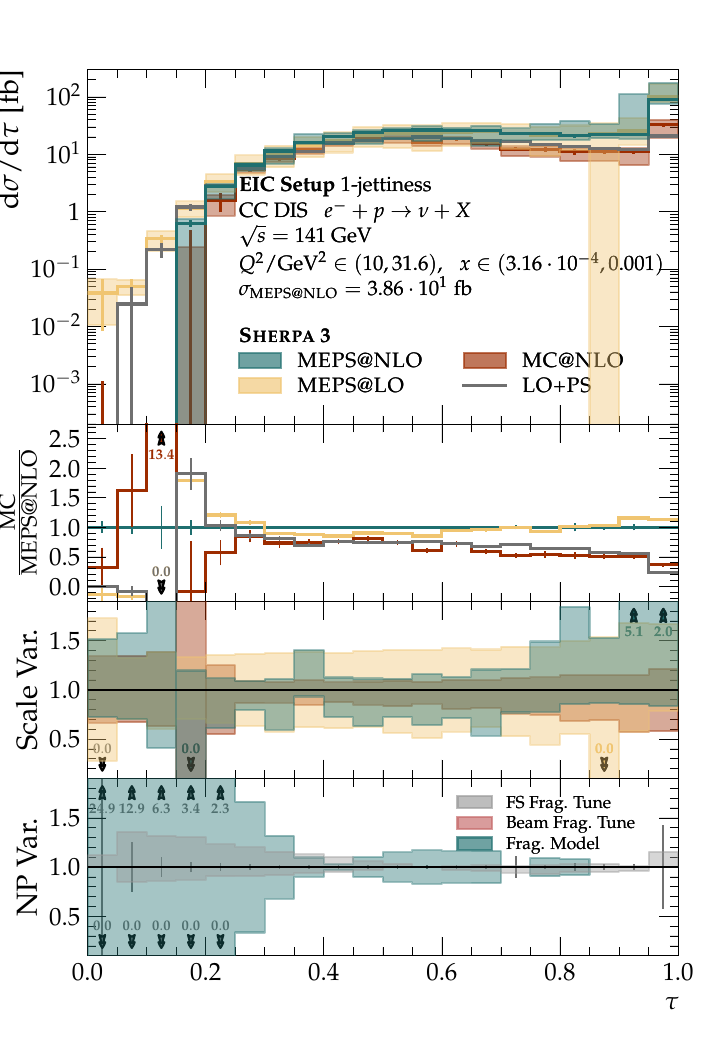}
    \includegraphics[width=0.333\linewidth]{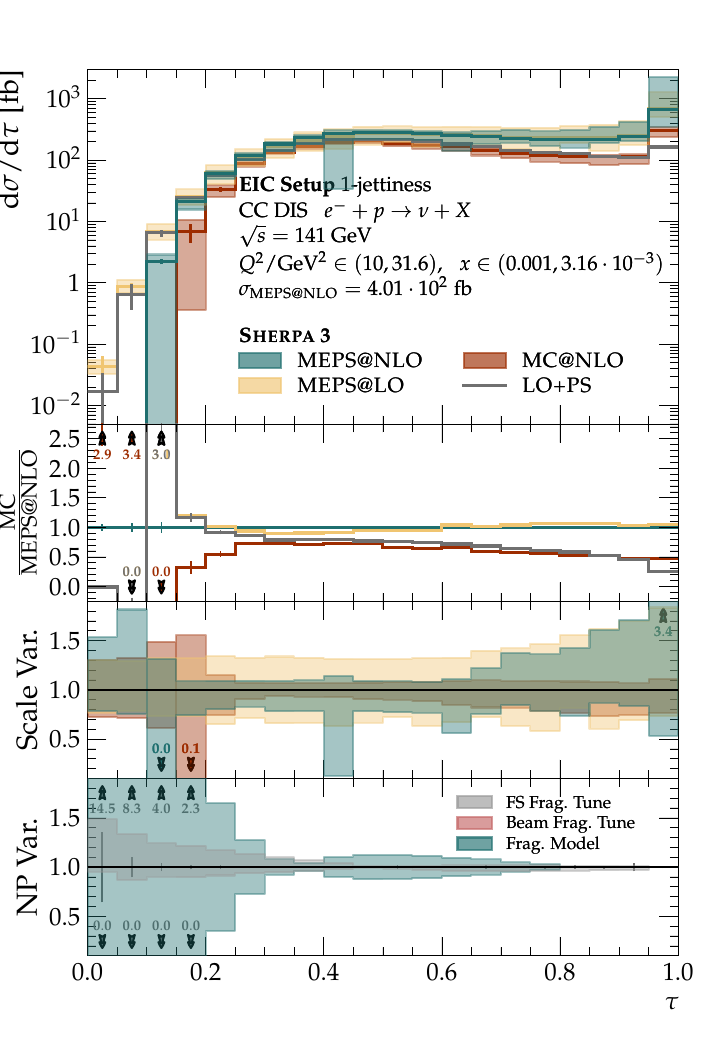}
    \includegraphics[width=0.333\linewidth]{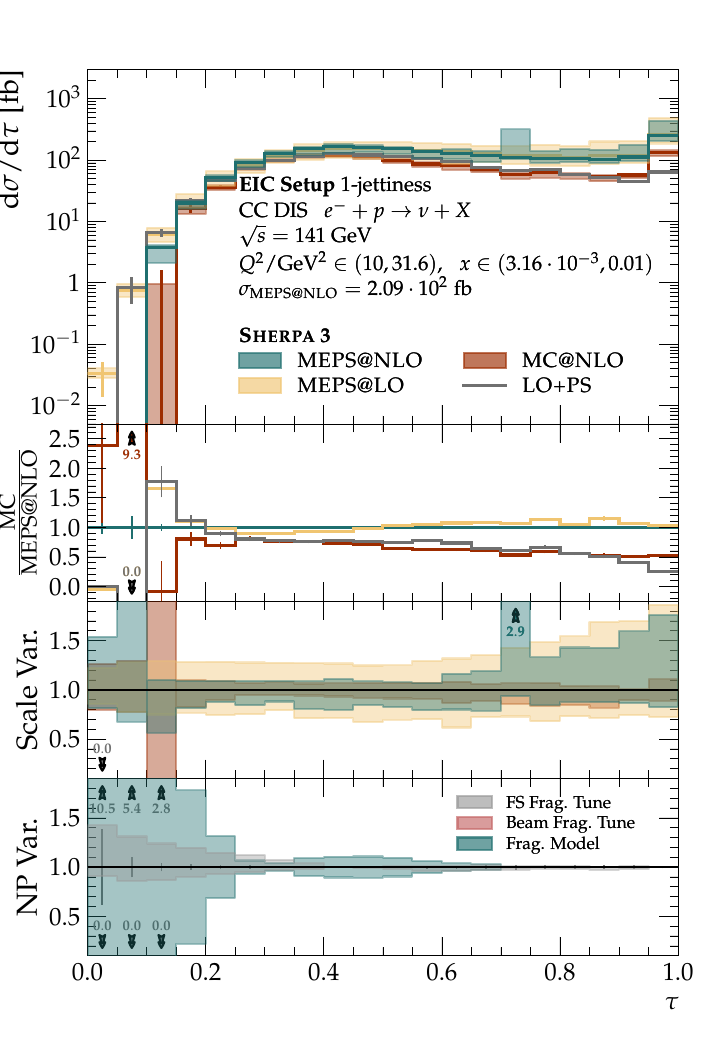}
  }
  \resizebox{\textwidth}{!}{
    \includegraphics[width=0.333\linewidth]{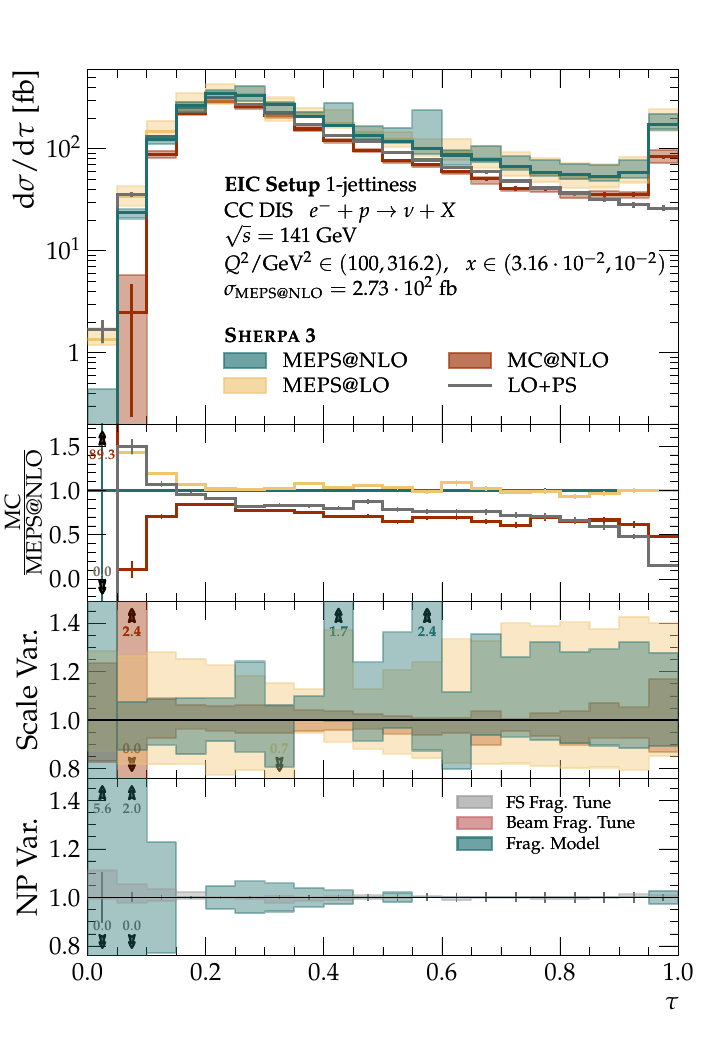}
    \includegraphics[width=0.333\linewidth]{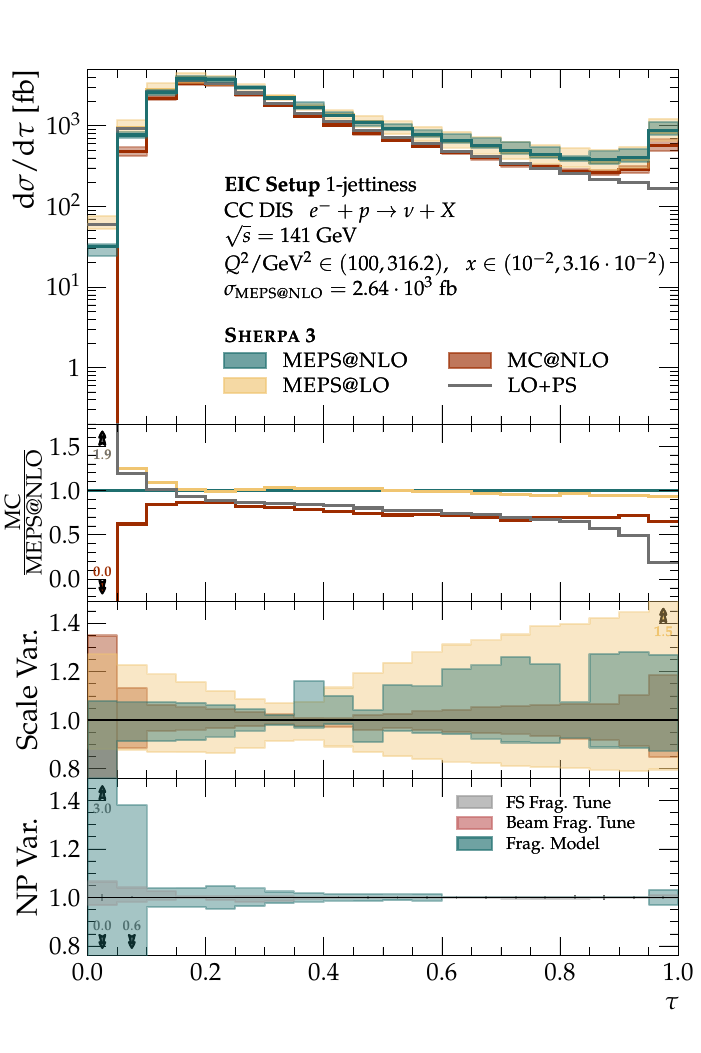}
    \includegraphics[width=0.333\linewidth]{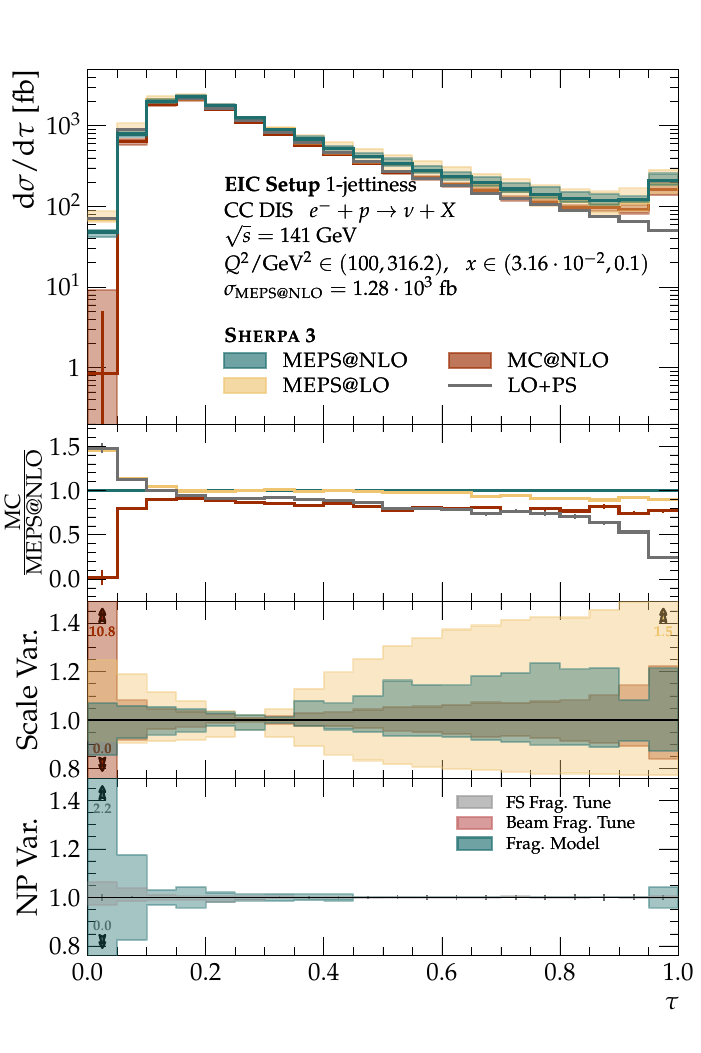}
  }
  \caption{Differential distributions of 1-jettiness $\tau$ in a fixed slice of
    $Q^2$ and different bins of $x$ in CC DIS at the \protect\EIC. The bottom ratios in each row
    indicate, from top to bottom, the ratio of each prediction to the
    \MEPSatNLO one, the uncertainty from scale variations and the uncertainty
    from nonperturbative model and tuning variations. The error bars in the
    lowest panel indicate the statistical uncertainty of the difference between
    cluster and string model.}\label{fig:app-2-1}
\end{figure}

\begin{figure}[ht]
  \centering
  
  \resizebox{\textwidth}{!}{
    \includegraphics[width=0.333\linewidth]{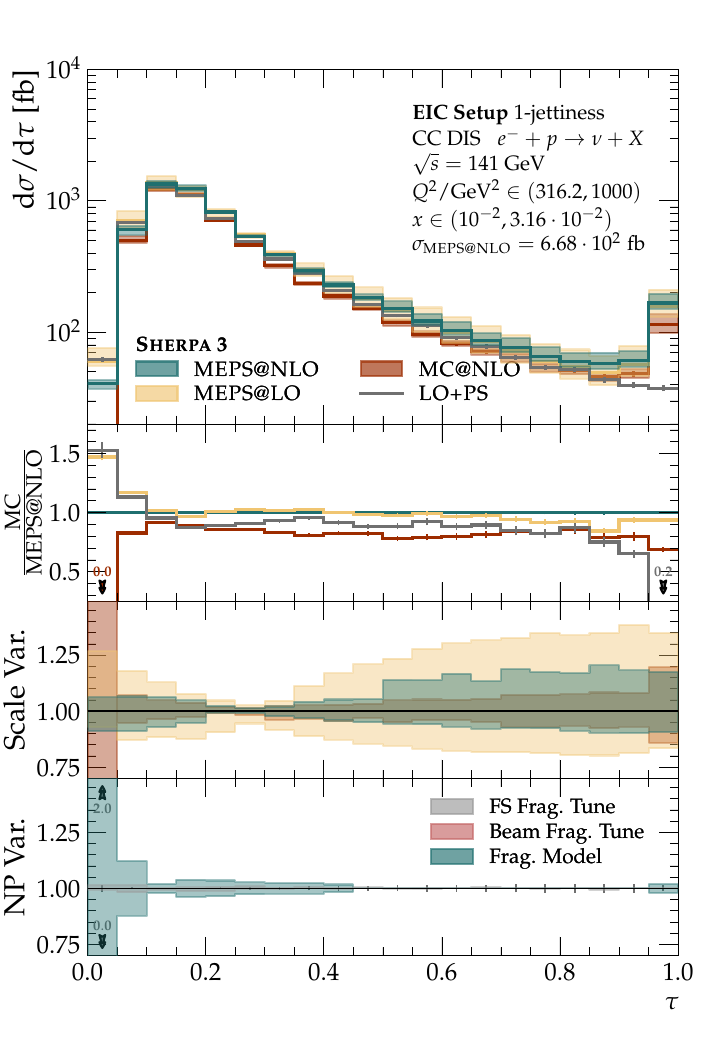}
    \includegraphics[width=0.333\linewidth]{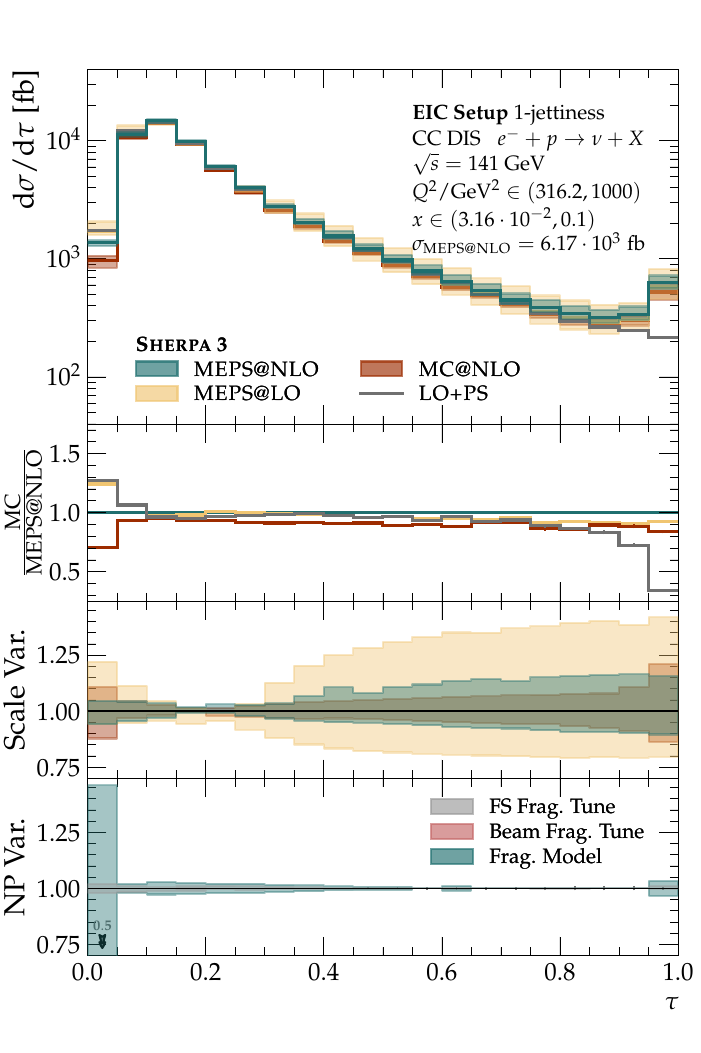}
    \includegraphics[width=0.333\linewidth]{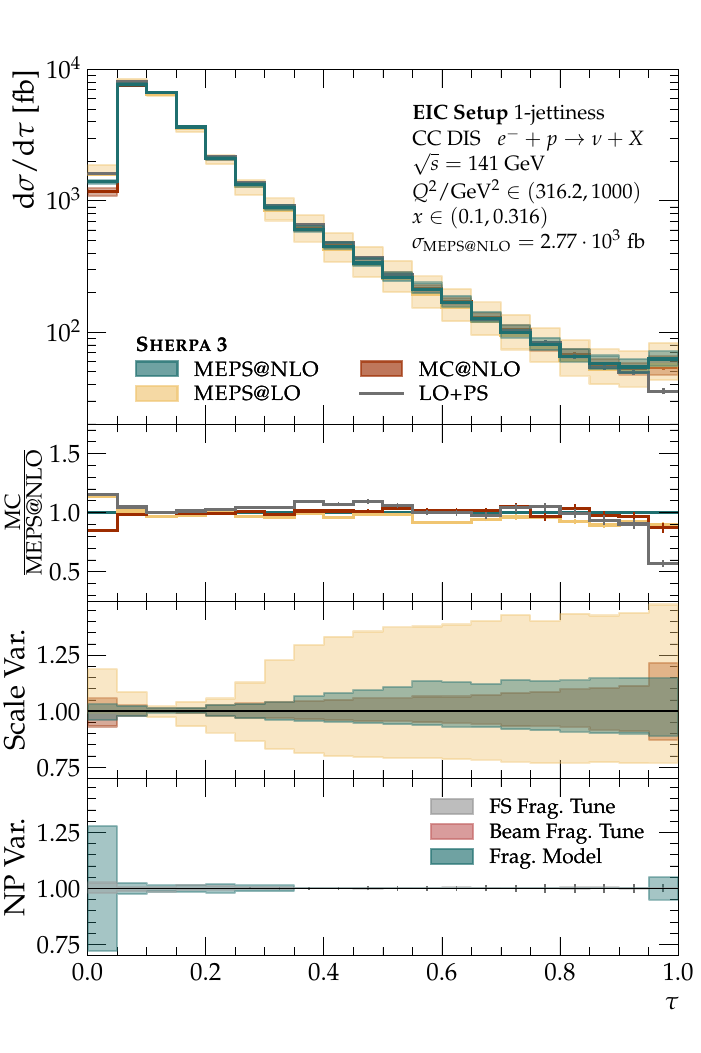}
  }
  \caption{Differential distributions of 1-jettiness $\tau$ in a fixed slice of
    $Q^2$ and different bins of $x$ in CC DIS at the \protect\EIC. The bottom ratios in each row
    indicate, from top to bottom, the ratio of each prediction to the
    \MEPSatNLO one, the uncertainty from scale variations and the uncertainty
    from nonperturbative model and tuning variations. The error bars in the
    lowest panel indicate the statistical uncertainty of the difference between
    cluster and string model.}\label{fig:app-2-2}

\end{figure}

Here, we show remaining the 1-jettiness observable in CC DIS in the $\xBj$ and
$Q^2$ ranges from Fig.~\ref{fig:cc-eic:thrust-in-bins} with the same ratio
panels as in Fig.~\ref{fig:cc-eic-tau-Q2-slice}.  These plots are shown in
Figs.~\ref{fig:app-2-1} and \ref{fig:app-2-2}.
\FloatBarrier
\clearpage
\subsection{Charged current groomed 1-jettiness}
\begin{figure}[ht]
  \centering
  \resizebox{\textwidth}{!}{
  \begin{tabular}{ccc}
      \includegraphics[width=0.333\linewidth]{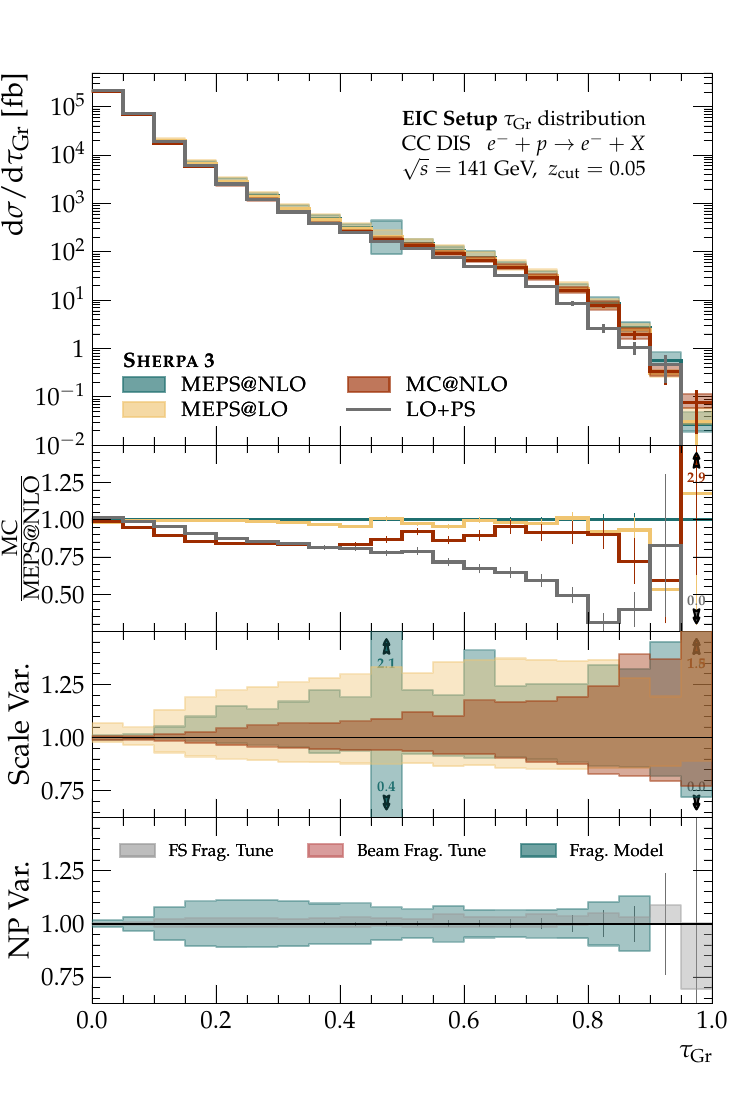} &
      \includegraphics[width=0.333\linewidth]{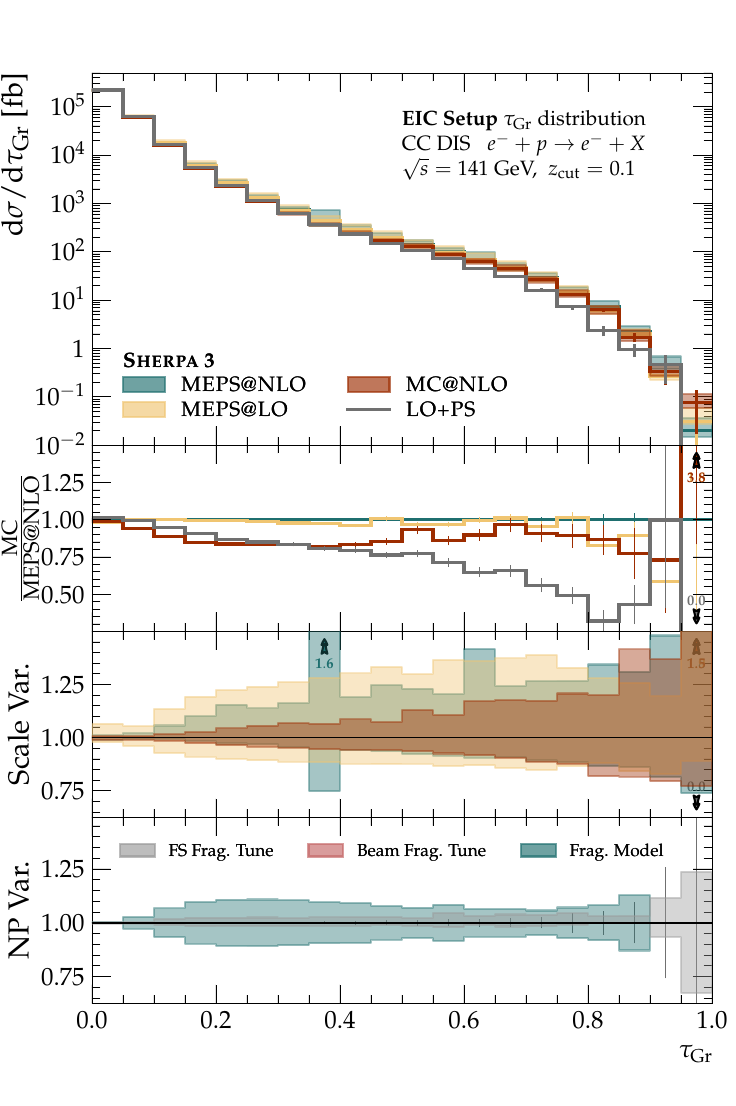} &
      \includegraphics[width=0.333\linewidth]{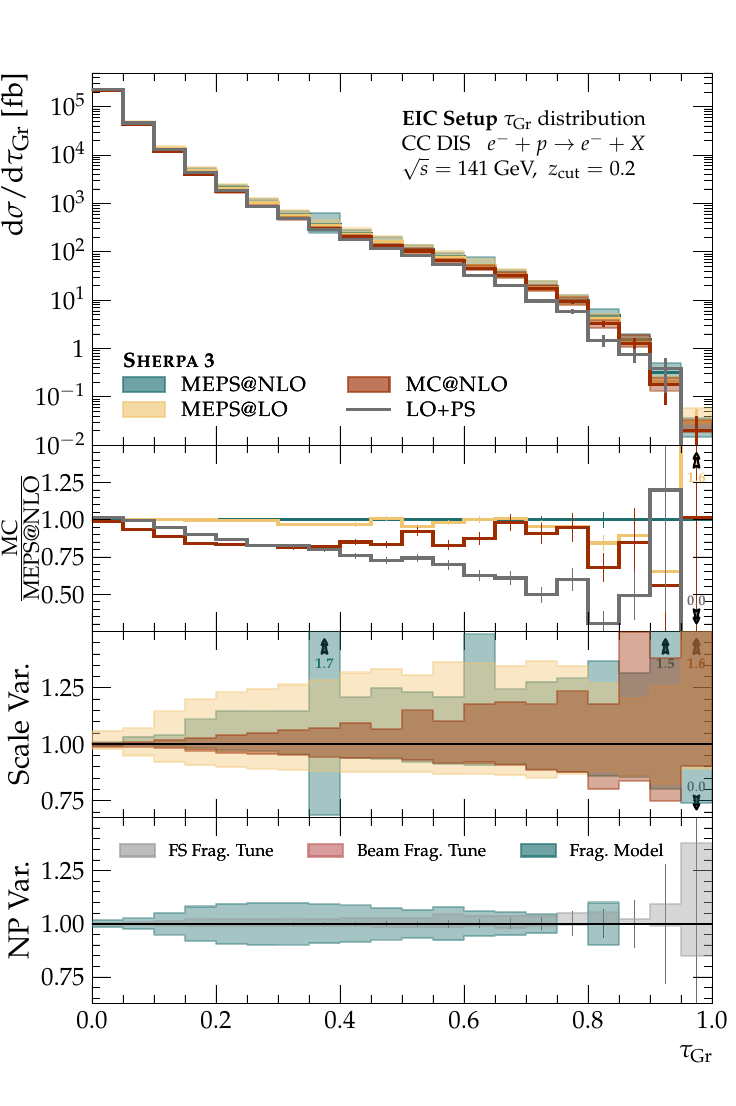}
  \end{tabular}
  }
  \caption{Groomed 1-jettiness in CC DIS at the \protect\EIC. The bottom ratios in each row
    indicate, from top to bottom, the ratio of each prediction to the
    \MEPSatNLO one, the uncertainty from scale variations and the uncertainty
    from nonperturbative model and tuning variations. The error bars in the
    lowest panel indicate the statistical uncertainty of the difference between the
    cluster and string model. \label{fig:cc-eic:event-shapes}}
\end{figure}

In this Appendix we include the equivalent of Fig.~\ref{fig:nc-eic:groomed-tau},
presenting 1-jettiness after soft-drop grooming for CC DIS, for
completeness.  These plots are shown in
Fig.~\ref{fig:cc-eic:event-shapes}.

\clearpage
\bibliographystyle{amsunsrt_mod}
\bibliography{journal,extra}

\end{document}